\documentclass[iop]{emulateapj}
\usepackage{array}
\usepackage{booktabs}
\usepackage{tabularx}
\usepackage{ragged2e}
\usepackage{lineno}
\usepackage{blindtext}
\usepackage{booktabs}
\usepackage{longtable}
\usepackage[referable]{threeparttablex}

\usepackage{longtable,tabu}
\usepackage{booktabs}

\usepackage{amsmath}
\usepackage{longtable}
\usepackage{tabularx}
\usepackage{amsmath}
\usepackage{array}
\usepackage{booktabs}
\usepackage{ragged2e}
\usepackage{lineno}
\usepackage{graphicx}
\newcommand\textlcsc[1]{\textsc{\MakeLowercase{#1}}}
\usepackage{natbib}
\usepackage{float}
\usepackage{tablefootnote}
\usepackage{apjfonts}
\usepackage{lipsum}

\newcommand\blfootnote[1]{%
  \begingroup
  \renewcommand\thefootnote{}\footnote{#1}%
  \addtocounter{footnote}{-1}%
  \endgroup
}
\usepackage[colorlinks=true, citecolor=blue, linkcolor=blue, urlcolor=blue]{hyperref}

\usepackage[T1]{fontenc}
\usepackage{beramono}
\usepackage{listings}
\received{}
\accepted{}
\shorttitle{HI Spectroscopy of RM AGN}
\shortauthors{Robinson, et al.}

\begin{document}
\title{HI SPECTROSCOPY OF REVERBERATION-MAPPED ACTIVE GALACTIC NUCLEI}

%%% Add names accordingly! Alphabetical order? %%%
\author{Justin H. Robinson$^{1}$}
\author{Misty C. Bentz$^{1}$}
\author{Megan C. Johnson$^{2}$}
\author{H\'el\`ene M. Courtois$^{3}$}
\author{Benjamin Ou-Yang$^{1}$}

\affiliation{\normalfont \centering $^{1}$Department of Physics and Astronomy, Georgia State University, Atlanta, GA 30303, USA; \url{jrob@astro.gsu.edu}}

\affiliation{\normalfont \centering $^{2}$United States Naval Observatory (USNO) 3450 Massachusetts Ave NW, Washington, DC 20392, USA}

\affiliation{\normalfont \centering $^{3}$University of Lyon; UCB Lyon 1/CNRS/IN2P3; IPN Lyon, France}

%% Mark off the abstract in the ``abstract'' environment. 
\begin{abstract}
We present HI 21 cm spectroscopy from the GBT for the host galaxies of 31 nearby AGNs with direct M$_{\textsc{BH}}$ measurements from reverberation mapping. These are the first published HI detections for 12 galaxies, and the spectral quality is generally an improvement over archival data for the remainder of the sample. We present measurements of emission-line fluxes, velocity widths, and recessional velocities from which we derive HI mass, total gas mass, and redshifts. Combining M$_{\textsc{GAS}}$ with constraints on M$_{\textsc{stars}}$ allows exploration of the baryonic content of these galaxies. We find a typical M$_{\textsc{GAS}}$/M$_{\textsc{stars}}$ fraction of 10\%, with a few reaching $\sim$30-50\%. We also examined several relationships between M$_{\textsc{stars}}$, M$_{\textsc{GAS}}$, M$_{\textsc{BH}}$, baryonic mass, and morphological type. We find a weak preference for galaxies with larger M$_{\textsc{GAS}}$ to host more massive black holes. We also find gas-to-stellar fractions to weakly correlate with later types in unbarred spirals, with an approximately constant fraction for barred spirals. Consistent with previous studies, we find declining M$_{\textsc{GAS}}$/M$_{\textsc{stars}}$ with increasing M$_{\textsc{stars}}$, with a slope suggesting the gas reservoirs have been replenished. Finally, we find a clear relationship for M$_{\textsc{BH}}$-M$_{\textsc{BARY}}$ with a similar slope as M$_{\textsc{BH}}$-M$_{\textsc{stars}}$ reported by \cite{misty2018}. The dwarf Seyfert NGC 4395 appears to follow this relationship as well, even though it has a significantly higher gas fraction and smaller M$_{\textsc{BH}}$ than the remainder of our sample.
\end{abstract}

%% Keywords should appear after the \end{abstract} command. 
%% See the online documentation for the full list of available subject
%% keywords and the rules for their use.

%% From the front matter, we move on to the body of the paper.
%% Sections are demarcated by \section and \subsection, respectively.
%% Observe the use of the LaTeX \label
%% command after the \subsection to give a symbolic KEY to the
%% subsection for cross-referencing in a \ref command.
%% You can use LaTeX's \ref and \label commands to keep track of
%% cross-references to sections, equations, tables, and figures.
%% That way, if you change the order of any elements, LaTeX will
%% automatically renumber them.

%% We recommend that authors also use the natbib \citep
%% and \citet commands to identify citations.  The citations are
%% tied to the reference list via symbolic KEYs. The KEY corresponds
%% to the KEY in the \bibitem in the reference list below. 

% Introduction
\section{Introduction} \label{sec:intro}
Hydrogen is the most abundant element in the universe \citep{h_abundance}, and is of fundamental importance in galactic and extragalactic studies. The spin-flip transition of electrons in neutral hydrogen (HI) atoms gives rise to the hyperfine 21\,cm line radiation, which is easily detected from gas-rich galaxies (usually late-type galaxies, e.g.\ \citealt{hg1984}, and references therein).

The HI emission line provides a number of interesting details about the host galaxy. First, the Doppler-shifted recessional velocity yields one of the most reliable redshift measurements of extragalactic sources. Since the neutral gas is spread throughout the galaxy, it follows that the center velocity of the emission profile acts as a systemic velocity indicator. HI 21\,cm emitting gas is also cold ($\sim$ 120\,K), and reflects the overall motion of the disk as opposed to gas at hotter temperatures ($\sim$ 10,000\,K) emitting in the optical (e.g.\ [O III]; \citealt{op1987}). These higher temperature emission lines can be affected by the internal motion of the regions in which they emit, thus affecting the radial velocity and hence the redshift measurement (e.g., nuclear, optical emission lines reflecting net outflow motion; \citealt{mw1984}).

Secondly, for inclined disk galaxies, the HI line width provides insight into their rotation speeds. The integrated emission profile is based on the distribution of radial velocities of the rotating disk, and correction for the disk's inclination provides a constraint on the maximum rotation rate \citep{ft1977}. Inclination-corrected widths of observed HI profiles are thus related to the rotation curves of disk galaxies \citep{roberts1969,epstein1964}. 

Finally, the total area under the integrated HI line provides an estimate of the total atomic gas content. For galaxies with angular extents smaller than the beam size of the telescope, the integrated HI flux is related to the total number of hydrogen atoms, and thus the mass in atomic hydrogen (M$_\textsc{{HI}}$; \citealt{HImass}). Atomic hydrogen is normally the dominant gas phase in disk galaxies, with molecular hydrogen (H$_{2}$) as the next significant component. HI has been observed to saturate and condense to H$_{2}$ above a threshold surface density of $\sim$ 10\,M$_\odot$\,pc$^{-2}$ \citep{mk2001,wb2002,bigiel2008}, and giant molecular clouds are the dominant locations for star formation in spiral galaxies (e.g., \citealt{leroy2008}). There have been many studies aimed at estimating the molecular gas content of disk galaxies (e.g., \citealt{cortese2017}), for example showing that M${_{\textsc{H}_{2}}}$/M$_\textsc{{HI}}$ scales as a function of morphology \citep{yk1989,md1997}. There is significant scatter in all of the distributions from \cite{yk1989}, but the mean of their result for late-type spirals is M${_{\textsc{H}_{2}}}$/M$_\textsc{{HI}}$ $\sim$ $0.2\pm0.1$. Constraints on the solar helium abundance, the next most abundant element and significant gas mass contributor, place the  M${_{\textsc{H}_\textsc{{E}}}}$/M$_\textsc{{HI}}$ fraction in the range of 0.274 $\pm$ 6\% \citep{cox2000}. The typical cosmic abundances of other elements such as carbon, nitrogen, and oxygen are only small fractions of the hydrogen abundance in spiral galaxies, including the Milky Way (e.g., \citealt{spitzer,obreschkow2009}). Yet, the mass contribution from non-HI gas is typically less than the uncertainty involved in constraining M$_\textsc{{HI}}$.  So the total gas mass of a galaxy (M$_\textsc{{GAS}}$) is often estimated by simply applying a scale factor to M$_\textsc{{HI}}$ to account for these contributions, the vast majority of which is helium (\citealt{mcgaugh2012}, and references therein).

The first significant HI study of galaxies hosting an active galactic nucleus (AGN) was an exploration of the relationship between the disk and the nucleus of Seyfert galaxies by \cite{heckman1978}. That initial study hinted that the host galaxies of AGNs have a relationship between UV excess outside of the nucleus and the ratio of atomic gas to galaxy luminosity (M$_\textsc{{HI}}$/L; luminosities from \citealt{heckman1978} are derived from the B-band magnitudes from \citealt{dv1976}), perhaps implying that feedback from nuclear activity triggers star formation in the larger galaxy disk. \cite{heckman1978} also mention the tendency for Seyferts in their study with peculiar HI properties (e.g., HI absorption, abnormal M$_{\textsc{HI}}$/L) to have peculiar morphological characteristics (e.g., double nucleus, one spiral arm, faint disk). \cite{pt1998} conducted an HI kinematic study near the active core of NGC 3984, finding that all the HI components were redshifted with respect to the stellar content of the galaxy, which they interpreted as the signature of central parsec-scale gas infalling and feeding the nucleus. \cite{fabello2011} used the Arecibo\footnote{The Arecibo Observatory is part of the National Astronomy and Ionosphere Center, which is operated by Cornell University under a cooperative agreement with the National Science Foundation.} Legacy Fast ALFA (ALFALFA) Survey \citep{alfalfa} to search for trends in the fraction of M$_{\textsc{HI}}$ to stellar mass (M$_{\textsc{stars}}$) and black hole accretion rate. For galaxies with low star formation rates (log\,SFR\,/\,M$_{\textsc{stars}}$\,<\,-\,11.0), the accretion rate scaled with increasing M$_{\textsc{HI}}$/M$_{\textsc{stars}}$. 
\begin{figure*}%\vspace{-0.1cm}
\gridline{\includegraphics[trim={2cm 12.5cm 2cm 1.8cm},clip,scale=0.25]{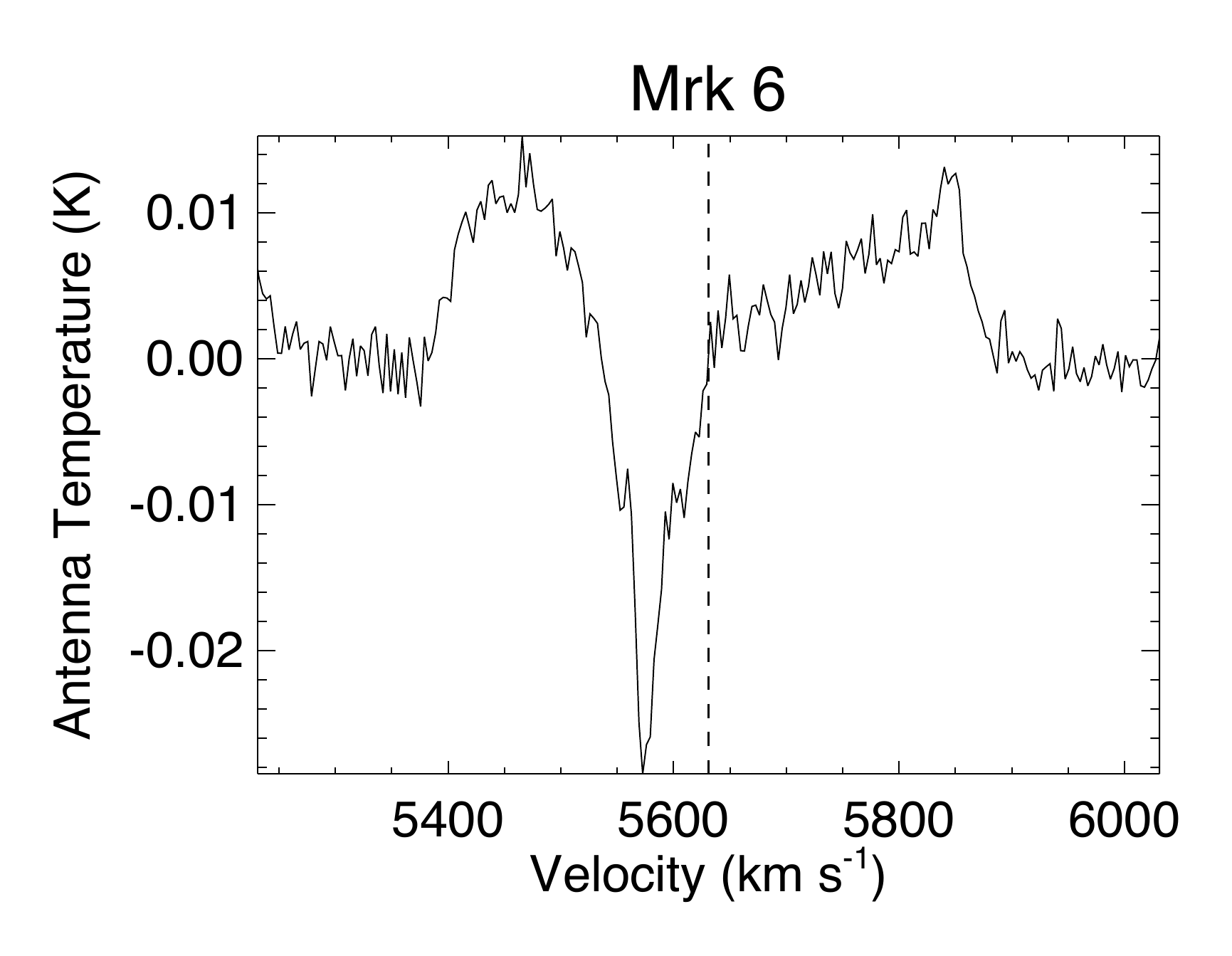}
\includegraphics[trim={2cm 12.5cm 2cm 1.8cm},clip,scale=0.25]{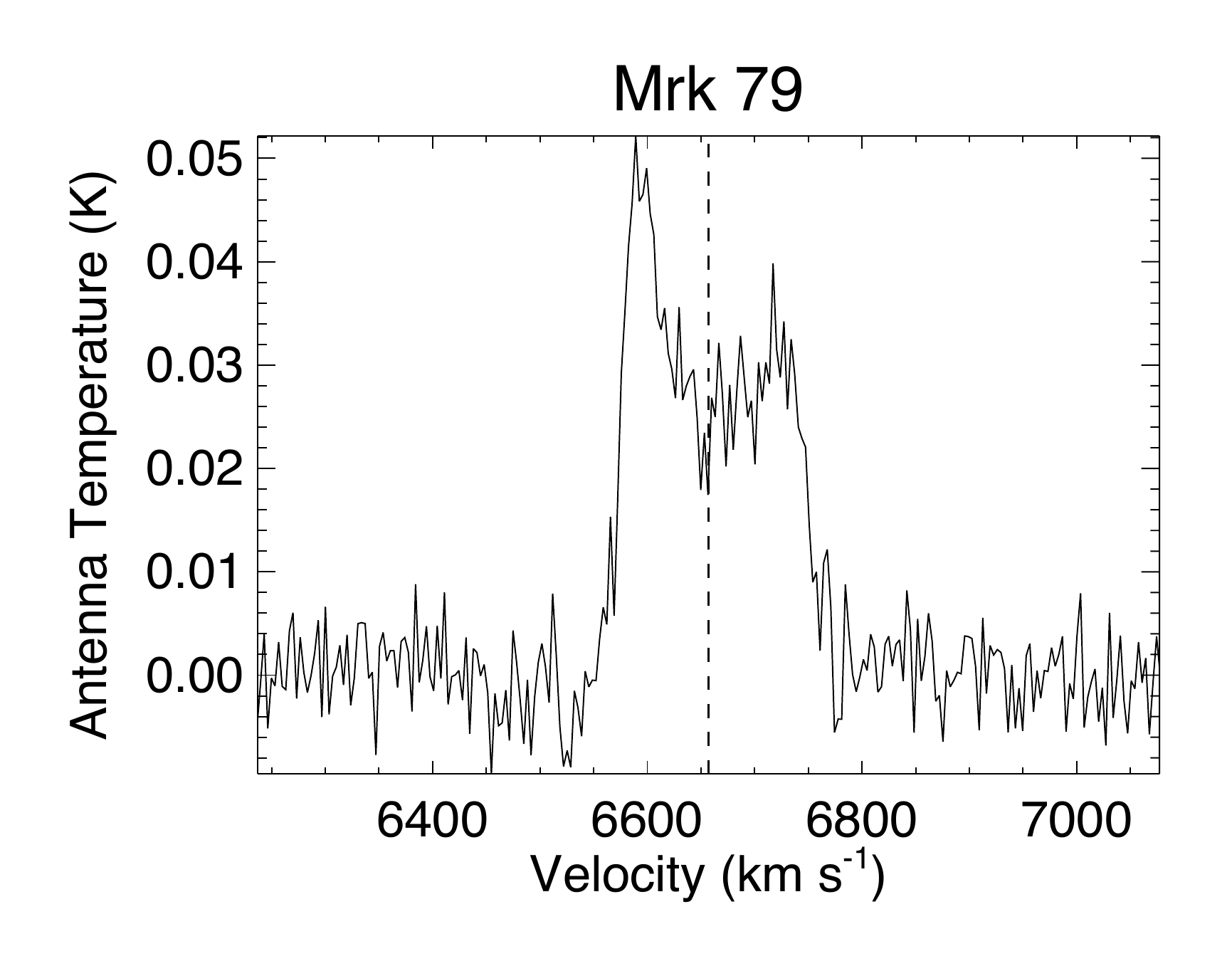}
\includegraphics[trim={2cm 12.5cm 2cm 1.8cm},clip,scale=0.25]{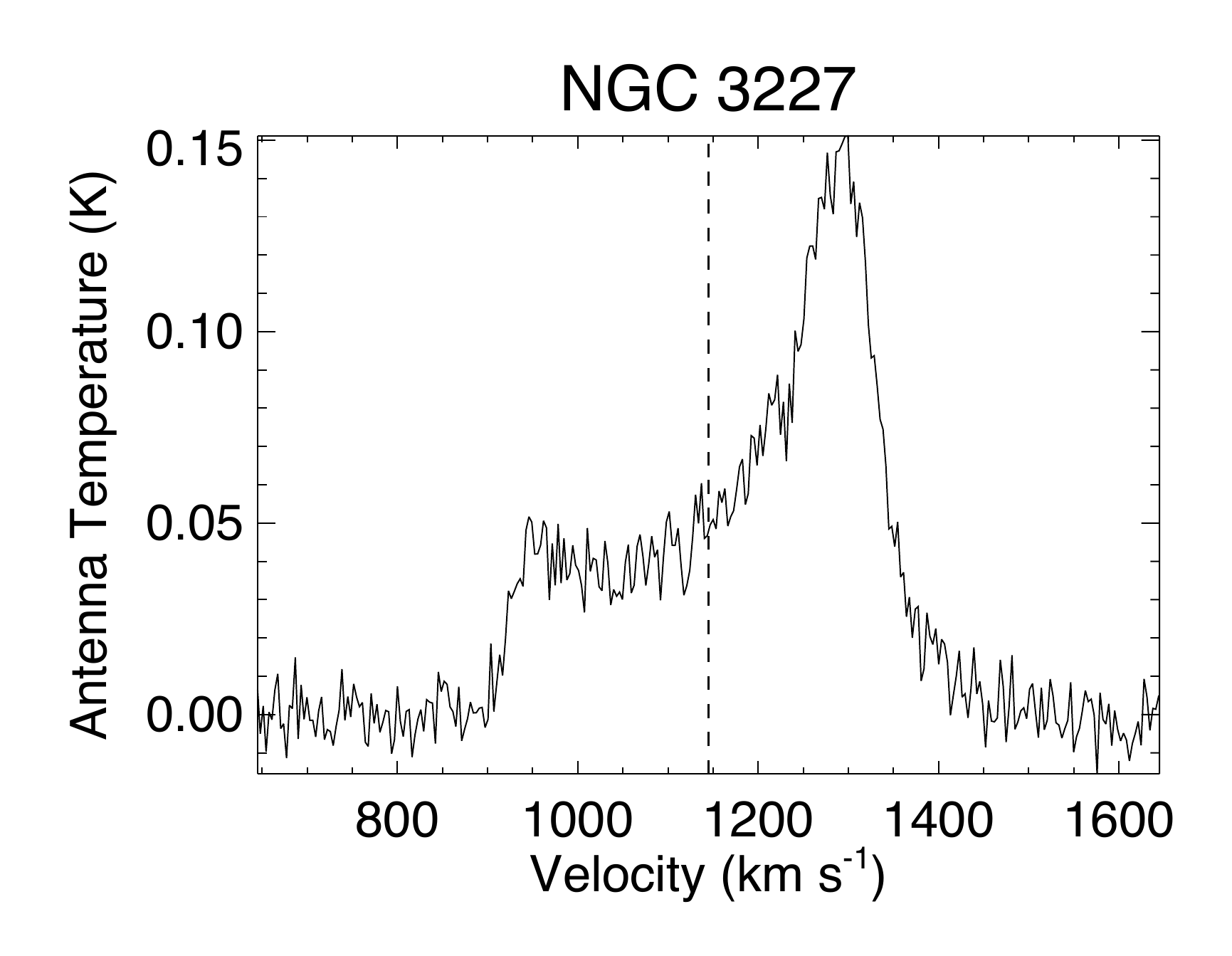}
\includegraphics[trim={2cm 12.5cm 2cm 1.8cm},clip,scale=0.25]{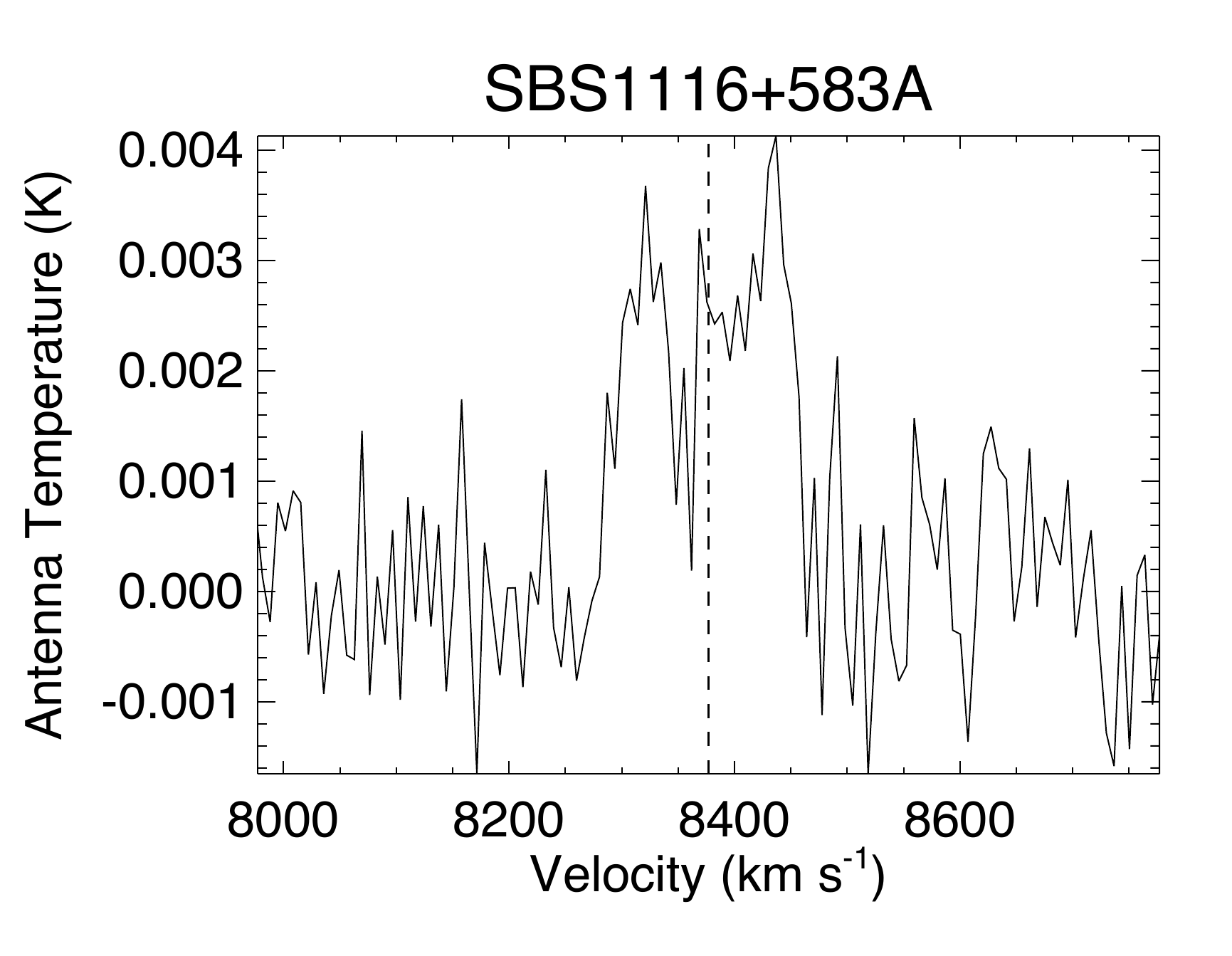}
}
\gridline{\includegraphics[trim={2cm 12.5cm 2cm 2cm},clip,scale=0.25]{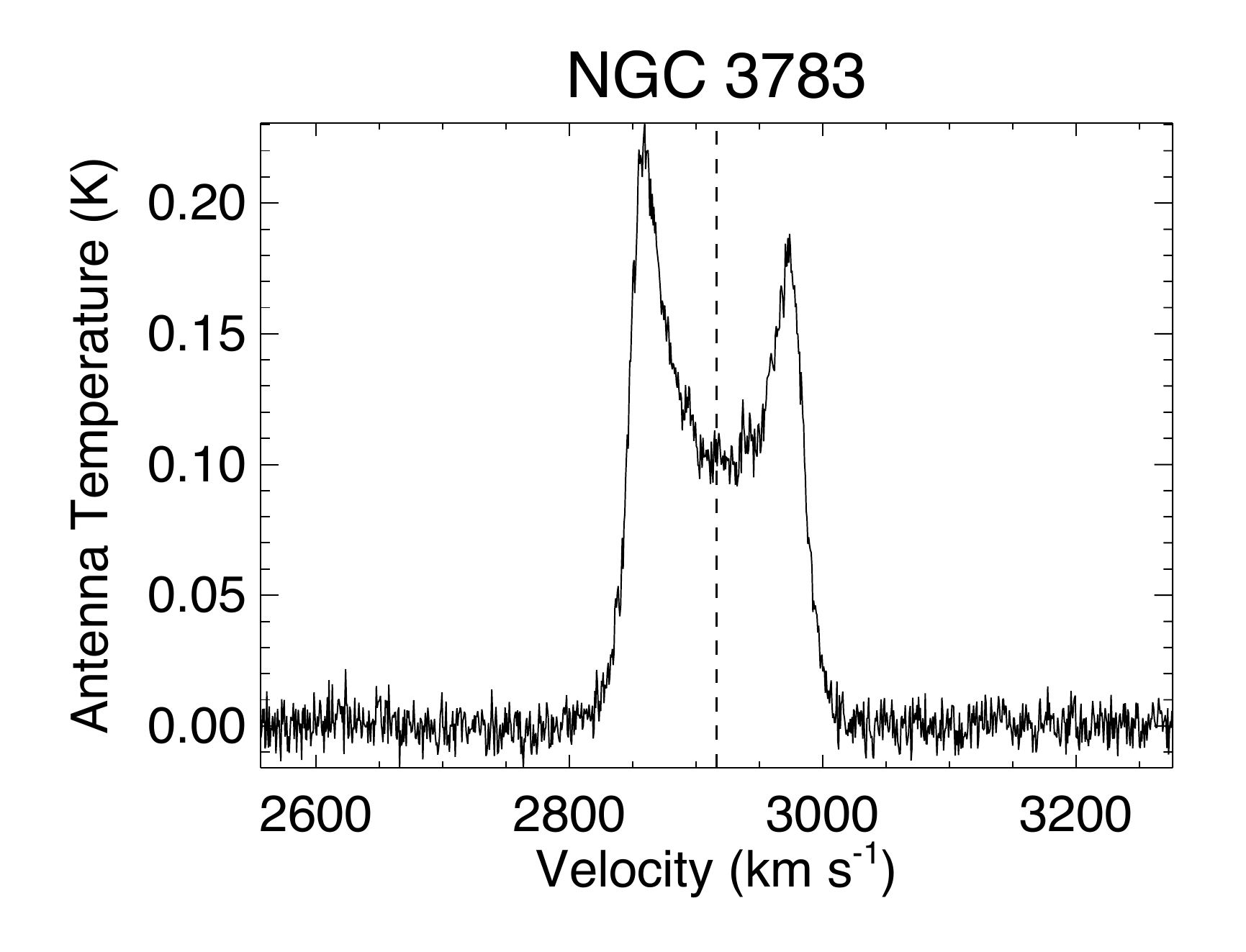}
\includegraphics[trim={2cm 12.5cm 2cm 2cm},clip,scale=0.25]{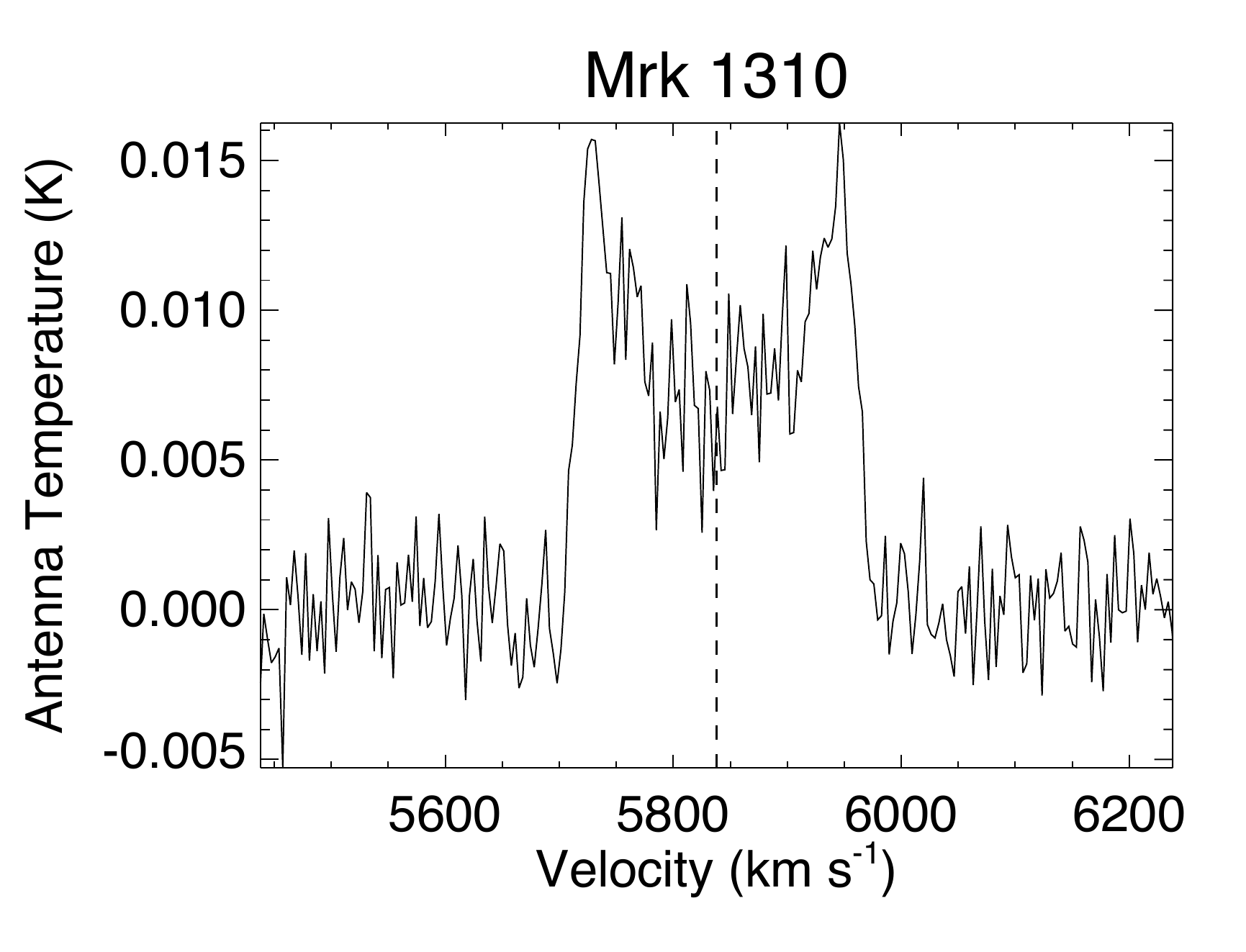}
\includegraphics[trim={2cm 12.5cm 2cm 2cm},clip,scale=0.25]{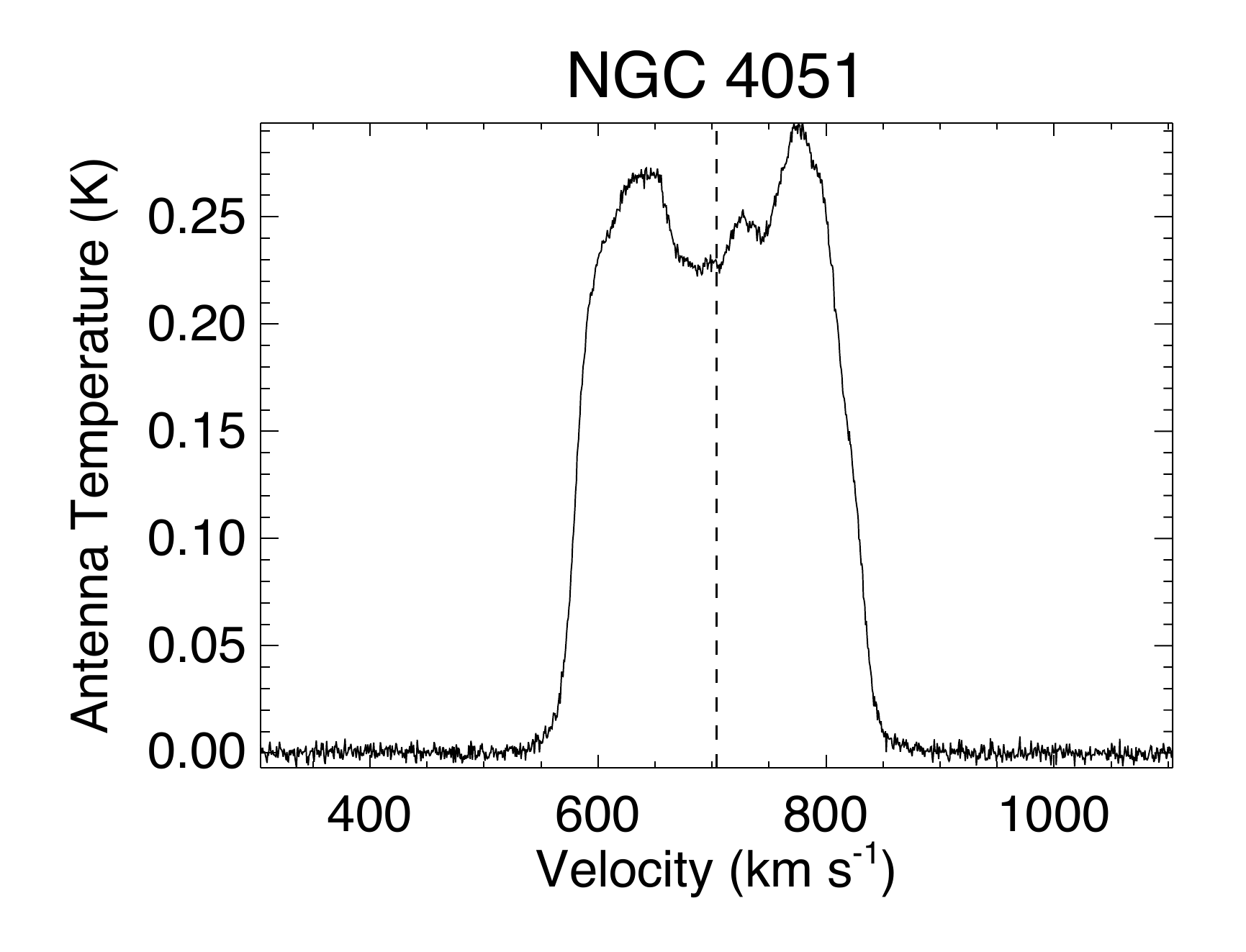}
\includegraphics[trim={2cm 12.5cm 2cm 2cm},clip,scale=0.25]{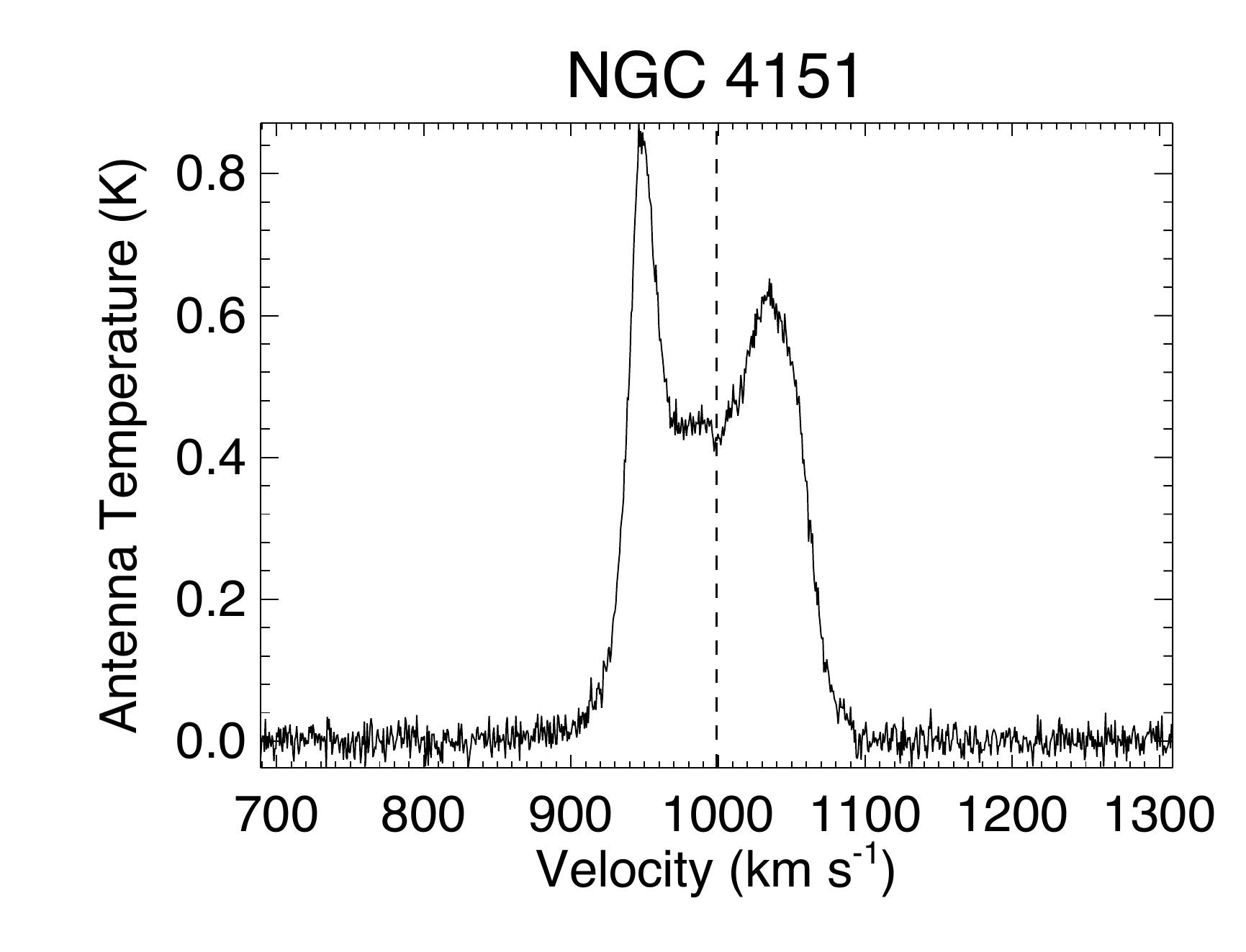}
}
\gridline{\includegraphics[trim={2cm 12.5cm 2cm 2cm},clip,scale=0.25]{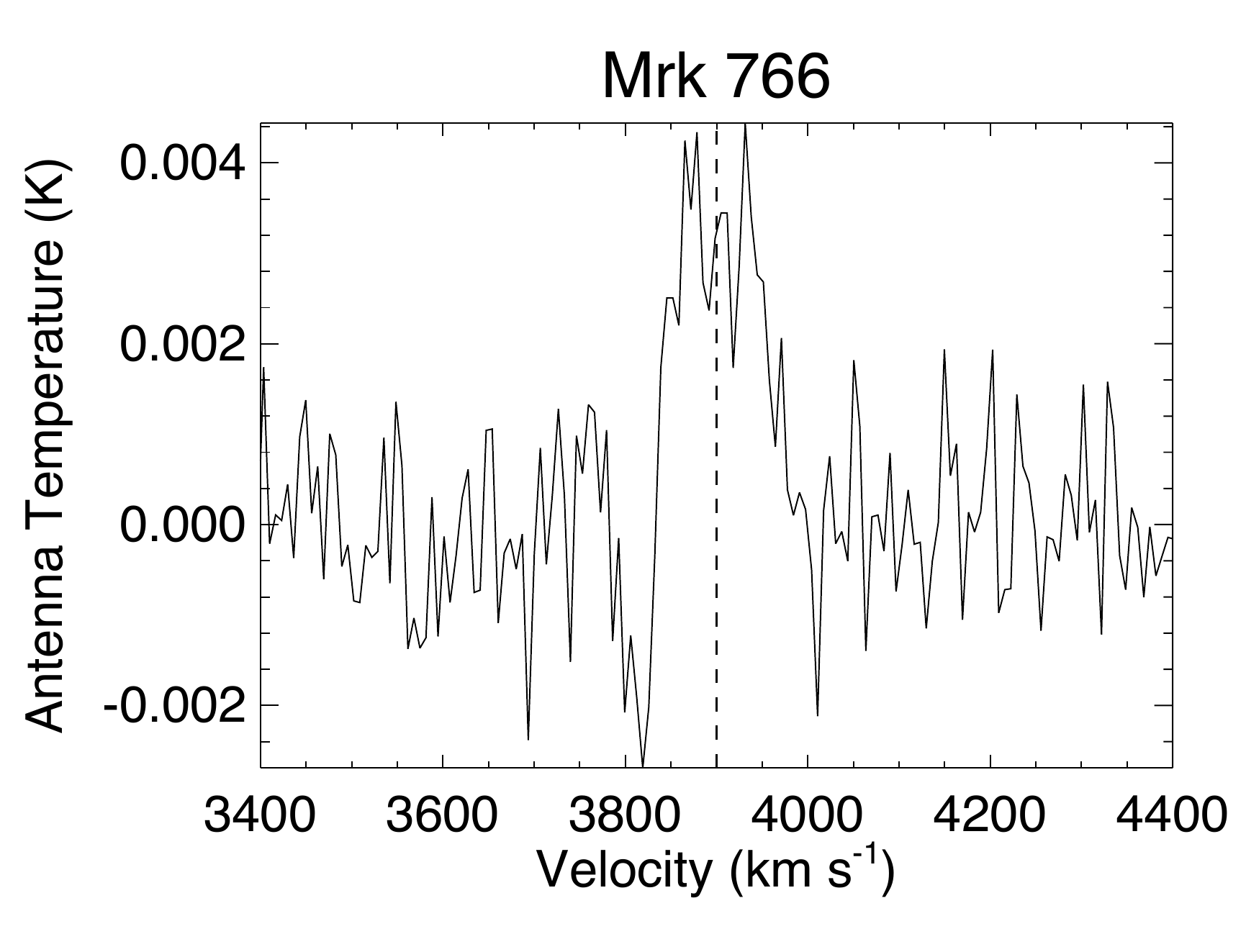}
\includegraphics[trim={2cm 12.5cm 2cm 2cm},clip,scale=0.25]{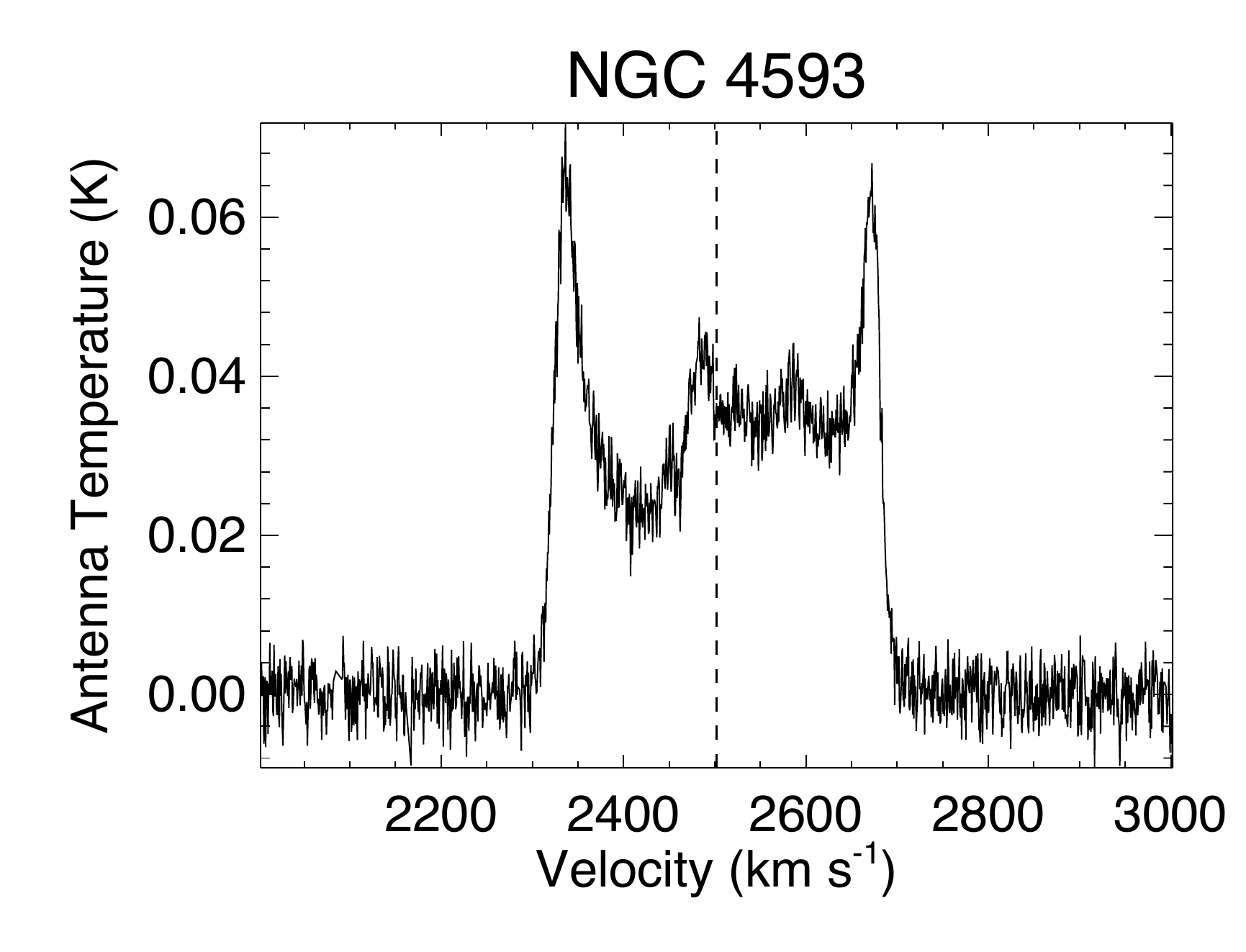}
\includegraphics[trim={2cm 12.5cm 2cm 2cm},clip,scale=0.25]{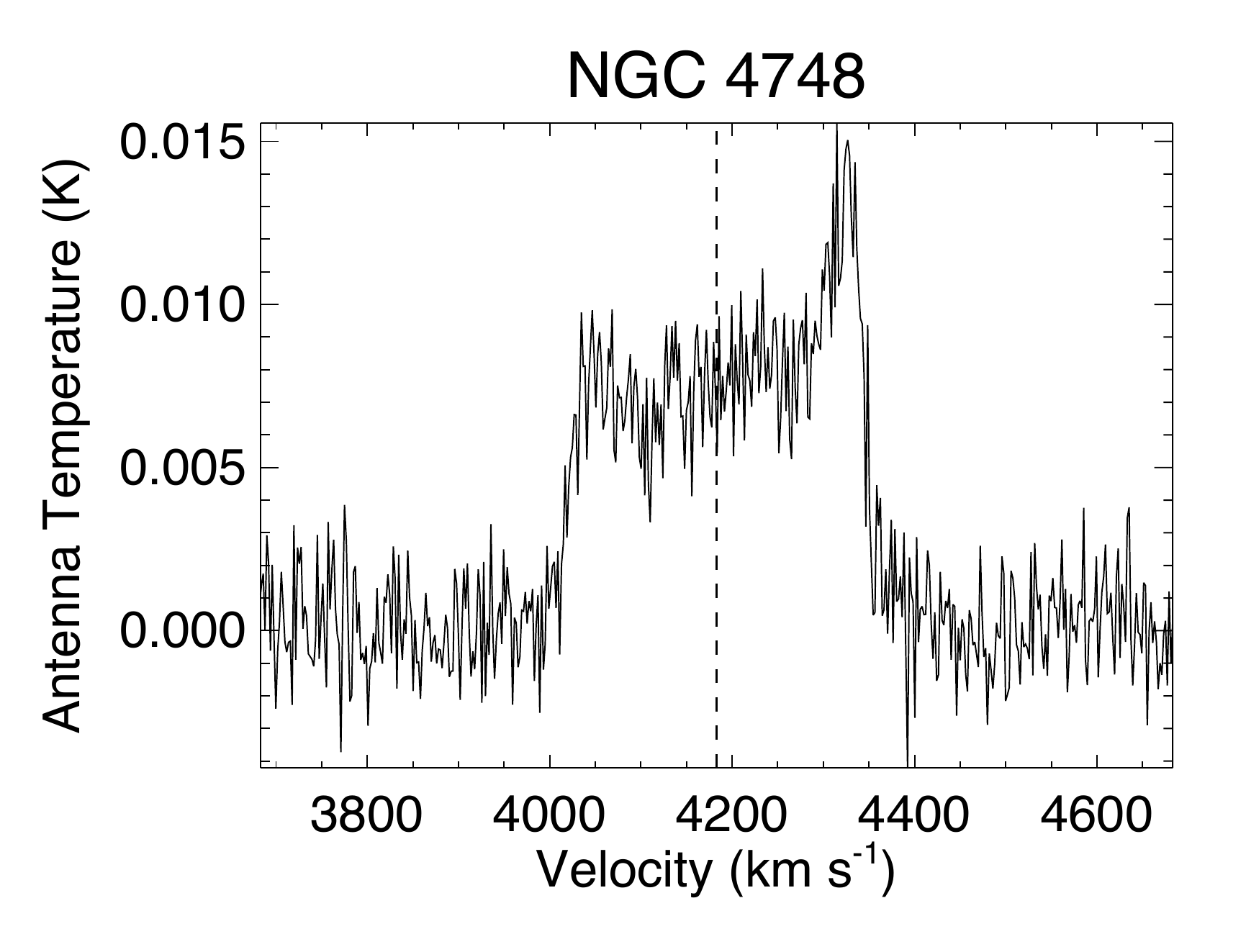}
\includegraphics[trim={2cm 12.5cm 2cm 2cm},clip,scale=0.25]{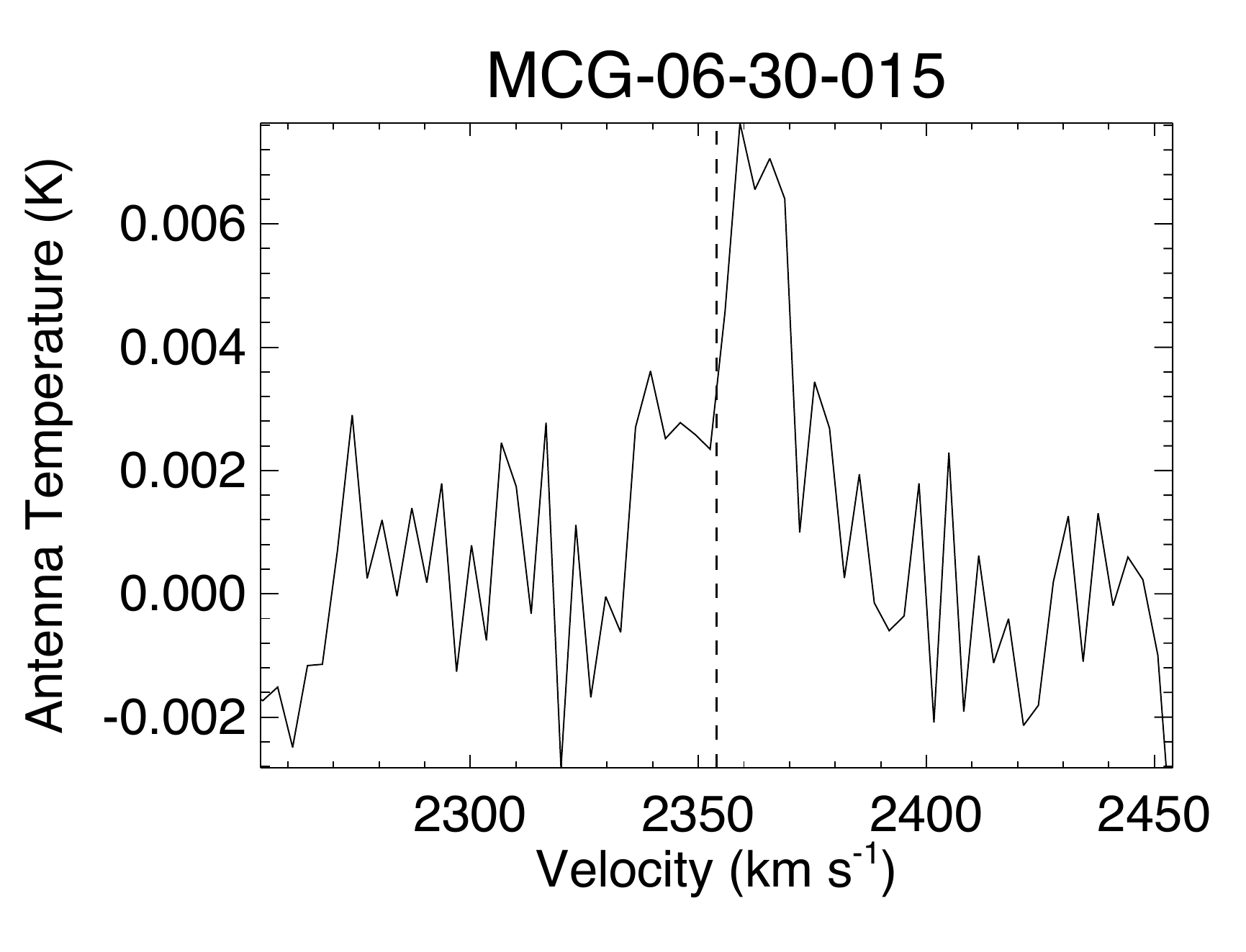}
}
\gridline{\includegraphics[trim={2cm 12.5cm 2cm 2cm},clip,scale=0.25]{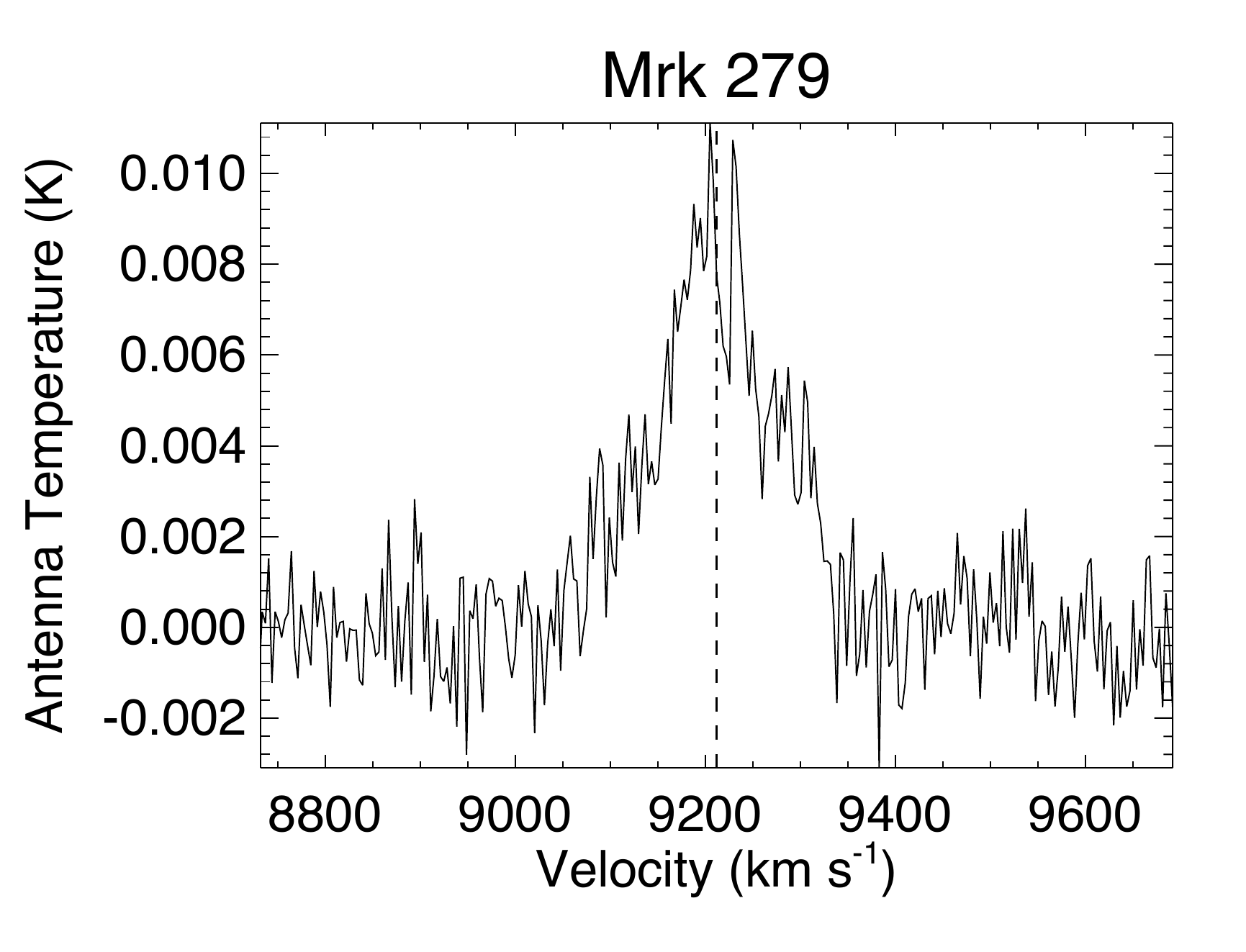}
\includegraphics[trim={2cm 12.5cm 2cm 2cm},clip,scale=0.25]{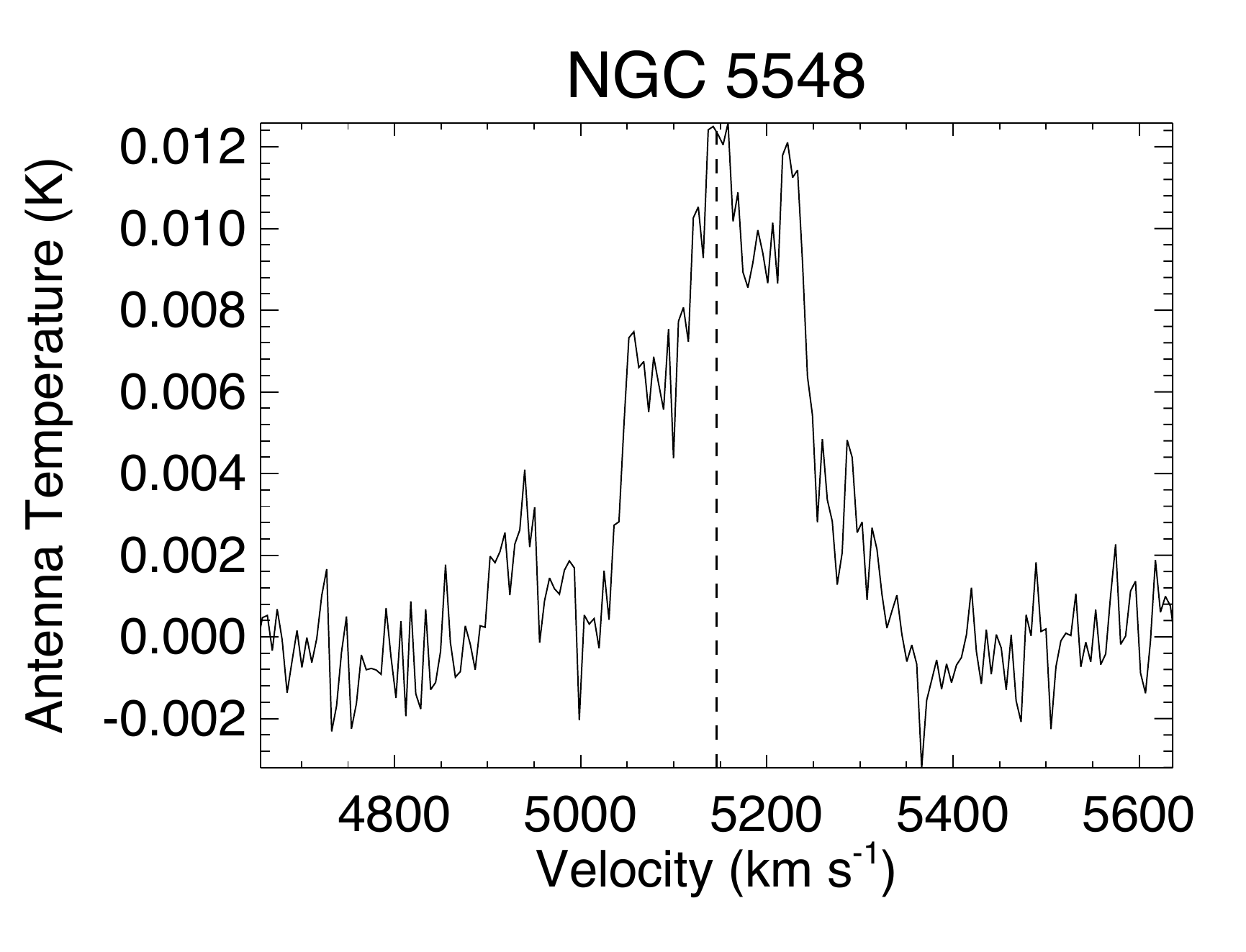}
\includegraphics[trim={2cm 12.5cm 2cm 2cm},clip,scale=0.25]{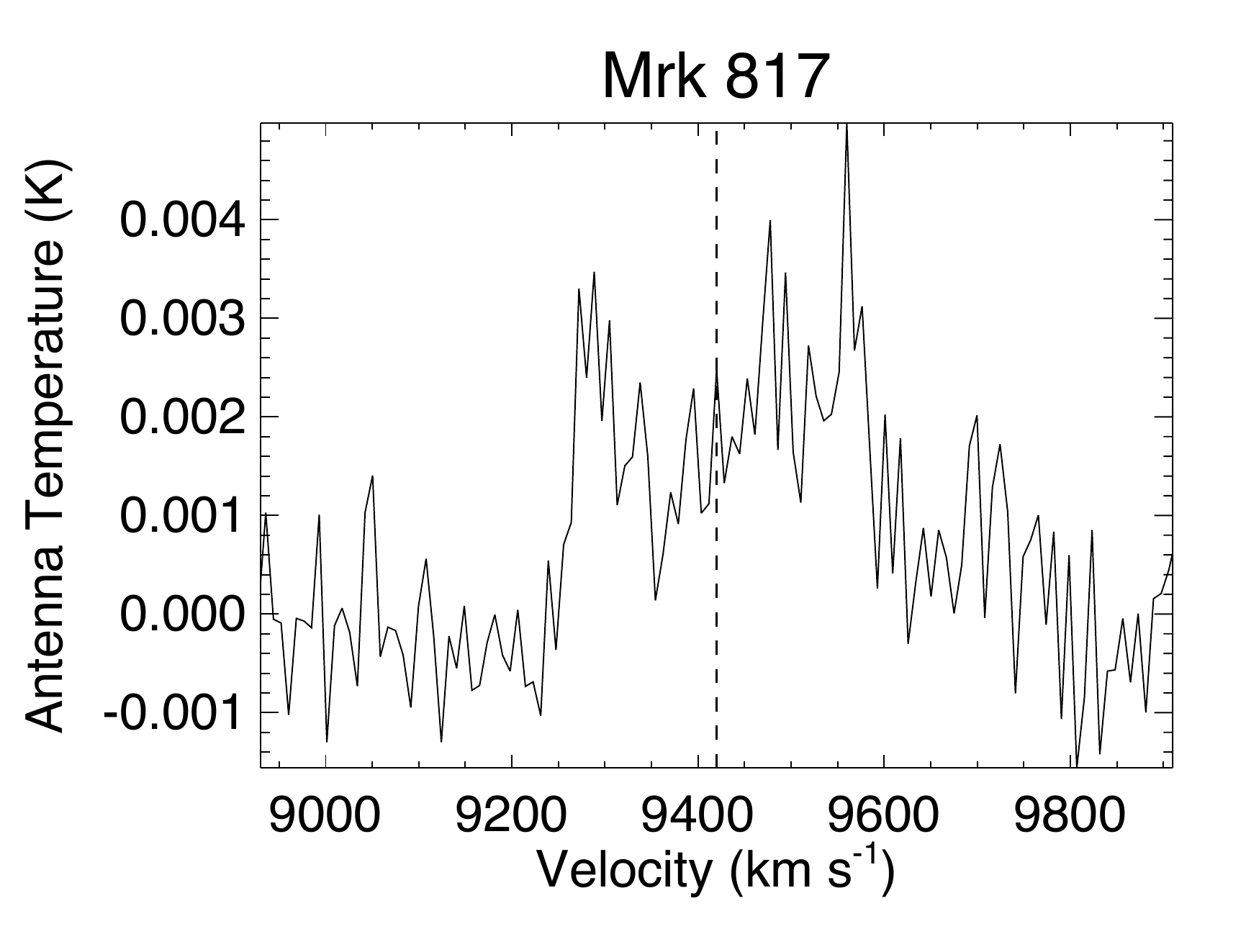}
\includegraphics[trim={2cm 12.5cm 2cm 2cm},clip,scale=0.25]{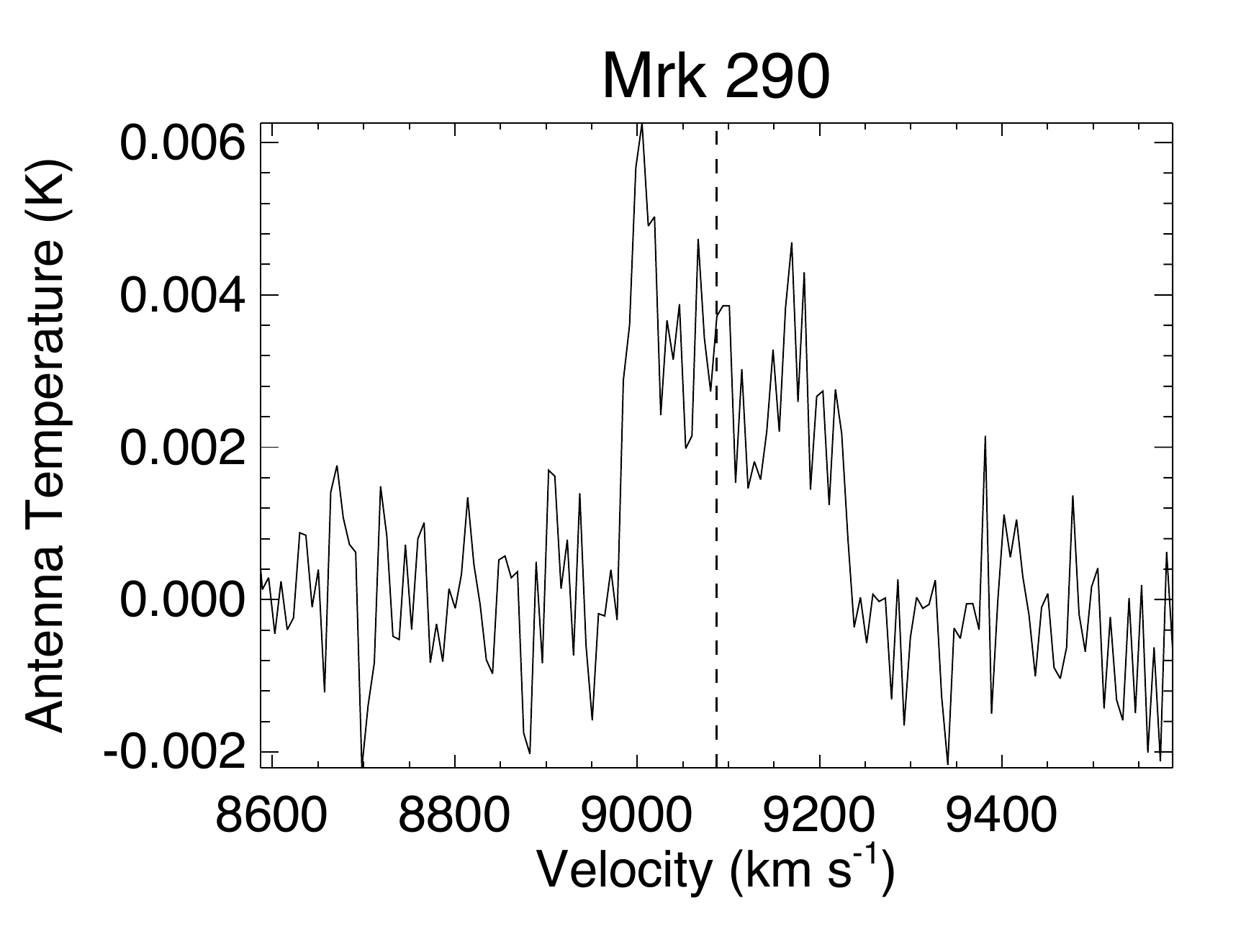}
}
\gridline{\hspace{3cm}\includegraphics[trim={2cm 12.5cm 2cm 2cm},clip,scale=0.25]{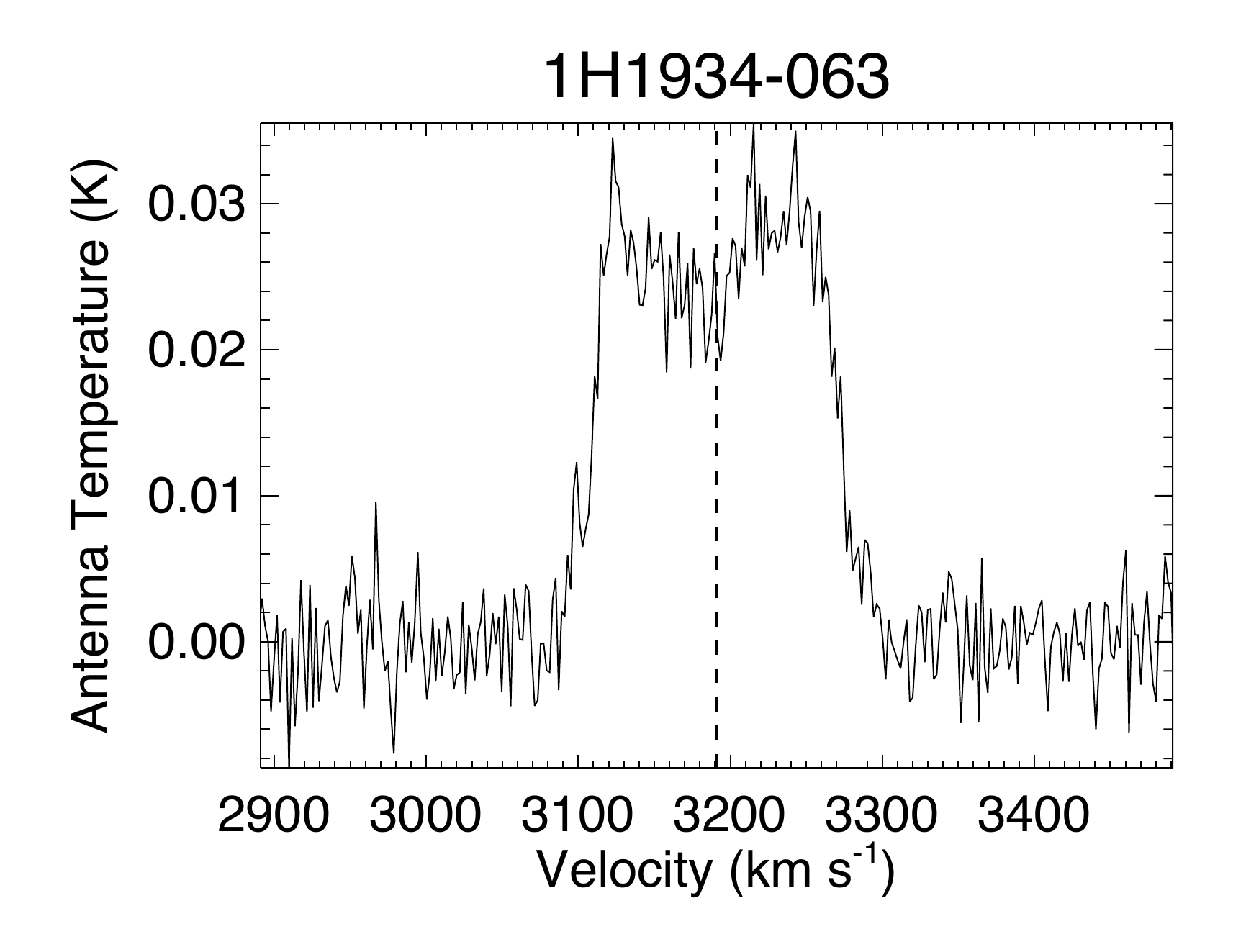}
\includegraphics[trim={2cm 12.5cm 2cm 2cm},clip,scale=0.25]{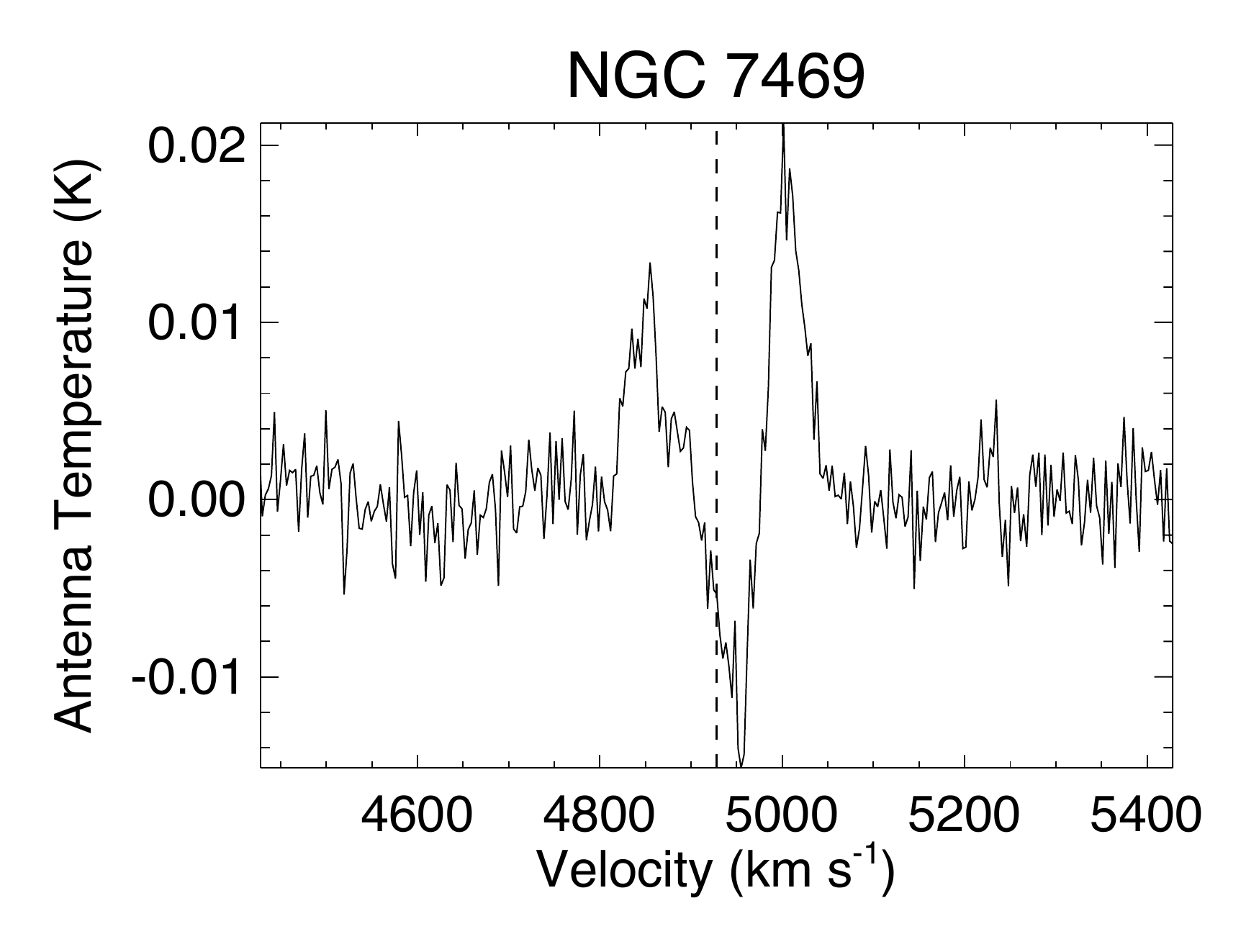}
}
\caption{HI emission line spectra from GBT13A-468 after reduction, baseline subtraction, and smoothing with $\textsc{GBTIDL}$ v2.8. Hanning smoothing was applied to all profiles, and further smoothing was dependent on the S/N (see Sec.\ \ref{red}). Note that Mrk 6 and NGC 7469 exhibit central absorption features. The vertical dashed lines indicate V$_\textsc{{R}}$ measurements from \texttt{gmeasure}.}\label{batmen}
\end{figure*}

Not all of the literature points to correlations between HI content and AGN activity, though. For galaxies with moderate star formation rates (log\,SFR\,/\,M$_{\textsc{stars}}$\,>\,-\,11.0) in the study by \cite{fabello2011}, no relationship was found between M$_{\textsc{HI}}$/M$_{\textsc{stars}}$ and accretion rate. \cite{bieging1983} conducted HI studies of active and interacting galaxies and compared their HI fluxes to the \cite{rieke1978} survey of 10.6\,$\mu$m emission from Seyfert nuclei and found no correlation. Their reasoning was that the infrared fluxes refer only to the nucleus as opposed to the HI flux which originates from the entire disk, therefore concluding no connection between gas and AGN luminosity. Finally, in their review of coevolution of black holes in AGNs and properties of their hosts, \cite{hb2014} conclude from a number of studies that HI is unlikely to reside within the central regions of AGN host galaxies. \cite{hb2014} also mention that HI surface density in all spiral galaxies, whether active or inactive, is usually lower or near zero at the center, and gas present near the central supermassive black hole (SMBH) is likely to be primarily molecular in the case of inactive galaxies, or ionized in the case of AGN. Thus, there seems to be no clear picture of how the overall gas content of a galaxy is related to AGN fueling.

However, we know that active galaxies have gas accreting onto their central SMBHs, and that the gas reservoir is large enough and/or replenished often enough to fuel the black hole for $\sim$ 10$^{7}$ years \citep{martini2004}, perhaps multiple times in the history of the galaxy. The growth of the SMBHs in AGNs via accretion also appears to be related to the growth of the bulges of their host galaxies (see reviews by \citealt{hb2014} and \citealt{kh2013}). The gas flows that fuel the accretion and growth of the SMBHs and bulges can be driven by mergers, with slower, gradual processes such as gravitational interactions with bar and spiral structures \citep{kk2004, athanassoula2008}, or rapid, gas-rich disk instabilities \citep{genzel2014, bournaud2010, elmegreen2008, dekel2009}. Furthermore, over the past two decades it has become clear that galaxies and SMBHs have a symbiotic relationship, even though their typical size scales are different by orders of magnitude. Empirical scaling relationships between the central SMBH mass (M$_\textsc{{BH}}$) and the host galaxy itself have been the subjects of many studies (e.g., \citealt{fm2000, gebhardt2000, kh2013}). It is therefore of interest to examine whether there is any relationship between M$_\textsc{{BH}}$ and the gas properties of their host galaxies.

\begin{figure*}
\gridline{\includegraphics[trim={2cm 12.5cm 2cm 2cm},clip,scale=0.25]{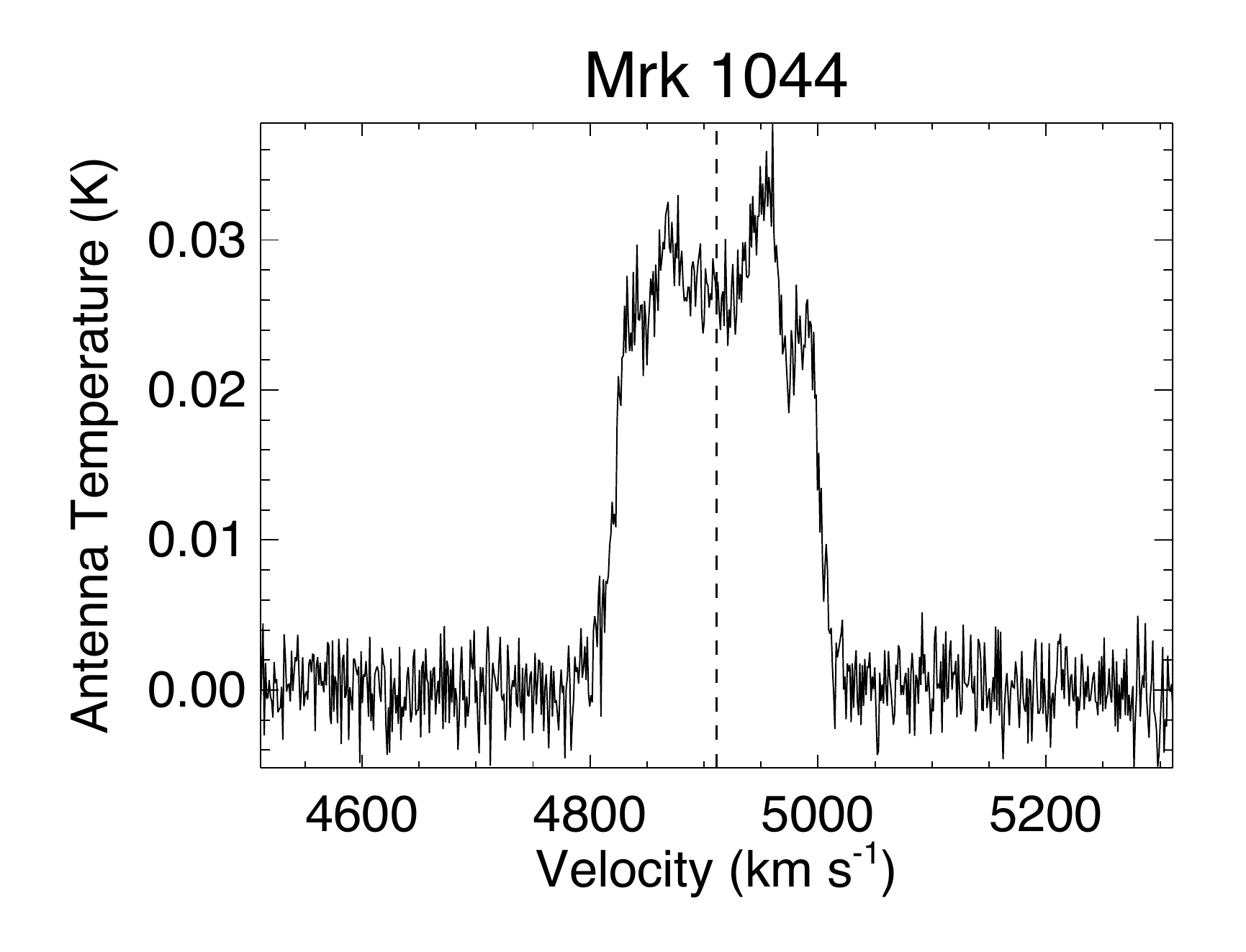}
\includegraphics[trim={2cm 12.5cm 2cm 2cm},clip,scale=0.25]{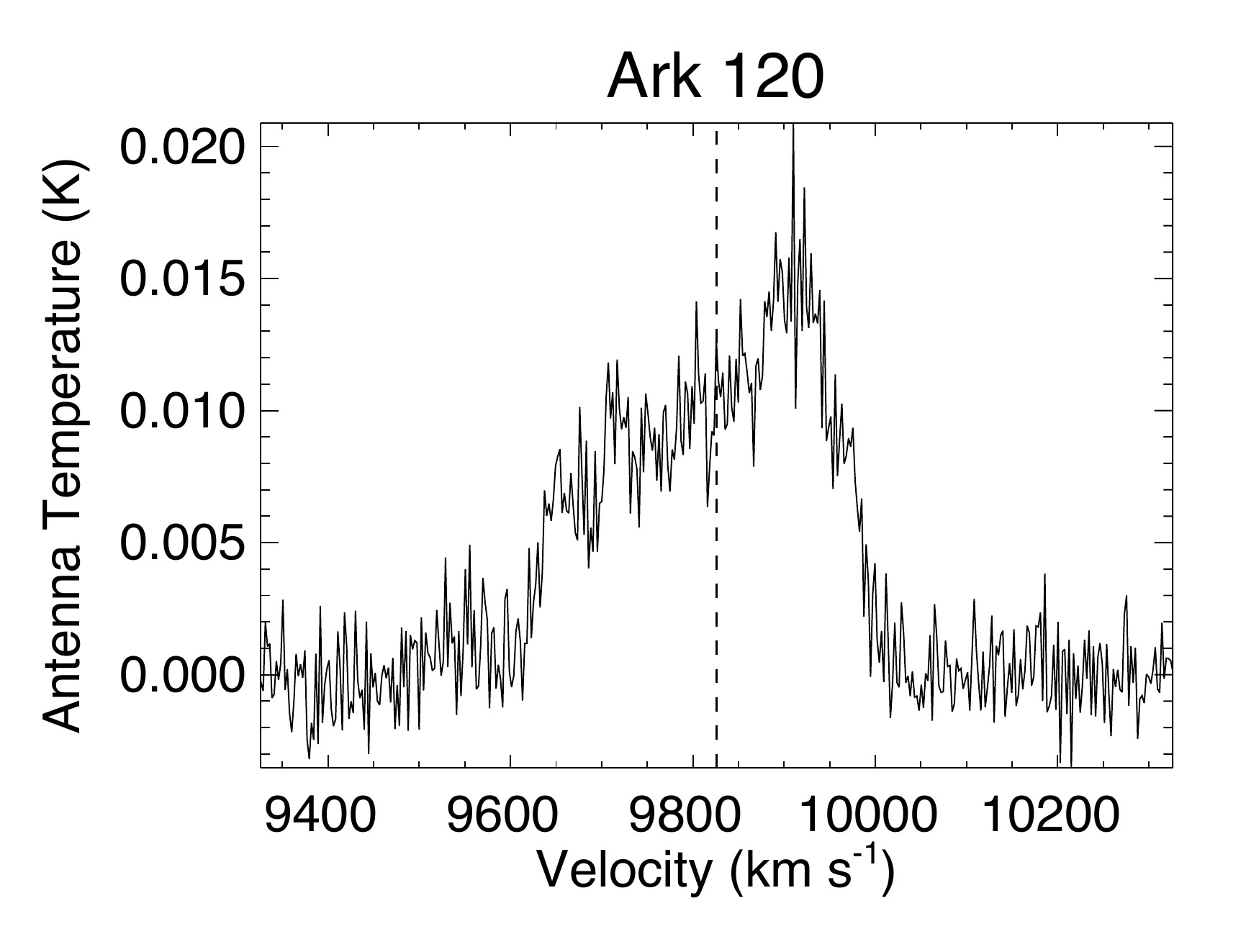}
\includegraphics[trim={2cm 12.5cm 2cm 2cm},clip,scale=0.25]{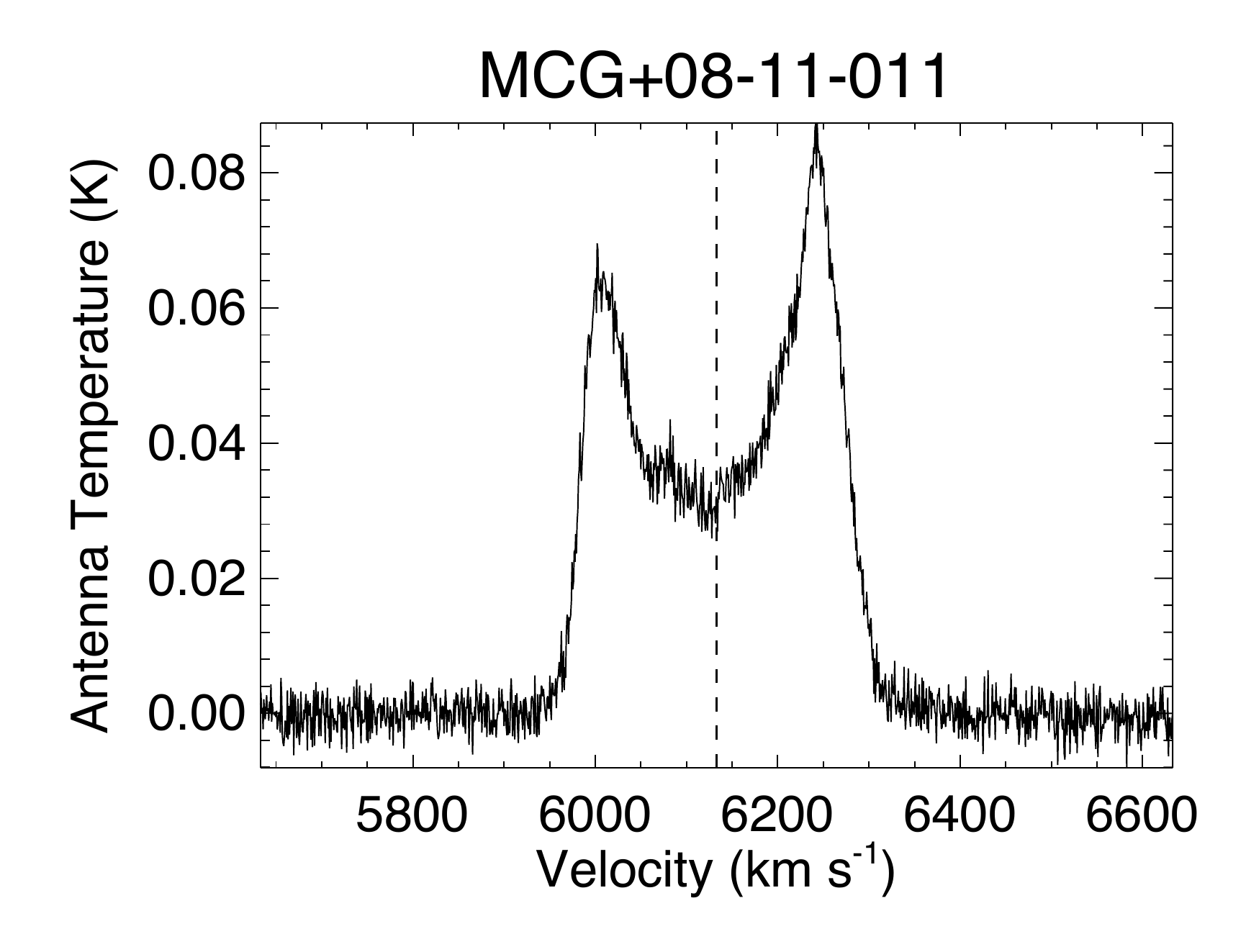}
\includegraphics[trim={2cm 12.5cm 2cm 2cm},clip,scale=0.25]{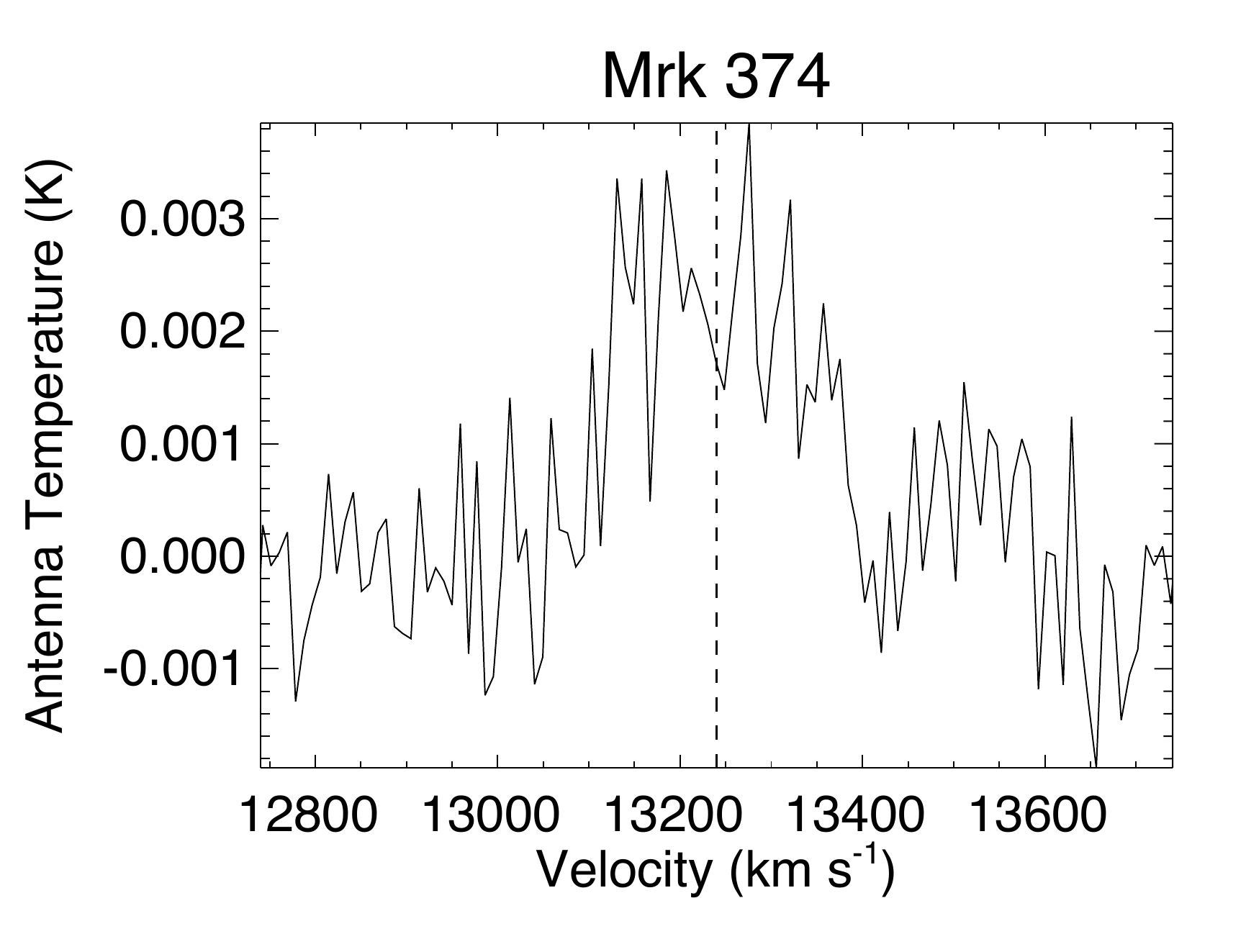}
}
\gridline{\includegraphics[trim={2cm 12.5cm 2cm 2cm},clip,scale=0.25]{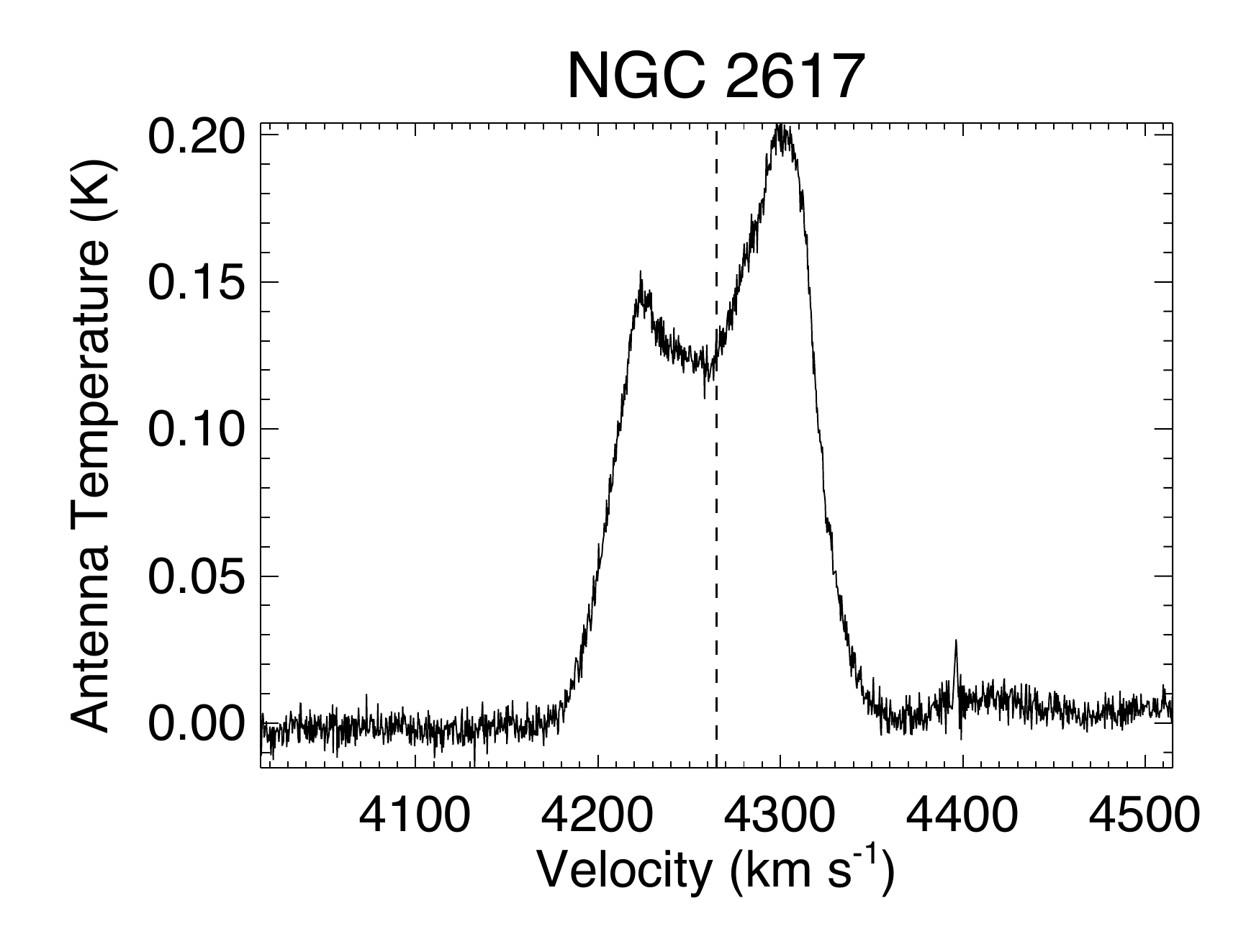}
\includegraphics[trim={2cm 12.5cm 2cm 2cm},clip,scale=0.25]{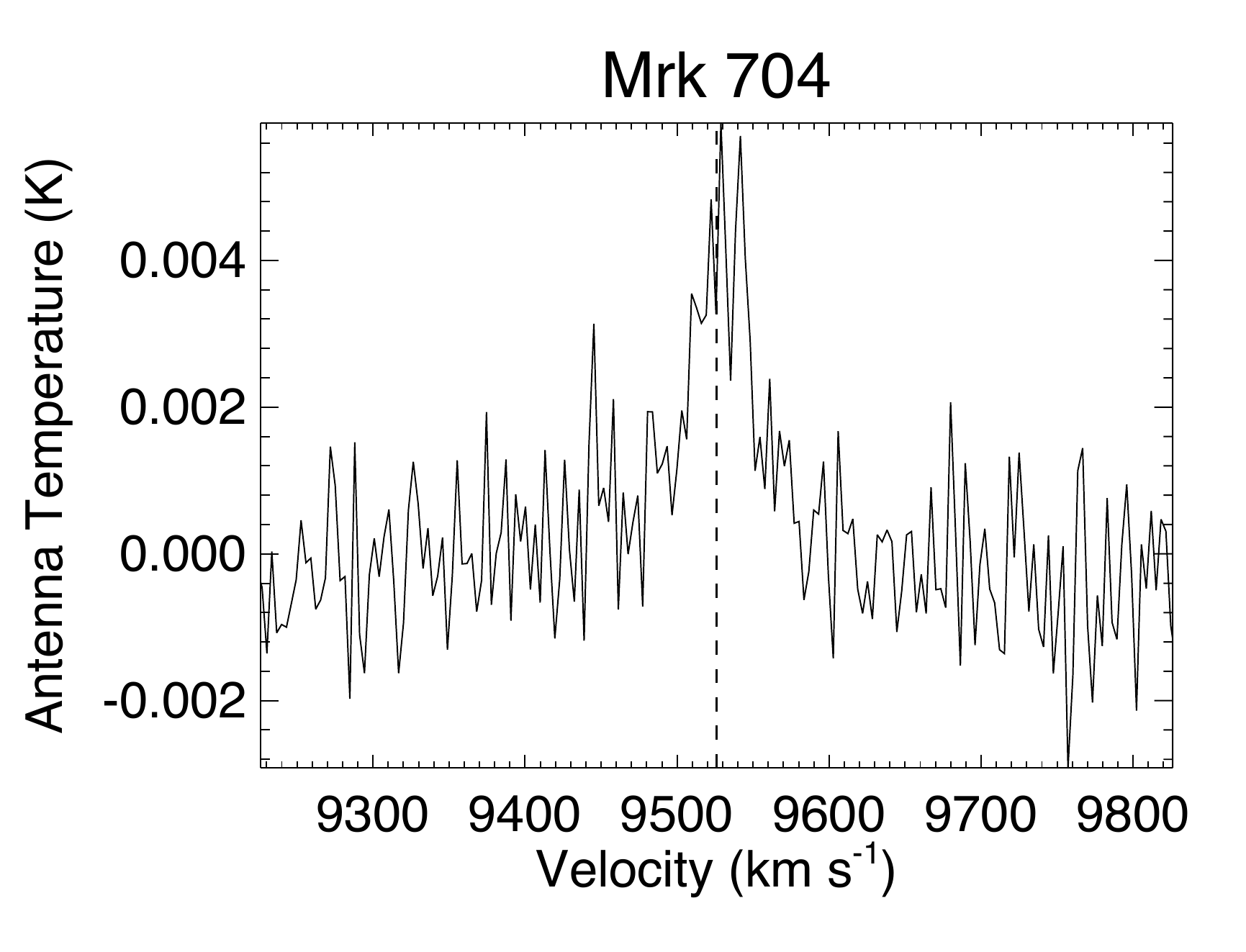}
\includegraphics[trim={2cm 12.5cm 2cm 2cm},clip,scale=0.25]{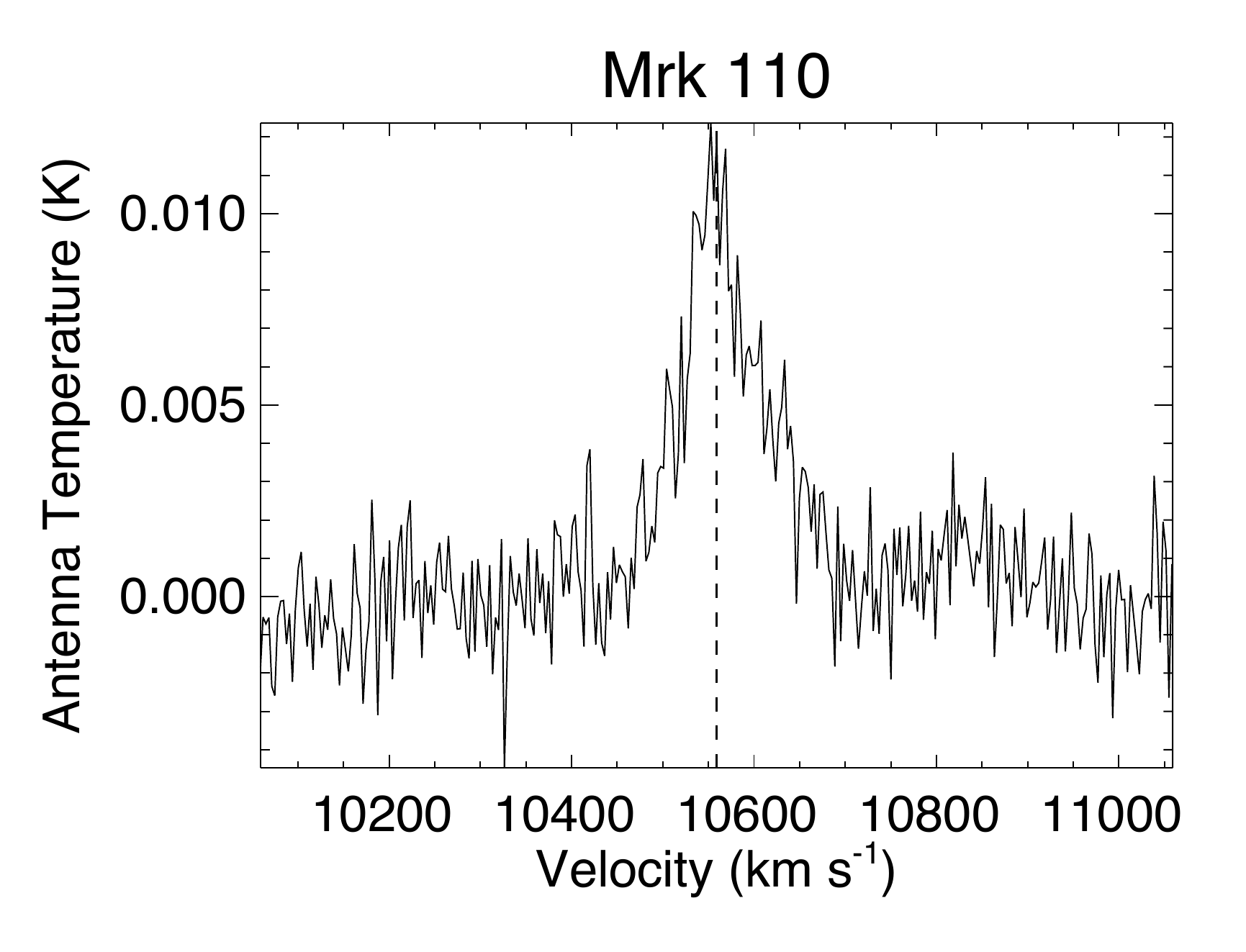}
\includegraphics[trim={2cm 12.5cm 2cm 2cm},clip,scale=0.25]{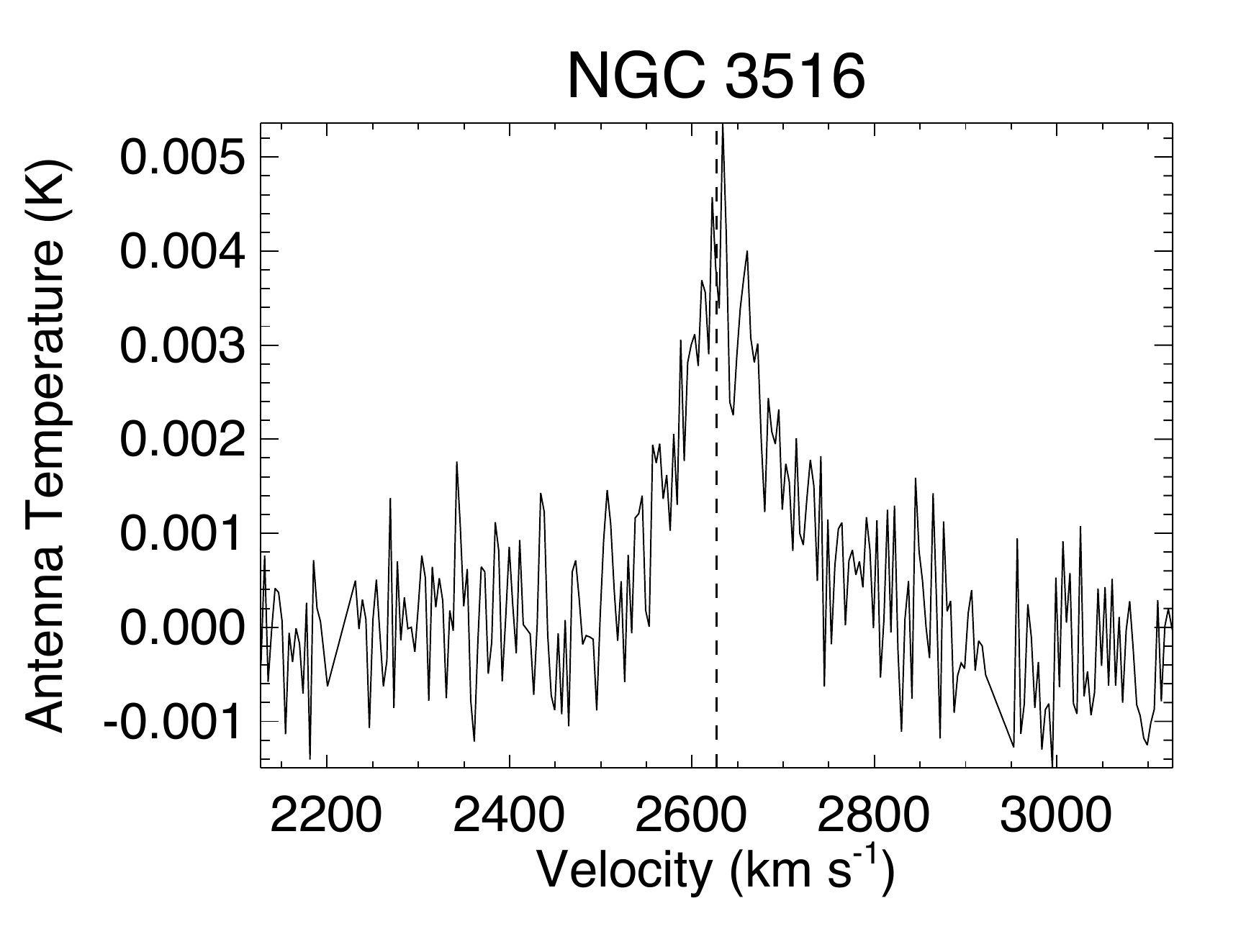}
}
\gridline{\includegraphics[trim={2cm 12.5cm 2cm 2cm},clip,scale=0.25]{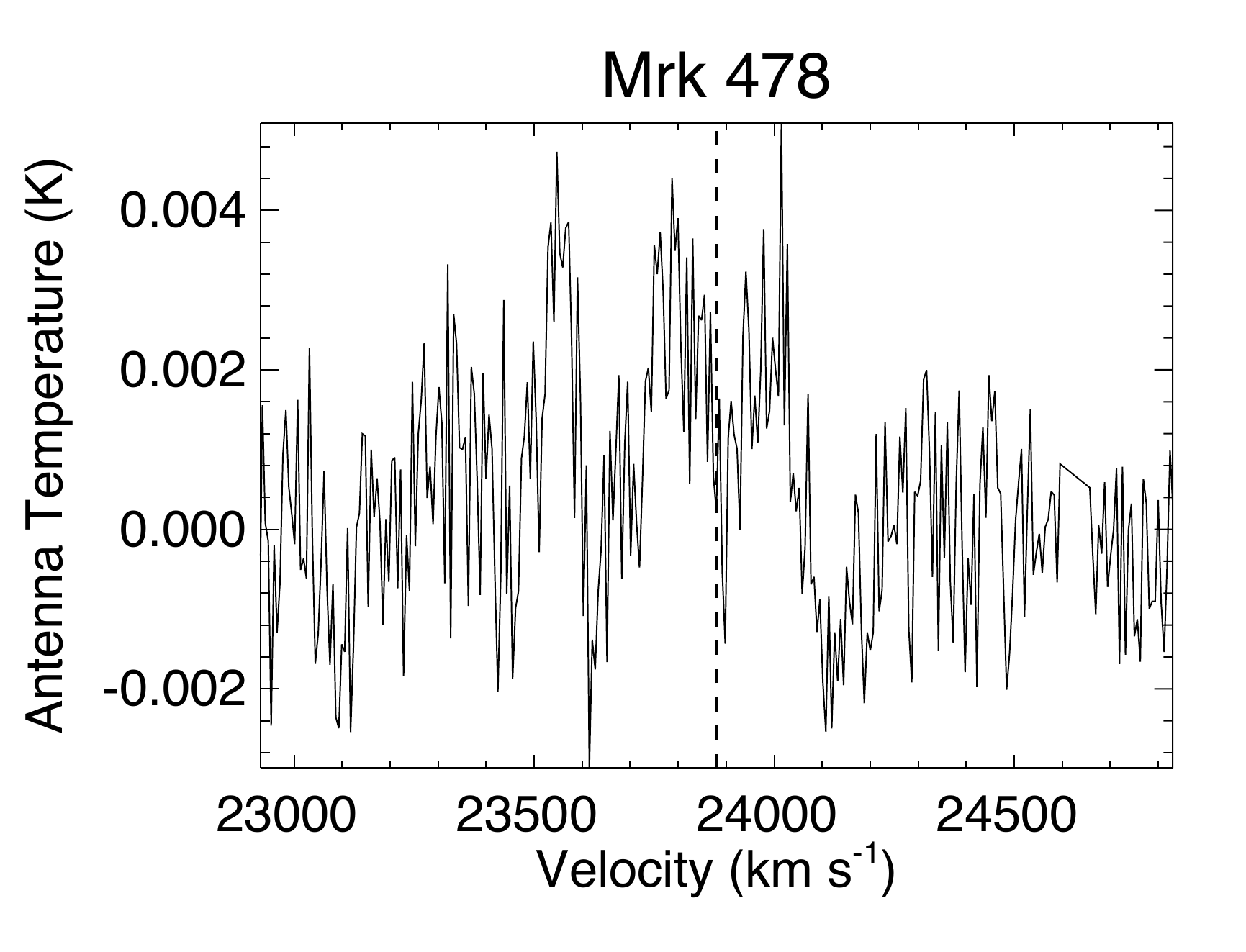}
\includegraphics[trim={2cm 12.5cm 2cm 2cm},clip,scale=0.25]{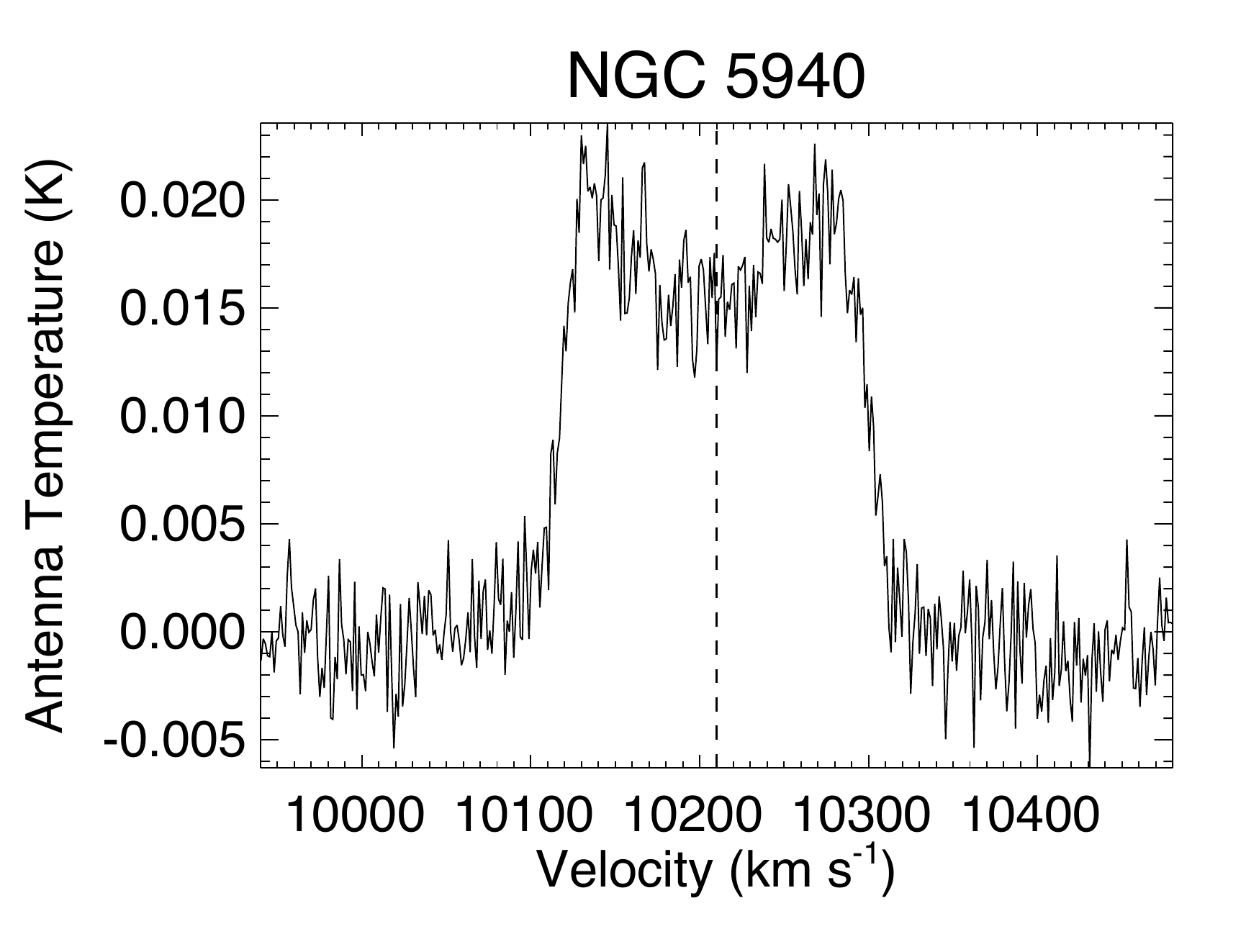}
\includegraphics[trim={2cm 12.5cm 2cm 2cm},clip,scale=0.25]{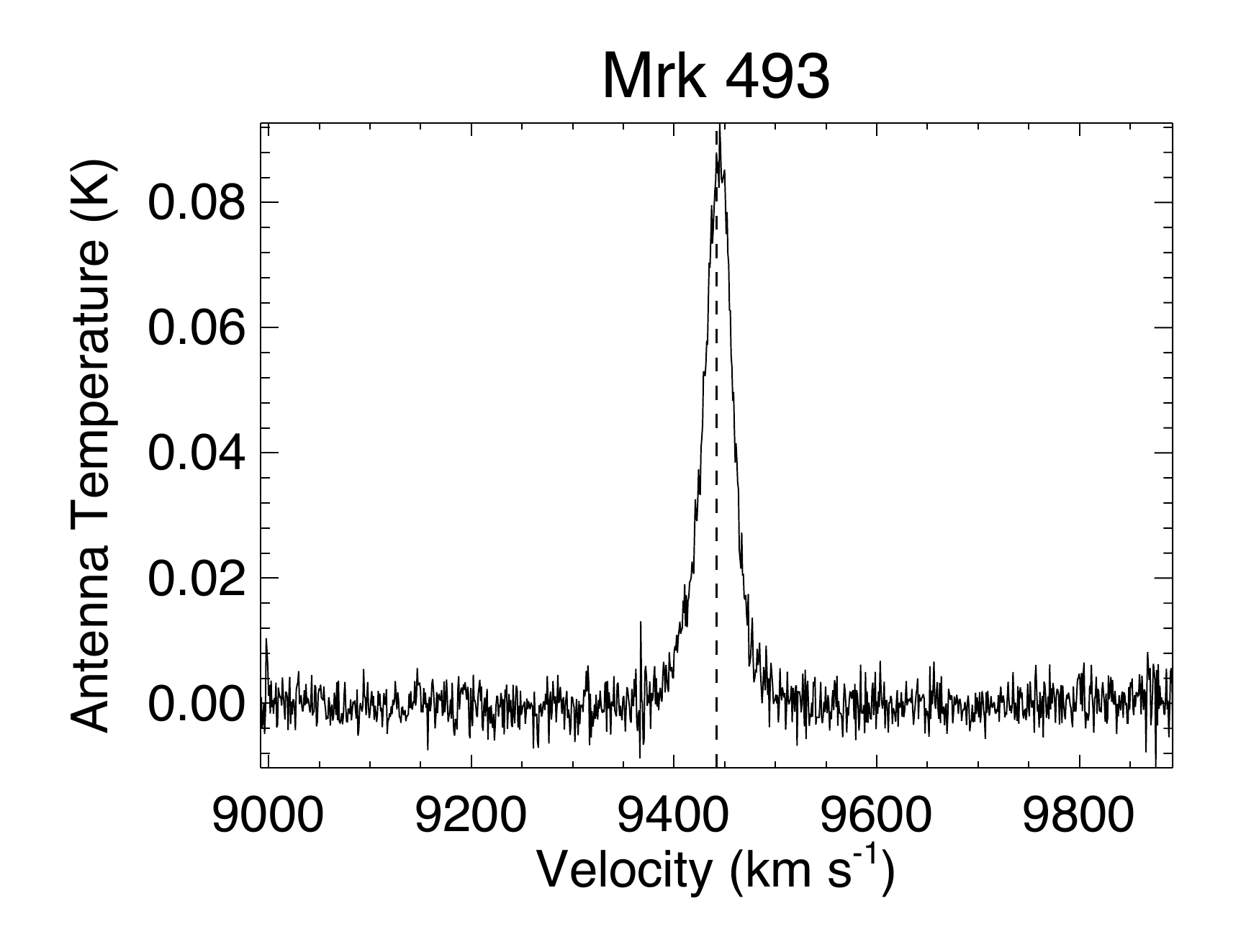}
\includegraphics[trim={2cm 12.5cm 2cm 2cm},clip,scale=0.25]{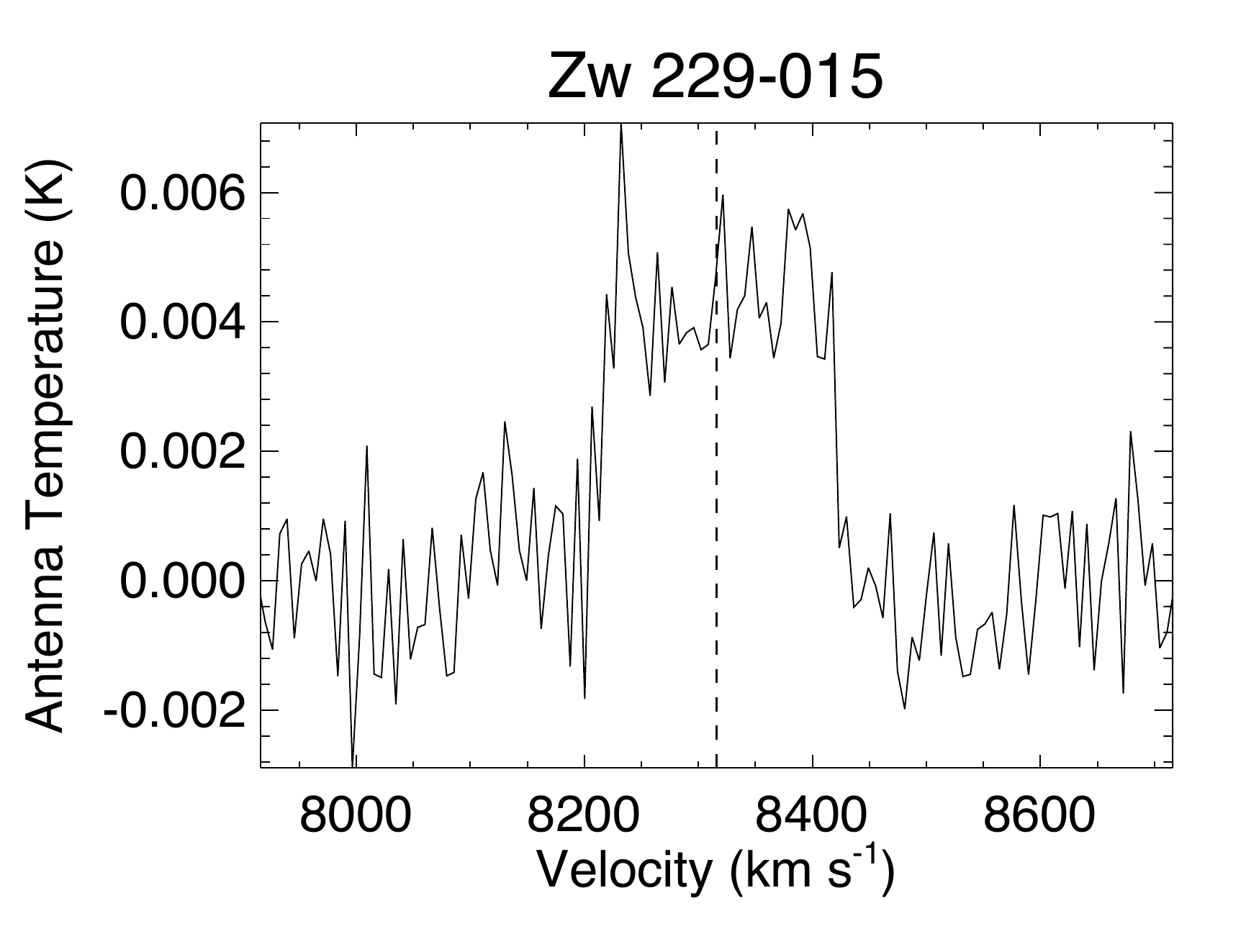}
}
\gridline{\hspace{6.8cm}\includegraphics[trim={2cm 12.5cm 2cm 2cm},clip,scale=0.25]{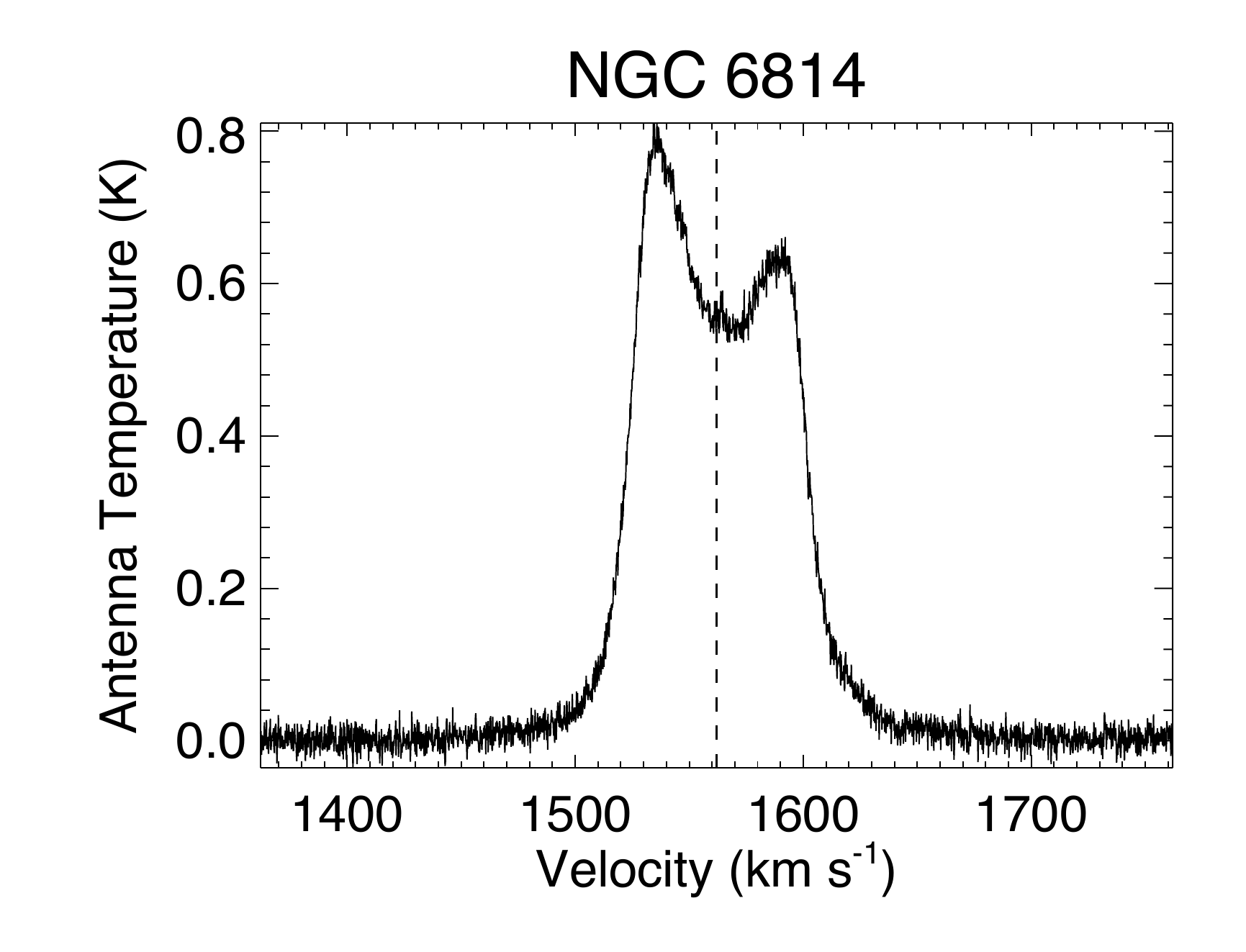}
}
\caption{HI emission line spectra from GBT18B-258 after reduction, baseline subtraction, and smoothing with $\textsc{GBTIDL}$ v2.10.1. Hanning smoothing was applied to all profiles, and further smoothing was dependent on the S/N (see Sec.\ \ref{red}). The vertical dashed lines indicate V$_\textsc{{R}}$ measurements from \texttt{gmeasure}.}\label{batmen2}
\end{figure*}

In this paper, we present the results of HI spectroscopy of 44 AGNs with direct M$_\textsc{{BH}}$ measurements from the reverberation mapping database\footnote{\url{http://www.astro.gsu.edu/AGNmass/}} of \cite{bhdatabase}. In Section \ref{sec:data}, for those galaxies where HI emission is detected, we provide measurements of profile widths, recessional velocities and thus redshifts, and HI flux. In Section \ref{sec:mass}, we detail our derived quantities of M$_\textsc{{HI}}$ and M$_\textsc{{GAS}}$ from the raw measurements, as well as other characteristics of the host galaxies and central black holes. In Section \ref{sec:discussion}, we explore the relationship between M$_\textsc{{BH}}$ and M$_\textsc{{GAS}}$, and we test relationships between M$_\textsc{{BH}}$ and baryonic mass (M$_\textsc{{BARY}}$), and M$_\textsc{{GAS}}$ and M$_\textsc{{stars}}$. 

Throughout this work we adopt a $\Lambda$CDM cosmology of H$_0 = 72$\,km\,s$^{-1}$\,Mpc$^{-1}$ \citep{hubble_constant}, $\Omega_{\textsc{M}}$=0.3, and $\Omega_{\normalfont{\Lambda}}$=0.7 \citep{bennett2014}.

\section{Data} \label{sec:data}
\subsection{Target Selection and Observations} \label{targets}
Our ultimate goal for these observations is to employ the Tully-Fisher distance measurement method \citep{tf1977} to provide a significant number of distances for galaxies in the reverberation mapping sample. In this paper, however, we focus on the HI properties of the galaxies. We began with the database of all broad-lined AGNs with black hole masses derived from reverberation mapping \citep{bhdatabase}.
Because the Tully-Fisher method requires spiral galaxies, the AGNs hosted by elliptical galaxies were removed from the sample. Potential targets were then removed if they were at $z > 0.1$, and therefore likely outside the reach of the Tully-Fisher method \citep{reyes2011}. Finally, the large, unblocked 100\,m dish of the Robert C. Byrd Green Bank Telescope\footnote{The Green Bank Observatory is a facility of the National Science Foundation operated under cooperative agreement by Associated Universities, Inc.} (GBT) and its access to a large fraction of the sky make it ideal for sensitive 21\,cm observations, therefore any remaining sources that were outside the pointing limits of the GBT were removed from the sample. The final selection  consisted of 27 active galaxies observed in the spring of 2013 (project ID GBT13A-468; PI: Ou-Yang) and 17 active galaxies observed in the fall/winter of 2018-2019 (project ID GBT18B-258; PI: Robinson).

The GBT Spectrometer backend was used for GBT13A-468 and employed a 12.5\,MHz bandwidth and 8,192 channels with velocity resolution of 0.3 km s$^{-1}$ channel$^{-1}$. The GBT Spectrometer has since been retired, so we employed the VErsatile GBT Astronomical Spectrometer (VEGAS) for GBT18B-258. The VEGAS mode we employed has an 11.72\,MHz bandwidth, 32,768 channels, and velocity resolution of 0.08 km s$^{-1}$ channel$^{-1}$. The large $9\farcm1$ GBT L-Band beamwidth, defined as the full-width at half-maximum of the beam, encompassed even the largest optical angular sizes of our target galaxies. 

Targets were observed in position-switched mode in pairs of equal on-off exposures, which were typically 60 second scans for GBT13A-468 and 120 second scans for GBT18B-258. All of the scans were divided into individual integrations of 3 seconds, so that if radio frequency interference (RFI) appeared, we could discard only the affected integrations rather than the whole scan. Off-source sky observations allowed for the removal of the frequency structure of the raw bandpass and an improvement in the signal-to-noise ratio (S/N). Total on-source exposure times were estimated from the expected gas content of the galaxy (based on its morphological type), its expected distance (based on redshift), and a goal of achieving S/N=10 in the peak flux of the HI emission line.  We capped our initial time requests at 9 hours per source, but for a few weak yet promising targets, we requested and received additional time to improve the S/N.  In total, our observations covered 184 hours for GBT13A-468 and 208.25 hours for GBT18B-258, with total on-source exposure times ranging from $\sim$ 6 minutes to 16 hours. Table \ref{1} lists the targets and the details of their observations. 

\subsection{Reduction} \label{red}
Data reduction was carried out with $\textsc{GBTIDL}$ \citep{gbtidl} v2.8 for GBT13A-468 and v2.10.1 for GBT18B-258. The updated software includes the ability to reduce spectra taken with the VEGAS backend, as well as bug fixes. Individual scans were visually inspected, and those which contained significant saturation from RFI were removed. The \verb|getps| $\textsc{GBTIDL}$ procedure retrieved the on-source and corresponding off-source data for each scan pairing and performed the (ON-OFF)/OFF operation. Scans were then accumulated and averaged over for each target. Weaker RFI spikes were manually removed by interpolating over the interference in the accumulated spectra. Targets that required several hours of exposure time were generally observed in separate blocks across a few days. Scans from separate observing sessions were managed in the same way; all scans from each observation block of the same source were accumulated and averaged at once into a single spectrum. 

\begin{figure*}
\gridline{\includegraphics[trim={2cm 13cm 2cm 2cm},clip,scale=0.35]{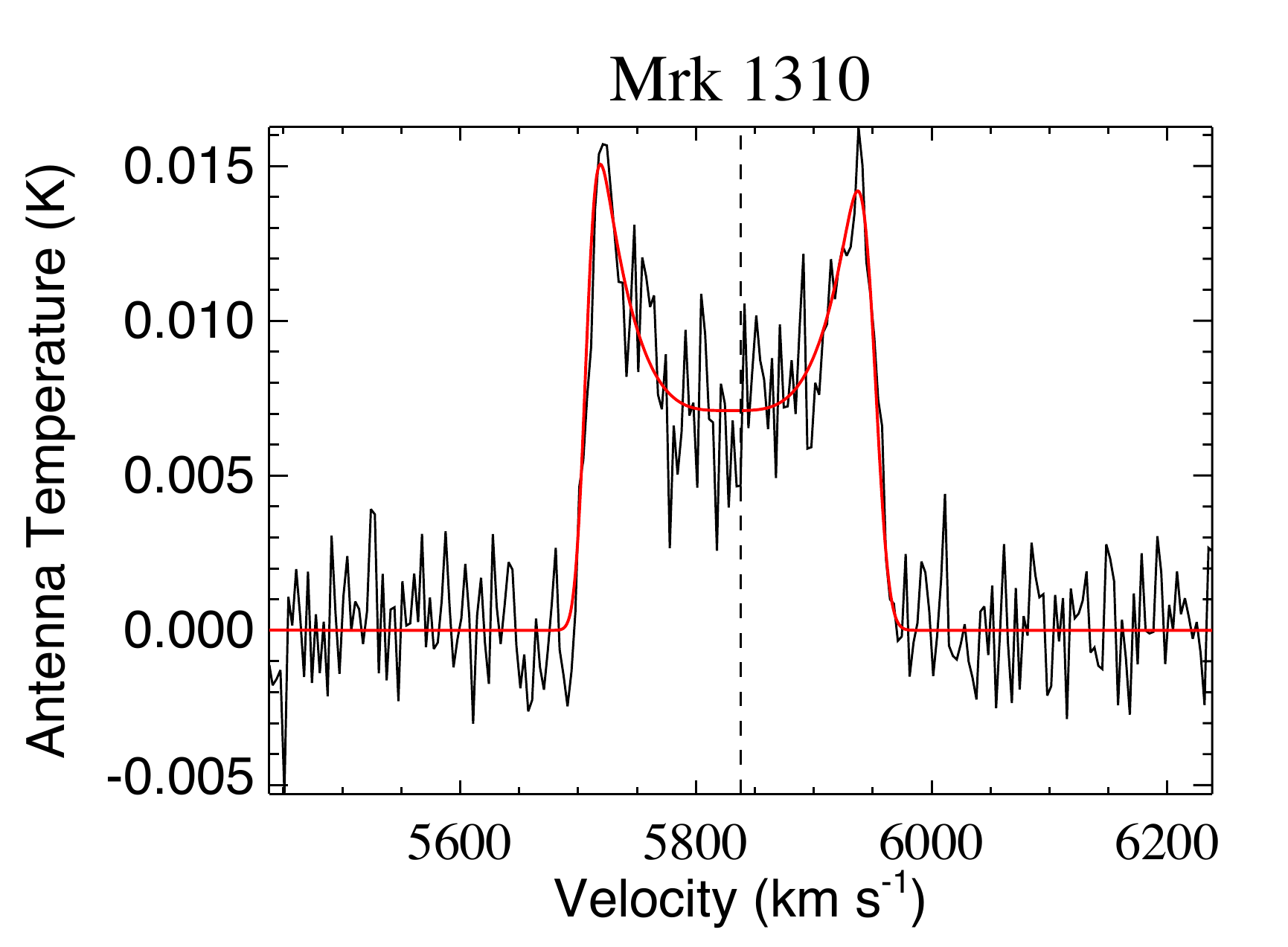}
\includegraphics[trim={2cm 13cm 2cm 2cm},clip,scale=0.35]{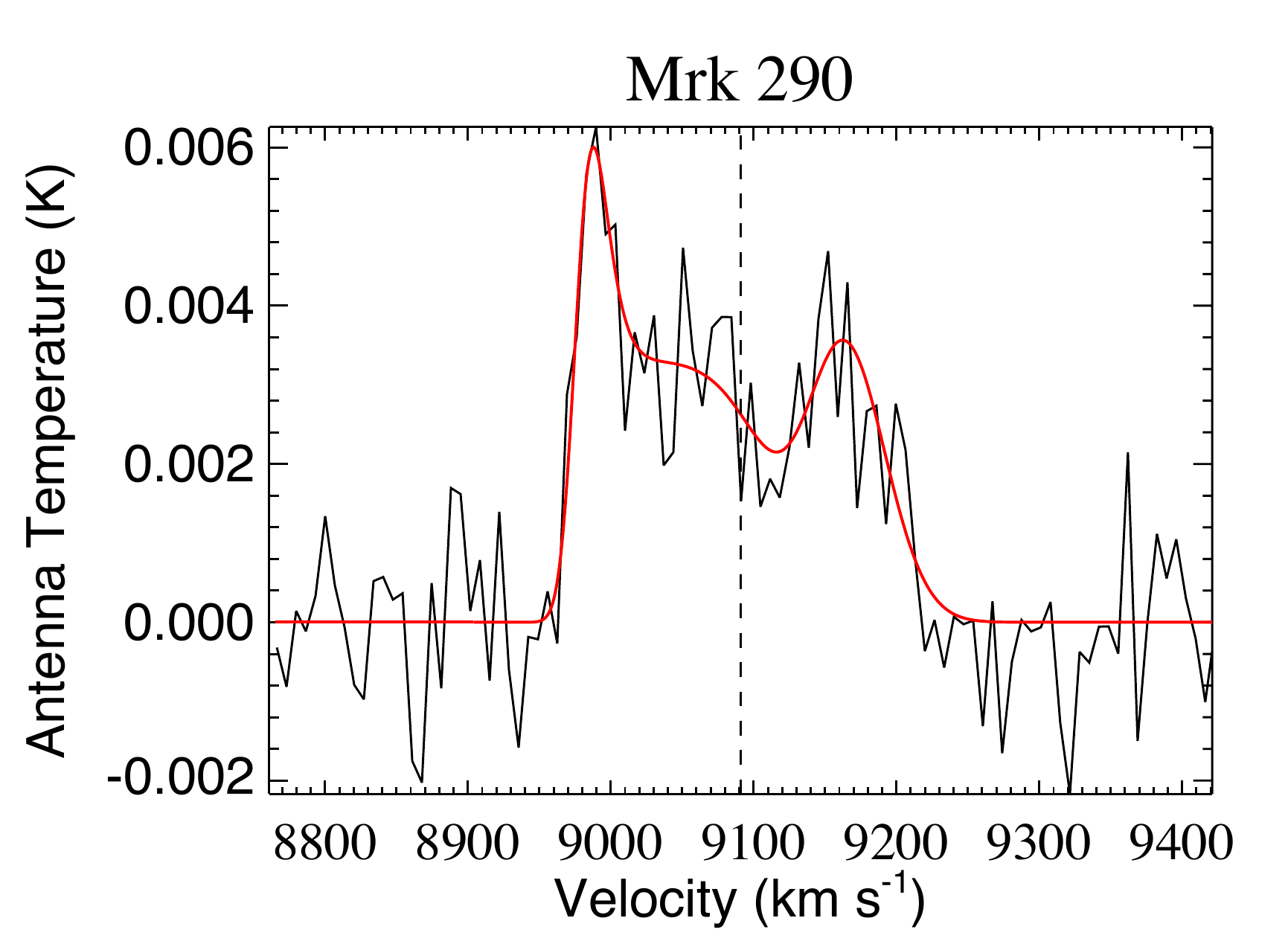}
\includegraphics[trim={2cm 13cm 2cm 2cm},clip,scale=0.35]{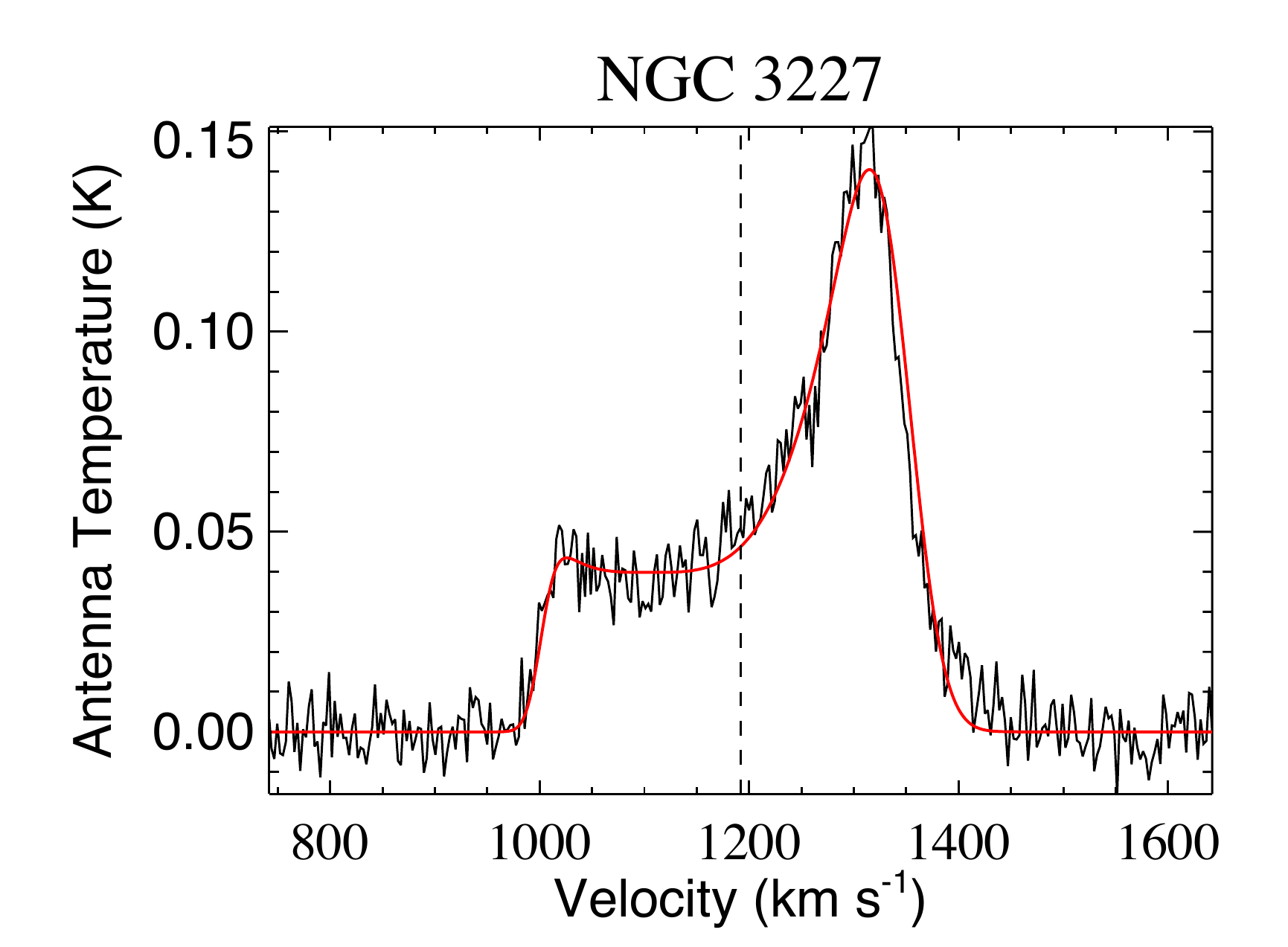}
}
\gridline{\includegraphics[trim={2cm 13cm 2cm 2cm},clip,scale=0.35]{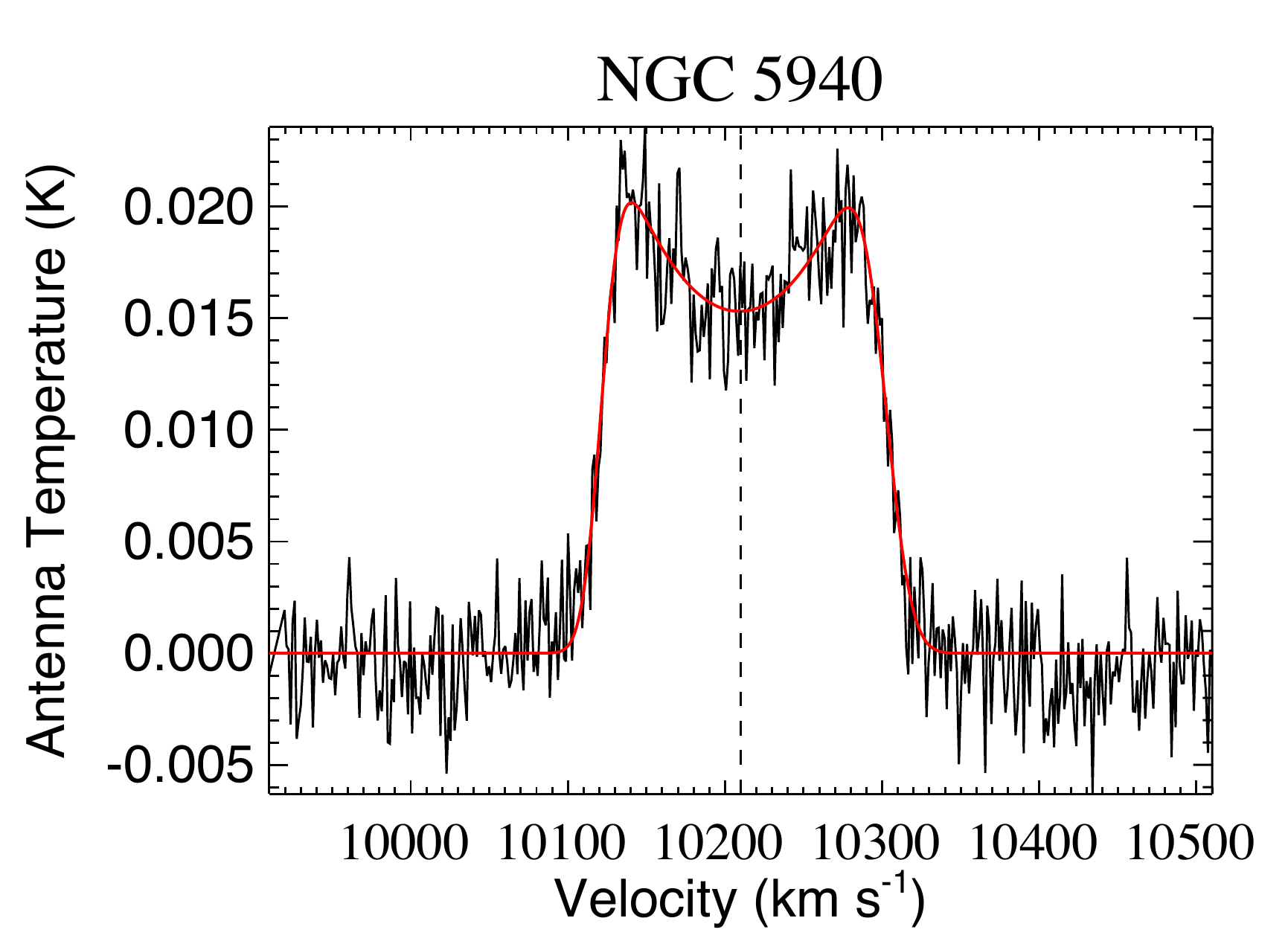}
\includegraphics[trim={2cm 13cm 2cm 2cm},clip,scale=0.35]{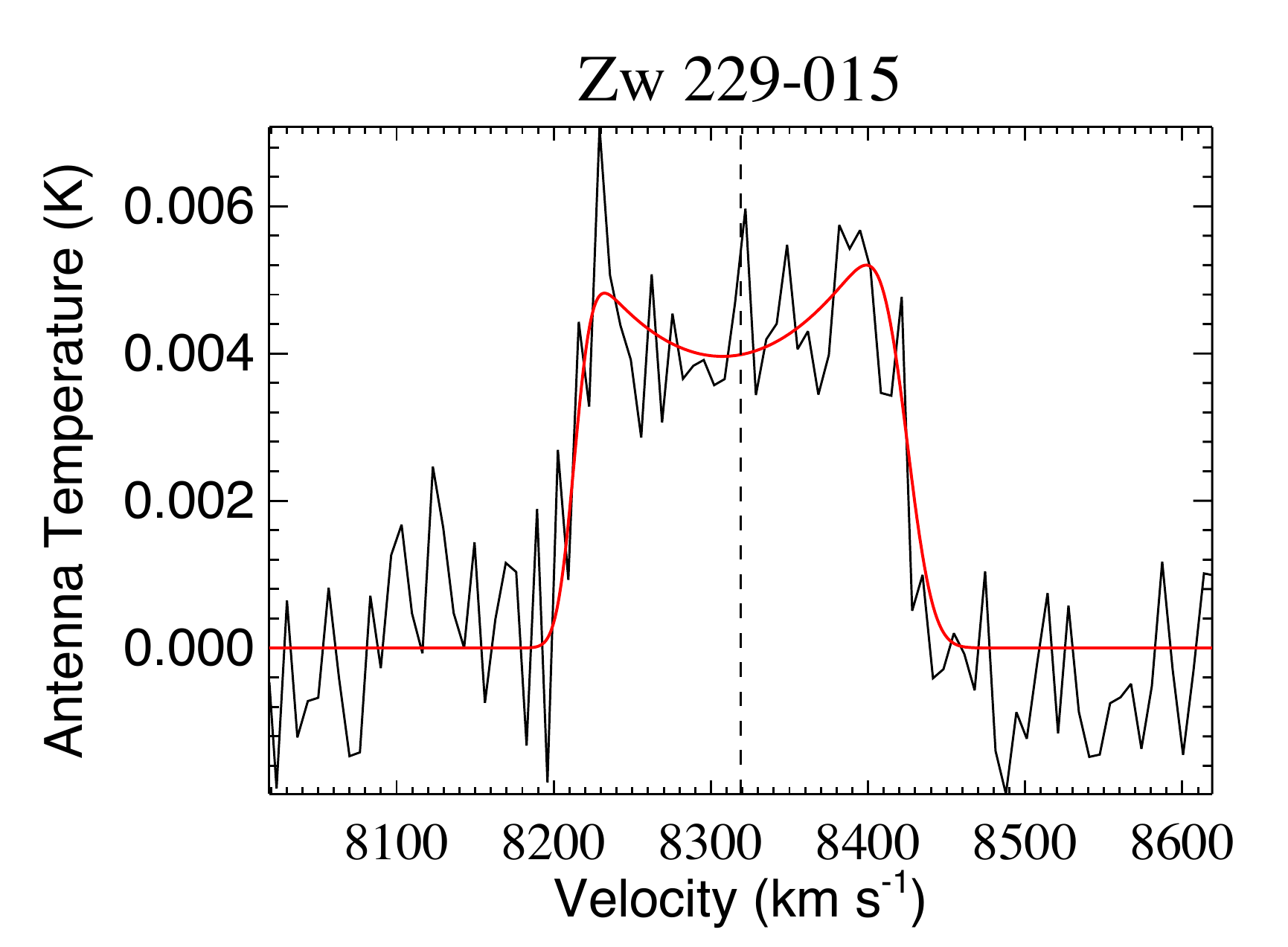}
\includegraphics[trim={2cm 13cm 2cm 2cm},clip,scale=0.35]{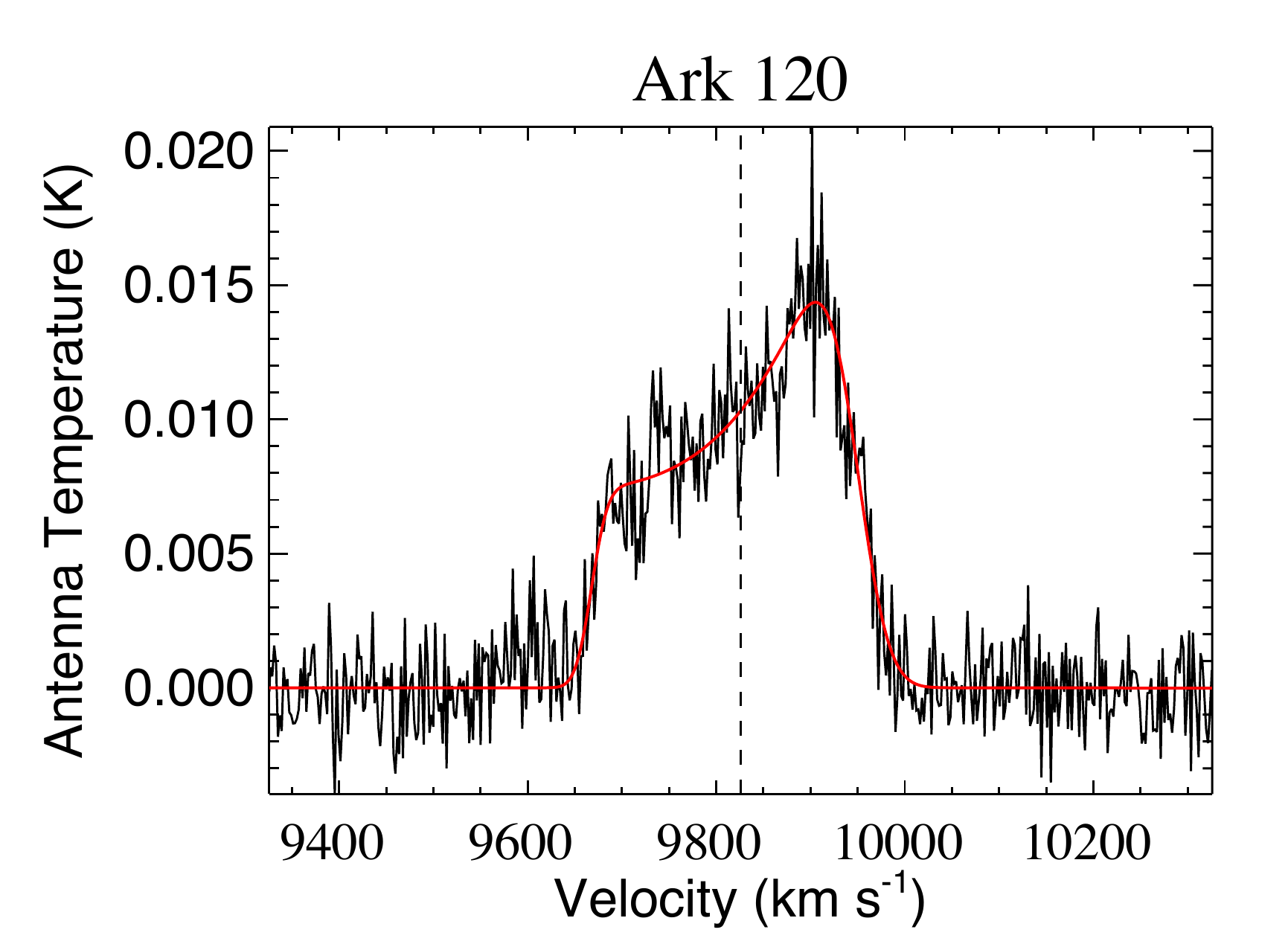}
}
\caption{Example $\textsc{BusyFit}$  \citep{busyfit} profiles for representative galaxies in our sample. The data is in black, the model fit is the solid red curve. The top row are profiles from GBT13A-468, the bottom are profiles from GBT18B-258. The vertical dashed lines indicate V$_\textsc{{R}}$ measurements from $\textsc{BusyFit}$.} \label{bf}
\end{figure*}

We fit a low-order polynomial (typically order 3) to the remaining baseline in each combined spectrum and subtracted it. Hanning smoothing \citep{hanning} was then applied to all spectra after accumulation and baseline subtraction. Hanning smoothing is a running mean across the spectrum that aids in reducing the ringing produced by strong RFI sources and reduces the spectral resolution by a factor of 2. Higher order smoothing was then applied to spectra with low apparent S/N to aid in the measurement of the emission line properties. 

We detected HI 21\,cm emission in 18 of the 27 targets from GBT13A-468 and in 13 out of 17 for GBT18B-258. Some of the more distant objects in the sample with $z > 0.05$ resulted in non-detections within the allotted exposure time. Figs.\ \ref{batmen} and \ref{batmen2} show the reduced, smoothed, and baseline subtracted spectra for targets where HI emission was detected from programs GBT13A-468 and GBT18B-258, respectively. We report in Table \ref{characteristics} the total resulting on-source exposure times after removal of contaminated scans, an approximate S/N for all spectra, values for the root-mean-square (RMS) of the noise in each spectrum, the final velocity resolution after smoothing, and the corresponding backend. For dual-horned profiles, approximate S/N values were calculated first by taking the average of the peak fluxes in each horn. We then averaged that with the mid-profile peak flux, and divided by the RMS of the noise to produce the approximate S/N. For Gaussian profiles, the approximate S/N was calculated as the peak flux value divided by the RMS of the noise.

\subsection{Analysis and Measurements} \label{analysis}
	
Measurements of the emission-line widths, center-line recessional velocities (V$_\textsc{{R}}$), and integrated line fluxes were determined with two methods. The first method utilized the \texttt{gmeasure} procedure available within $\textsc{GBTIDL}$, which calculates line widths, fluxes, and recessional velocities directly from the data. For width measurements, the \texttt{gmeasure} procedure determines the edges of an emission profile by linear interpolation over channels containing the profile until the data are greater than the provided threshold. The threshold is normally $50\%$ or $20\%$ times the mean flux over the range of channels containing the HI signal. W$_{50}$ and W$_{20}$ values (line widths at $50\%$ and $20\%$, respectively) as well as V$_\textsc{{R}}$ values are provided in km s$^{-1}$. We choose the mean flux, rather than the peak flux, for determination of W$_{50}$ and W$_{20}$ because it is less sensitive to the noise level, especially when a profile consists of significantly asymmetric peak horn fluxes. Calculated line fluxes are given in terms of antenna temperature (T$_\textsc{{L}}$; see Section \ref{sec:mass}) in K km s$^{-1}$. Uncertainties on the \texttt{gmeasure} measurements were achieved using a bootstrap method. We began by defining beginning and ending spectral channel windows on either side of the line profile. The designated range of channels for each window on each side was unique to each profile and was mainly dependent on the noise properties (see Table \ref{characteristics}), but was typically $\sim$ 50 channels in width. We then performed 100,000 iterations in which a starting and ending wavelength were randomly drawn from the defined windows, and a line width, flux, and central velocity were calculated using \texttt{gmeasure}.  A distribution of each measurement was built up in this way, and we report the median of the distribution as the measurement value and the measurement uncertainty as the 1-$\sigma$ deviation from the median on each side of the distribution, allowing for asymmetric distributions. 

The second method employed the $\textsc{BusyFit}$ software \citep{busyfit}. For well-defined profiles, $\textsc{BusyFit}$ is able to automatically determine the best-fit parameterization, but for noisy or poorly-defined profiles, additional user intervention is necessary. Measurements of W$_{50}$ and W$_{20}$, (which in this case are defined as line widths at 50\% and 20\% of the peak flux density, which differs slightly from the definition employed by \texttt{gmeasure}), T$_\textsc{{L}}$, and V$_\textsc{{R}}$ are derived from the $\textsc{BusyFit}$ profiles. For both methods, we define V$_\textsc{{R}}$ in the optical convention (c($\lambda$-$\lambda_{0}$)/$\lambda_{0}$). The fitting function has the form 
\begin{equation}
\begin{split}
B(x) = \frac{a}{4} \times (erf[b_{1}\{w + x - x_{e}\}]+1)\\
& \hspace{-4cm} \times (erf[b_{2}\{w - x + x_{e}\}]+1) \times (c|x-x_{p}|^{n}+1)
\end{split}
\end{equation}
where $x$ denotes the spectral axis input, $a$ is the amplitude scaling factor, the error functions fit the sides of the HI profile (flanks), $b_{1}$ and $b_{2}$ are the independent slopes of the flanks allowing for asymmetric shapes of the lines to be fit, $w$ is the half-width of the HI profile, $x_{e}$ and $x_{p}$ are separate offsets also aiding in fitting asymmetric profile shapes, and $c$ denotes the amplitude of the central trough of the profile relative to the flanks which is fit with a polynomial of degree $n$. 

The majority of the dual-horned profile fits converged without the need to hold any free parameter values fixed. Most well-defined, Gaussian-shaped profiles also achieved convergence in the fit. For these, $\textsc{BusyFit}$ automatically fixes the parameters included in the central trough factor of the fitting function ($c$, $x_{p}$, and $n$) at 0. We found it common that low S/N or weakly defined dual-horned profiles required holding the $c$ and $n$ values fixed because the initial fits often converged to Gaussian shapes. We also found common that narrow, low S/N Gaussian profiles required holding at least one of the flank slopes fixed as the initial fits resulted in either extremely high uncertainties in these parameters or did not reach convergence. Generally in the cases which necessitated parameters to be held fixed, we inferred the values from the fits for well-defined profiles which converged automatically. We assessed by eye the accuracy of the shape of the fit relative to the inherent shape of the emission line, and we estimate the following additional uncertainties on each parameter: $n$: $\pm$15\%, $c$: $\pm$29\%, $w$: $\pm$4\%, $b_{1}$: $\pm$21\%, $b_{2}$: $\pm$22\%, $x_{p}$: $\pm$8.5\%.

$\textsc{BusyFit}$ offers a number of methods for determining uncertainties. We employed the Monte Carlo method which generates 10,000 best fits by randomly varying the free parameters in each iteration. The variations are dependent on the covariance matrix of the values of the free parameters, with each parameter's random distribution centered on the initial fit value and standard deviation derived from the square root of the diagonal elements of the covariance matrix. The uncertainties are assumed to be symmetric and are reported as the standard deviation from the mean of the resulting measurement distribution. For uncertainty determinations on fits in which parameters were held fixed, there is a tendency for underestimations due to the fixed parameters not contributing to uncertainty propagation. To account for this, we conducted best fits on SBS1116+583A (the lowest S/N HI profile in which no parameters were held fixed for the reported $\textsc{BusyFit}$ measurements), fixing each free parameter and each combination of fixed free parameters to calculate the differences in resulting uncertainties from the initial fit's uncertainties. We then inflated the measurement uncertainties by a corresponding amount to match the differences on SBS1116+583A for objects with parameters that were held fixed in the fitting process. 

\begin{figure*}%left lower right upper
\gridline{\includegraphics[trim={1cm 8.5cm 1cm 9cm},clip,scale=0.47]{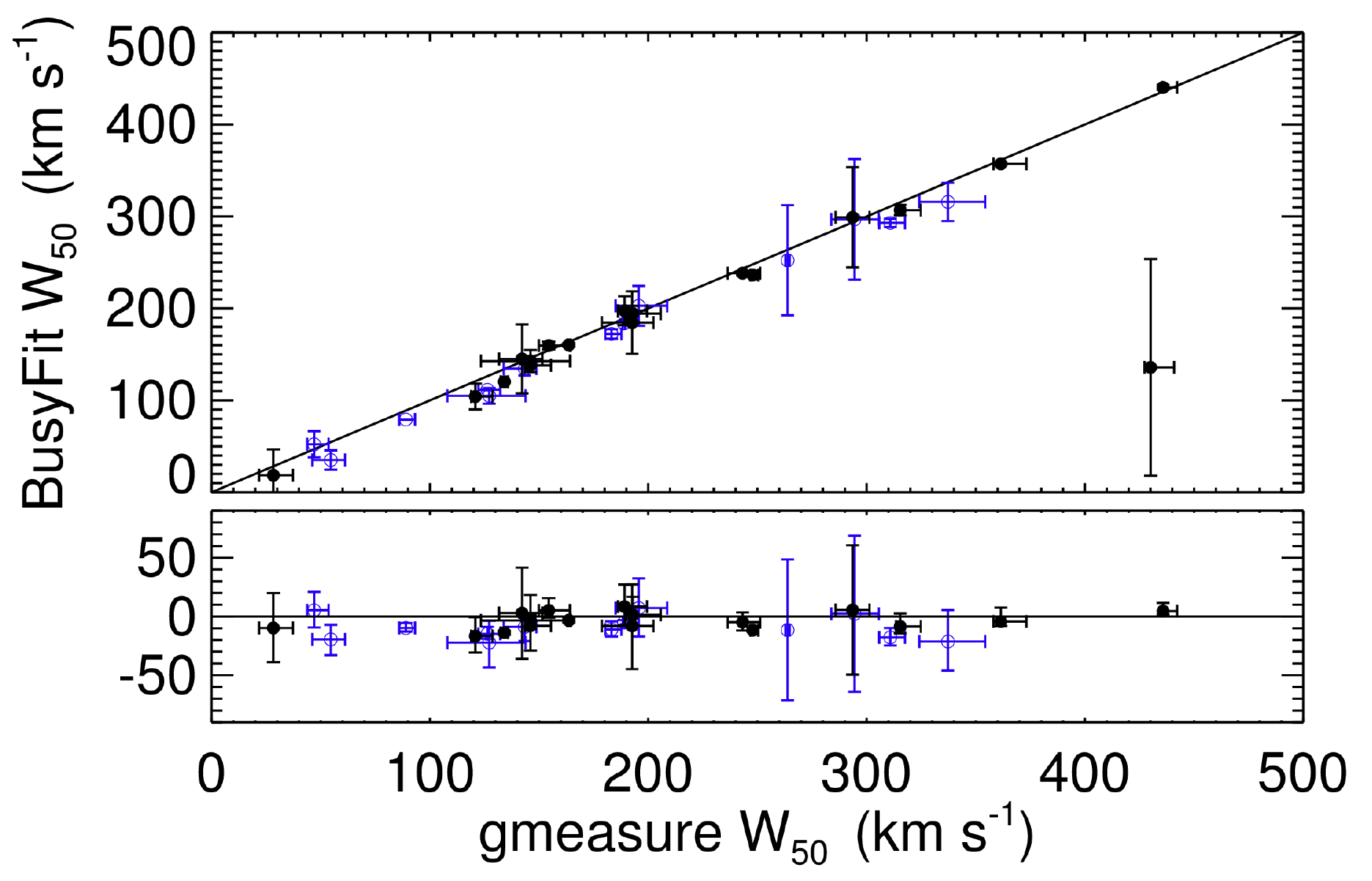}
\includegraphics[trim={1cm 8.5cm 1cm 9cm},clip,scale=0.47]{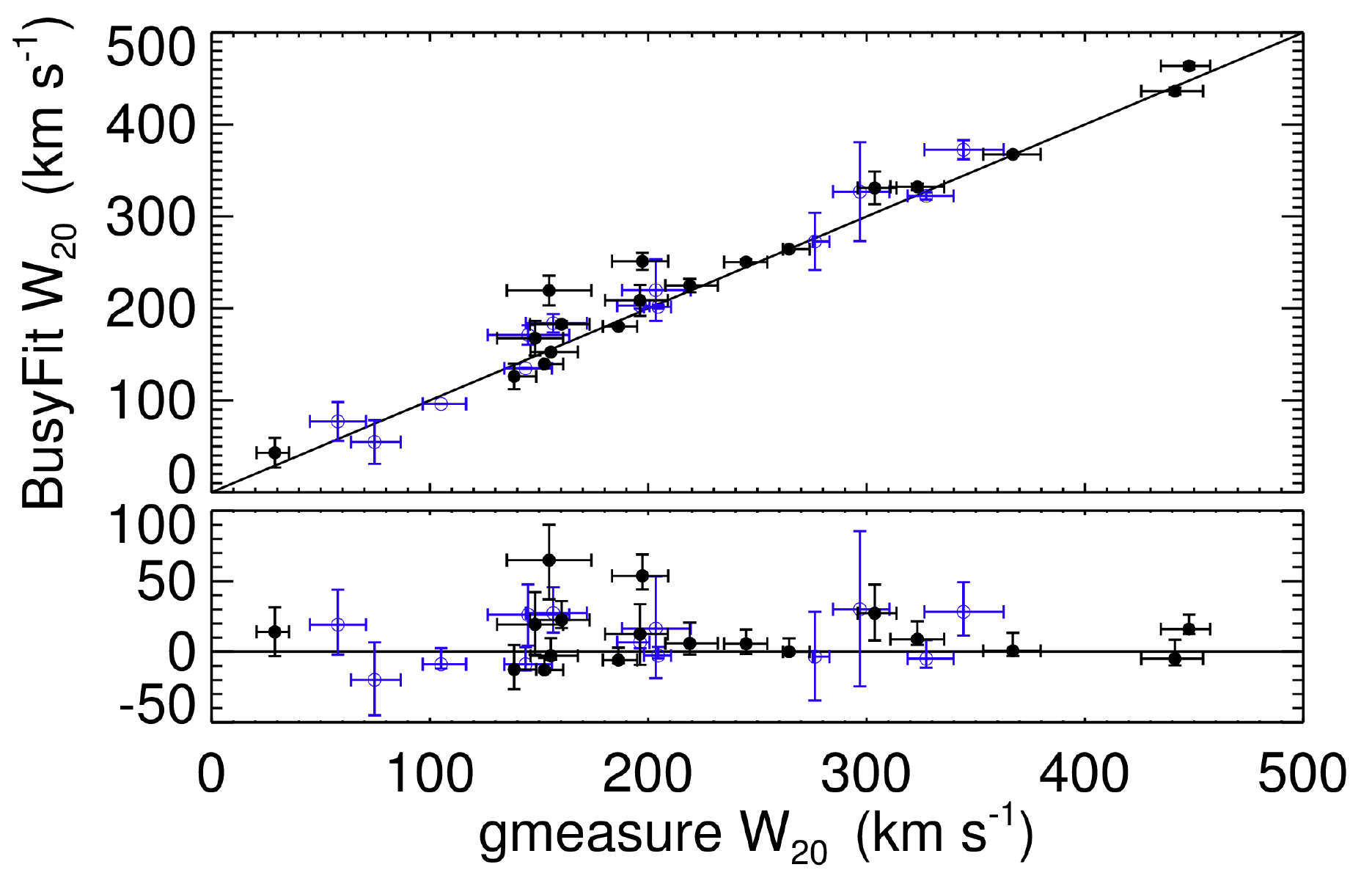}
}
\gridline{\includegraphics[trim={1cm 8.5cm 1cm 9.3cm},clip,scale=0.49]{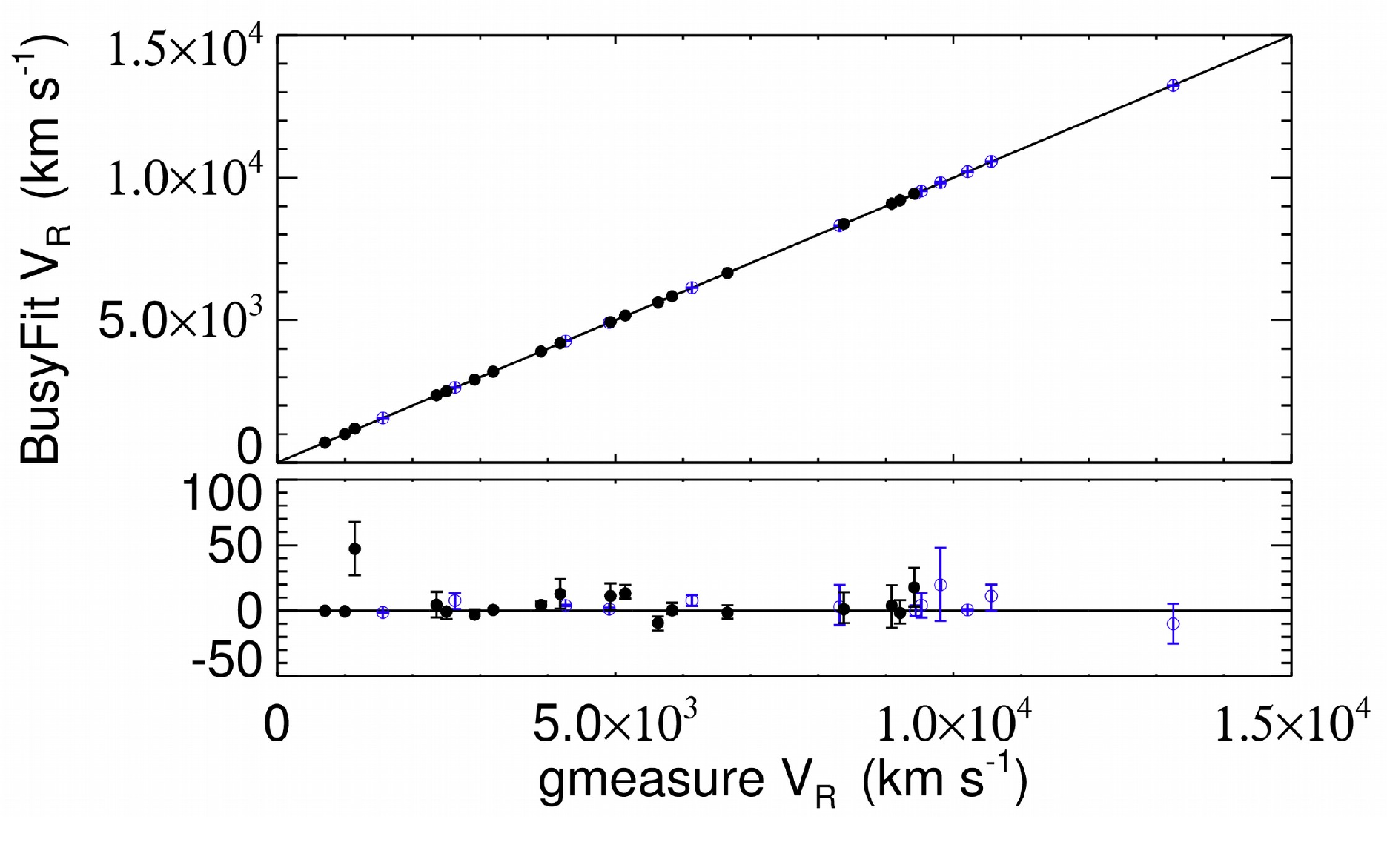}
\includegraphics[trim={1.5cm 8cm 1cm 9cm},clip,scale=0.45]{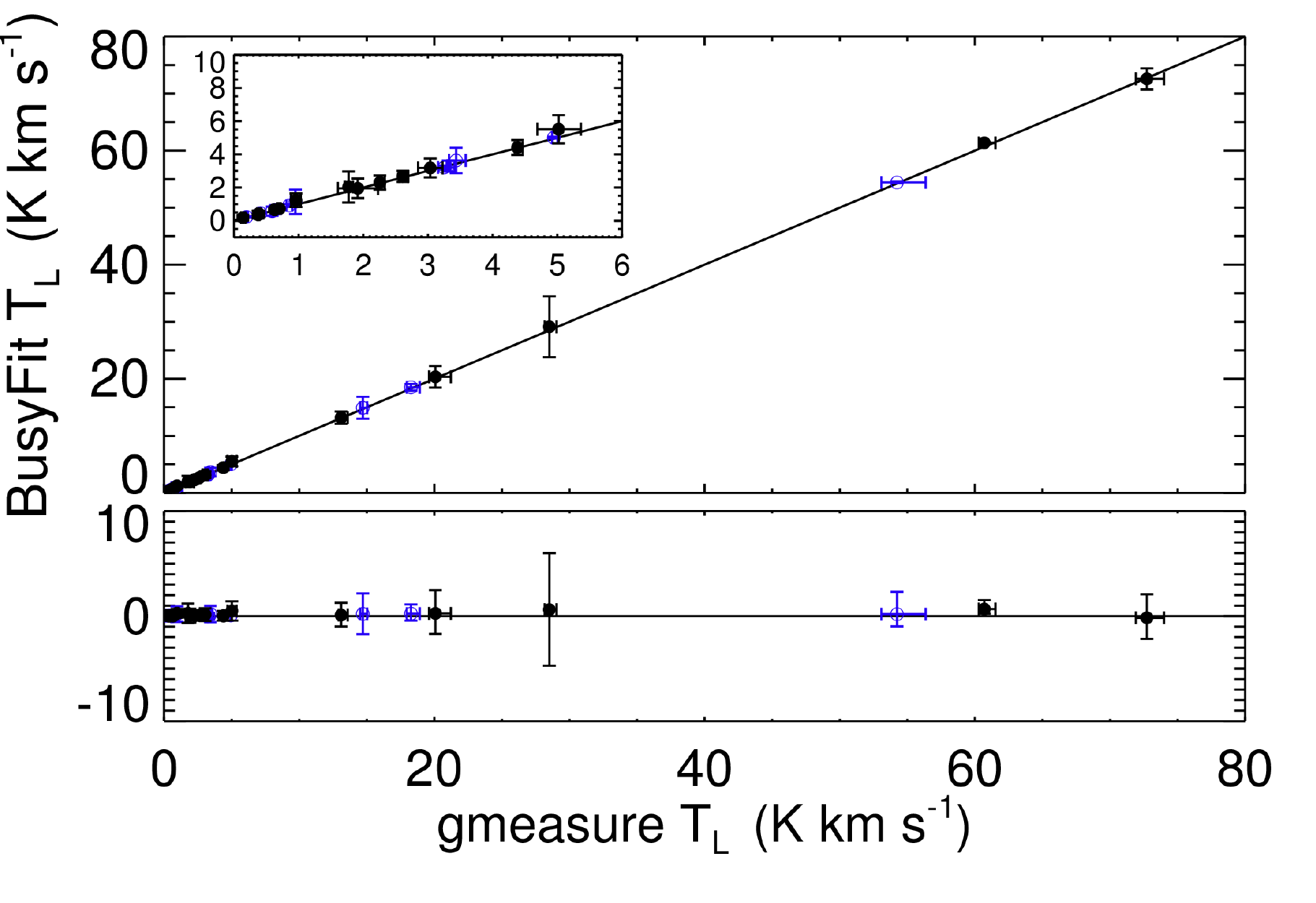}
}
\caption{Comparison of the measurements from \texttt{gmeasure} and $\textsc{BusyFit}$: W$_{\textsc{50}}$ (top left), W$_{\textsc{20}}$ (top right), V$_{\textsc{R}}$ (bottom left), and T$_{\textsc{L}}$ (bottom right). The solid line in each panel is a line of unity, and $\textsc{BusyFit}$ minus \texttt{gmeasure} is plotted below each panel. Measurements of profiles from GBT13A-468 are shown in solid, black circles, and measurements from GBT18B-258 are show in open, blue circles. The outlier in the top left plot is NGC 3227 (see Fig.\ \ref{batmen}, Sec.\ \ref{previous_meas}), which exhibits a lopsided profile, resulting in significant uncertainty in the W$_{\textsc{50}}$ line width measurement due to the uncertainty on the exact value of 50\% peak flux.}\label{comparisons}
\end{figure*}

Fig.\ \ref{bf} displays the best-fit $\textsc{BusyFit}$ profiles overlaid on the HI spectra for a few representative galaxies. Profiles from GBT13A-468 are displayed in the top row, and profiles from GBT18B-258 in the bottom row. Table \ref{spectra} reports the measured values of T$_\textsc{{L}}$, W$_{50}$, W$_{20}$, and V$_\textsc{{R}}$ from \texttt{gmeasure} and from $\textsc{BusyFit}$. Comparisons between all measurements from \texttt{gmeasure} and $\textsc{BusyFit}$ are shown in Fig.\ \ref{comparisons}. Measurements from GBT13A-468 are shown in solid black circles, and GBT18B-258 are show in open blue circles. A line of unity is shown in each panel, and the differences between the two methods' measurements are shown below each panel. The results are generally in close agreement even though the definitions of the line widths are slightly different, with only a few objects showing large discrepancies. NGC 3227, in particular, has a highly asymmetric line profile with a low central trough and blueshifted horn.  This makes the determination of W$_{50}$ quite sensitive to the noise in the spectrum, and whether 50\% of the peak flux is below or above the blueshifted horn. If we choose the peak flux definition for \texttt{gmeasure}'s W$_{50}$ measurement, we find a more consisent value with $\textsc{BusyFit}$. Both MCG-06-30-15 and Mrk 478 show some disparities in their line widths due to the low S/N in their spectra.  And for Gaussian-shaped profiles, like those seen in Mrk 279, NGC 5548, Mrk 704, Mrk 110, NGC 3516, and Mrk 493, the two methods are more likely to disagree due to the difficulty in determining the true edges of the profiles. We also note that three of our spectra (NGC 4051, NGC 4593, NGC 5548) exhibit the surprising feature of excess HI emission at or near the systemic velocity. For the remaining analysis, we prefer the measurement values from \texttt{gmeasure} due to their asymmetric distributions more accurately reflecting the asymmetric nature of the majority of the profiles. 

\subsection{Notes on Individual Objects}\label{previous_meas}
	Of the 31 galaxies with HI detections in our sample, 12 have not been previously studied in HI. For the remaining objects, we have tabulated their previous measurements for comparison with our own in Table \ref{previous}. Below, we include a short discussion of the different measurements for each object. We also include discussion on the best-fit models produced by $\textsc{BusyFit}$ whenever user intervention was necessary. 
	
	Two of the objects in this study, Mrk 6 and NGC 7469 (see Fig.\ \ref{batmen}) exhibit a strong center-line absorption feature in their HI profiles. While the absorption does not affect their line widths or recessional velocities, it does affect the line flux.  In order to estimate a reasonable range of values for the unabsorbed line flux, we used a bootstrapping interpolation method. We first determined the ratios of horn-height to mid-profile-height from all of the unabsorbed dual-horned profiles in our sample. We then designated the lowest and highest ratios (0.16 and 0.52, respectively) as the acceptable range of flux values for the underlying unabsorbed central trough in the line profiles of Mrk 6 and NGC 7469. We then linearly interpolated over the central absorption with 100,000 random draws between the minimum and maximum allowed trough height.  For each iteration, the total line flux was recorded, thereby building up a distribution of likely unabsorbed line flux measurements. The median of the resulting distribution is reported as the final T$_\textsc{{L}}$ value, and the uncertainties reflect the 1-$\sigma$ deviation from the median on either side of the distribution.
	
	While the optical angular extents of all galaxies with HI detections in our sample are encompassed by the $9\farcm1$ GBT beam, the total extent of the HI distribution of the closest objects may not be. We examined resolved HI maps to verify the angular HI extent of the nearest galaxies in our sample, whenever such maps were available in the literature.  %Resolved neutral hydrogen studies have shown that accreted, extraplanar gas can have slower rotation speeds than the disk gas (e.g., \citealt{fraternali2008,vargas2017}), thus could cause excess 21\,cm emission near systemic velocity. Gas expended from the disk and mixed with the hotter material in the halo of spiral galaxies can also decrease the cooling time of the coronal gas, leading to condensation and accretion onto the disk which is reflected as a lag in rotational velocity \citep{fraternali2017}. 
	
   Mrk 1044: The emission profile for this galaxy contains several flux peaks between the flanks (see Fig.\ \ref{batmen2}). This caused the initial $\textsc{BusyFit}$ model to fit the flank slopes while converging with a value of 0 for the central trough amplitude. In order to fit a weak dual-horn signature, we held the trough amplitude ($c$) fixed at 0.0001, the trough offset ($x_{p}$) fixed at 235, and the polynomial degree ($n$) fixed at 2. For this object, \cite{mw1984} employed the National Radio Astronomy Observatory (NRAO) 91\,m telescope's 384 channel autocorrelation spectrometer with 11 km s$^{-1}$ channel spacing and velocity resolution of 22 km s$^{-1}$. The spectrum reported by \cite{mw1984} contains a noise spike close to the redshifted side of the HI profile. The inclusion of the noise spike as part of the HI profile, due to low S/N and low resolution, is the most probable cause for the large discrepancy between their W$_{20}$ measurement of 489 km s$^{-1}$ and ours of 196.2$^{+4.3}_{-10.4}$ km s$^{-1}$ and the slight offset in their V$_{\textsc{R}}$ value of 4932 km s$^{-1}$ compared to ours of 4910.77$^{+0.69}_{-1.35}$ km s$^{-1}$. Measurements made by \cite{kh2009} based on observations with the Effelsberg 100\,m telescope's 8,192 channel autocorrelater are consistent with our measurements.
   
   Ark 120: The profile of Ark 120 has a slightly skewed dual-horn shape (see Fig.\ \ref{batmen2}). \cite{theureau2005} conducted an observation using the Nan\c cay 94\,m telescope with an 8,192 channel autocorrelation spectrometer, which resulted in an HI detection with S/N=3.8. The low S/N in the \cite{theureau2005} observations would make it difficult to fully characterize the weaker, blueshifted side of the line profile, and likely accounts for the discrepancies between their reported measurements and ours.  Where they find values of W$_{50}$=194 $\pm$ 33 km s$^{-1}$, W$_{20}$=233 $\pm$ 50 km s$^{-1}$, and V$_\textsc{{R}}$=9807 $\pm$ 17 km s$^{-1}$, we report values of W$_{50}$=337.3$^{+17.0}_{-13.1}$ km s$^{-1}$, W$_{20}$=344.4$^{+18.4}_{-17.9}$ km s$^{-1}$, and V$_\textsc{{R}}$=9806.38$^{+9.22}_{-4.51}$ km s$^{-1}$.  On the other hand, \cite{ho2008_data} report values of W$_{20}$=370.3 $\pm$ 6.8 km s$^{-1}$ and V$_\textsc{{R}}$=9809.2 $\pm$ 3.4 km s$^{-1}$ that are fairly consistent with our measurements within the uncertainties.
   
   MCG+08-11-011: \cite{mw1984} observed this target with the NRAO 91\,m telescope, which was then reprocessed \citep{edd2005} for inclusion in the Extragalactic Distance Database (EDD; \citealt{tully2009}). Both V$_\textsc{{R}}$ measurements (6146 and 6133 km s$^{-1}$, respectively) and the reprocessed width measurement (310$\pm$15 km s$^{-1}$) are consistent with our measurements of V$_\textsc{{R}}$=6133.26$^{+1.31}_{-1.04}$ and W$_{50}$=310.8$^{+6.7}_{-5.1}$ km s$^{-1}$.

   Mrk 374: Mrk 374 has a relatively faint emission line (see Fig.\ \ref{batmen2}) and we were only able to achieve a S/N of 4.7. The initial $\textsc{BusyFit}$ profile converged into a box-shaped best fit with a value of 0 for the central trough amplitude. In order to generate a trough feature while keeping the slopes of the flanks accurate, we held fixed the initial best fits to the flank slopes ($b_{1}$ and $b_{2}$) at 0.8, 0.5, respectively. We then fixed the central trough amplitude ($c$) at 0.01 and polynomial degree ($n$) at 1.5, which allowed a double horned profile shape to converge. Mrk 374 was previously observed by \cite{dc2004} with the 512 channel autocorrelator on the Nan\c cay telescope. They defined their recessional velocities in the radio convention 
(c($\lambda$$-$$\lambda_{0}$)/$\lambda$) instead of the optical convention (c($\lambda$$-$$\lambda_{0}$)/$\lambda_{0}$), which we have used in this work. The recessional velocity reported for Mrk 374 by \cite{dc2004} of 12780 $\pm$ 8 km s$^{-1}$ is equivalent to 13349 km s$^{-1}$ in the optical convention, as compared to our measurement of 13250.00$^{+0.07}_{-0.06}$ km s$^{-1}$ in the optical. They report a S/N of 6.9 with an integrated flux of 8.54 Jy km s$^{-1}$, which is equivalent to 17.08 K km s$^{-1}$, much higher than our flux of 0.59$^{+0.02}_{-0.01}$ K km s$^{-1}$. Their width measurements also consist of significant discrepancies relative to ours; they list W$_{50}$=74$\pm$16 km s$^{-1}$ and W$_{20}$=121$\pm$24 km s$^{-1}$, as opposed to our reported values of W$_{50}$=263.8$^{+0.9}_{-1.0}$ and W$_{20}$=276.3$^{+6.6}_{-0.9}$ km s$^{-1}$. However, their spectrum contains a strong Gaussian-shaped signal unlike the faint dual-horned profile which we report. \cite{dc2004} discuss that observations at recessional velocities near $\sim$ 12500 km s$^{-1}$ (in the radio convention) contained significant interference from radar signals, and that galaxies in their sample near those velocities are possibly unreliable. Therefore, it is possible that the target was misidentified in their work.

   Mrk 79: Two previous observations of this object with the GBT are reported in the HI Digital Catalog of \cite{edd2005}. Each observation reports a W$_{50}$ value (169$\pm$15, 155$\pm$7 km s$^{-1}$) and a recessional velocity (6657, 6659$\pm$5 km s$^{-1}$), and they are consistent with our measurements of W$_{50}$=154.4$^{+9.7}_{-4.5}$ and V$_\textsc{{R}}$=6657.41$^{+4.76}_{-3.88}$ km s$^{-1}$ within the uncertainties.
   
   NGC 2617: Previous observations include the Nan\c cay telescope \citep{hyperleda} and the Parkes telescope including two measurements from the HI Parkes All Sky Survey Catalogue \citep{meyer2004,doyle2005}. The HI data for NGC 2617 was also reprocessed \citep{theureau2006} for inclusion in the EDD. All previous reported measurements are in agreement with our measurements.
   
   Mrk 704: Mrk 704 has a narrow Gaussian emission profile shape (see Fig.\ \ref{batmen2}). The $\textsc{BusyFit}$ central trough, offset, and polynomial parameters ($c$, $x_{p}$, and $n$) were thus automatically fixed at 0, and we also held the left flank slope ($b_{1}$) fixed at 0.15 to allow the profile to converge. The spectrum reported by \cite{hutchings1989} from observations with the Arecibo telescope contains a very low S/N emission line blended with a noise spike on the blueshifted side. The larger W$_{20}$ value they report of 250 km s$^{-1}$ compared to ours of 57.8$^{+12.8}_{-12.9}$ km s$^{-1}$ is possibly due to the nearby noise spike's inclusion in the profile measurement. This would also explain the slight offset in their V$_\textsc{{R}}$ value of 9510 km s$^{-1}$ compared to our measurement of 9525.87$^{+1.60}_{-2.55}$ km s$^{-1}$.
    
	NGC 3227: The blueshifted side of the profile of NGC 3227 is significantly weaker in flux than the redshifted side, resulting in a dramatically asymmetric shape (see Fig.\ \ref{batmen}). It is interacting with its neighboring galaxy NGC 3226 \citep{tonry2001}, which is an elliptical galaxy, and in the resolved HI study by \cite{mundell1995}, they detected no HI emission from it. So while the interaction might be a possible explanation as to NGC 3227's skewed profile shape, NGC 3226 most likely does not contribute to the emission profile we have detected. The spectral resolution of the previously published spectra range from 6.6 km s$^{-1}$ \citep{dr1978} from observations which used the 91\,m telescope at the NRAO to 30 km s$^{-1}$ \citep{bf1979} for observations with the 305\,m Arecibo telescope. The V$_{\textsc{R}}$ measurement of 1284 $\pm$ 9 km s$^{-1}$ reported by \cite{dr1978} presents the biggest discrepancy with our V$_{\textsc{R}}$ measurement of 1144.74$^{+4.33}_{-0.82}$ km s$^{-1}$. The baseline in their spectrum contains significant residual fluctuations and possible source confusion; it appears the S/N in their spectrum is too low for identification of the blueshifted side of the profile. Our spectral resolution of 3.0 km s$^{-1}$ is higher than all previous spectra, and our higher S/N of 13.1 allows for clear identification of the entire profile. Our measurements of W$_{20}$ and V$_\textsc{{R}}$ are consistent with \cite{ho2008_data} within the uncertainties. \cite{martin1998} report a maximum HI angular diameter of NGC 3227 of 5$'$ based on the resolved study by \cite{mundell1995}, therefore the total 21\,cm emission is most likely enclosed by the $9\farcm1$ GBT beamwidth.
	
	NGC 3516: This is the first HI spectrum for NGC 3516, a relatively nearby galaxy, due to the extreme faintness of its HI emission. The total on-source observing time spent on this object was longer than any of our other targets by a large margin ($\sim$ 15.6 hours). The HI profile of NGC 3516 exhibits a Gaussian shape, and as such the $\textsc{BusyFit}$ model held the central trough amplitude, offset, and polynomial degree  parameters ($c$, $x_{p}$, and $n$) fixed at 0. We also held the right flank slope ($b_{2}$) fixed at 0.1 to allow the profile fit to converge.
    
    NGC 3783: Previous HI line widths are derived from observations with the Nan\c cay telescope \citep{theureau2006} and reprocessed for inclusion in the EDD \citep{edd}. Our values for W$_{\textsc{50}}$, W$_{\textsc{20}}$, and V$_{\textsc{R}}$ are consistent within the uncertainties.
    
    NGC 4051: Our reported measurements of W$_{50}$ and W$_{20}$ are consistent with those of \cite{tf1981}, obtained from the NRAO 91\,m telescope and reanalyzed for inclusion in the EDD. \cite{dr1978}, who used the same instrument as \cite{tf1981}, defined their profile widths as the half-width between points at one-quarter peak intensity corrected for the spectral resolution of the instrument of 6.6 km s$^{-1}$. Their reported value (doubled to achieve a full width at quarter intensity) of 268 km s$^{-1}$ is consistent with our similar measurement of W$_{20}$=264.5$^{+9.5}_{-2.9}$ km s$^{-1}$. All previous V$_{\textsc{R}}$ measurements are consistent with our measurements. The resolved HI study of \cite{liszt1995} reveals the diameter of the major axis is similar to that of the optical diameter at $5\farcm2$, therefore it is expected that all the HI emission is contained within the $9\farcm1$ GBT beam. 

    NGC 4151: Our measurements of W$_{50}$ and W$_{20}$ are approximately consistent with the values reported by \cite{tc1988}. The small discrepancies of $\sim$ 3 $-$ 10 km s$^{-1}$ most likely come from the difference in spectral resolution, channel span, and channel spacing. For relatively flat baselines such as that present in our spectrum of NGC 4151, low-order polynomial fits can possibly introduce low-level sinusoidal structure in the baseline. This can affect subsequent measurements and/or fits to the emission profile, and can thus result in small discrepancies in reported measurements. The NRAO 91\,m telescope employed by \cite{tc1988} produced a spectrum for NGC 4151 with a resolution of 11 km s$^{-1}$ over 192 channels. As with NGC 4051, \cite{dr1978} defined their width as the half-width between points at quarter-intensity corrected for a spectral resolution of 6.6 km s$^{-1}$. Their reported value (doubled to match a full width) of 156 km s$^{-1}$ is consistent with our measurement of W$_{20}$=152.5$^{+8.5}_{-0.7}$ km s$^{-1}$. The previous V$_{\textsc{R}}$ measurements are consistent with our measurements. In their neutral hydrogen study of NGC 4151, \cite{pedlar1992} report the extent of the spiral arms reach $\sim$ 6$'$ from its center, and the reanalysis of the same study by \cite{martin1998} from their compiled catalog of HI maps report the largest angular extent of the neutral hydrogen as $10\farcm4$. The $9\farcm1$ beam of the GBT may not fully enclose the total extent of the HI emission from NGC 4151, but only a small fraction is likely to have been missed.
    
    Mrk 766: Mrk 766 has a low S/N emission line (5.1; see Fig.\ \ref{batmen}) with a very faint dual-horn signature. With all eight $\textsc{BusyFit}$ parameters free, the initial fit was Gaussian in shape. In order to fit the weak horns, the right flank slope ($b_{2}$) was held fixed at 0.45 in addition to the polynomial degree ($n$) which was fixed at 2. There are no previous measurements of the HI emission from this galaxy.
    
    NGC 4593: Observations conducted by \cite{sd1987} and \cite{kh2009} employed the Jodrell Bank 76\,m MklA radio telescope's 1024 channel autocorrelation spectrometer with a velocity resolution of 7.3 km s$^{-1}$ and the Effelsberg 100\,m telescope's 8,192 channel autocorrelator with a velocity resolution of 4.1 km s$^{-1}$, respectively. Our reported resolution is 0.6 km s$^{-1}$, and our width and velocity measurements are consistent with theirs. NGC 4593 is also composed of many morphological components including an outer ring and a bar, which is a possible explanation for the fluctuating HI emission between the horns of its profile.
    
    MCG-06-30-015: The HI measurements in this work are the first reported for this galaxy. MCG-06-30-015 has one of the faintest emission lines that was detected (S/N=3.4) in our sample. To fit the asymmetric dual-horned profile, we held the central trough amplitude ($c$) fixed at 0.011, the half-width ($w$) fixed at 5, and the polynomial degree ($n$) fixed at 2.8.
    
    NGC 5548: Within the uncertainties, our measurement of W$_{50}$ is consistent with that of \cite{stierwalt2005} based on observations with the Arecibo telescope, with a S/N of 4.1 and spectral resolution of 8.5 km s$^{-1}$. The spectrum reported by \cite{bf1979}, also from Arecibo, has a low resolution of 30 km s$^{-1}$, as opposed to our smoothed velocity resolution of 4.8 km s$^{-1}$, leading to significant ambiguity in identification of the edges of their profile and their subsequent W$_{50}$ measurement of 110 km s$^{-1}$, compared to our W$_{50}$ measurement of 189.1$^{+10.3}_{-3.0}$ km s$^{-1}$. In the spectrum reported by \cite{mw1984} from Arecibo, with a velocity resolution of 22 km s$^{-1}$ and channel spacing of 11 km s$^{-1}$, the profile exhibits an extended, low-amplitude blueshifted wing, possibly leading to the discrepancy in their W$_{20}$ measurement of 472 km s$^{-1}$ in comparison to our W$_{20}$ measurement of 197.3$^{+11.8}_{-14.0}$ km s$^{-1}$. The same issue is present in the spectrum reported by \cite{ho2008_data} (velocity resolution of 5.15 km s$^{-1}$), leading to disagreement between their W$_{20}$ measurement of 321.1 $\pm$ 6.8 km s$^{-1}$ and our W$_{20}$ measurement. After smoothing our spectrum to match the lower velocity resolution of 22 km s$^{-1}$, we arrive at a W$_{20}$ measurement of 270 km s$^{-1}$, closer to the larger values of \cite{mw1984} and \cite{ho2008_data}. Our V$_{\textsc{R}}$ value is within the range of reported values. The deep optical imaging of NGC 5548 by \cite{tyson1998} reveals a low surface brightness arm wrapping around the galaxy, an extended tail, and ripples in the inner disk, all of which could contribute to the highly turbulent HI flux distribution present in our spectrum.
    
    Mrk 478: There are multiple emission peaks near the expected location of HI emission from Mrk 478 (see Fig.\ \ref{batmen2}), which is between 22484 $-$ 23700 km s$^{-1}$ \citep{richards2009,dv1991} from recessional velocities measured from optical emission lines. The systemic velocity of Mrk 478 is not well constrained, and we do not detect emission at or near the low end of its range of optical velocities. We fit independent Gaussians to the three peaks present in our spectrum to compare to the systemic velocities of galaxies in the NASA/IPAC Extragalactic Database (NED) in a $9\farcm1$ diameter neighbor search (equal to the GBT L-Band beam size). We measure the following velocities for each peak (from left to right): 23540$\pm$11, 23800$\pm$15, and 23980$\pm$4 km s$^{-1}$. The left peak's velocity is comparable to the nearby galaxy 2MASX J14415920+3527489 (V$_\textsc{{R}}$=23554 km s$^{-1}$; $2\farcm2$ to the NW). Our V$_\textsc{{R}}$ measurement for the center peak is near the reported velocities of neighboring galaxies 2MASX J14421361+3524459 (V$_\textsc{{R}}$=23738 km s$^{-1}$; $2\farcm3$ to the SW) and 2MASX J14421426+3528139 (V$_\textsc{{R}}$=23763 km s$^{-1}$; $2\farcm3$ to the NE). Lastly, within a $9\farcm$1 diameter neighbor search, there are no objects classified as galaxies near our measured velocity of the right peak. From the present information, it is unclear whether Mrk 478 exhibits a dual-horned shape, and it is also unclear as to which of the peaks represent emission from Mrk 478. Based on the similarity of the center and right peak's shape with that of the measured shapes of low S/N dual-horned profiles, as seen with other objects in our sample (SBS1116+583A, Mrk 817, Mrk 290), we have assumed the center and right peaks belong to the emission profile of Mrk 478. For analysis with $\textsc{BusyFit}$, we fixed the slope of the right flank ($b_{2}$) at 0.26, the half-width parameter ($w$) at 23, and the polynomial degree ($n$) at 3 in order to achieve a characteristic dual-horned fit. \cite{teng2013} observed Mrk 478 with the 100\,m GBT (identified as PG 1440+356 in their work), producing a spectrum with a resolution of $\sim$ 6 km s$^{-1}$ channel$^{-1}$ and S/N of 4.81. The large absorption feature present in their spectrum at $\sim$ 24000 km s$^{-1}$ is absent from ours. \cite{teng2013} note that the feature has a dramatic variability over short timescales and is also dependent on polarization, and their Figure 6 shows strong continuum fluctuations on month-long
periods. They report measurements of W$_{50}$=395$\pm$26, W$_{20}$=477$\pm$39, and V$_\textsc{{R}}$=23406$\pm$13 km s$^{-1}$, which differ significantly from our values of W$_{50}$=294.5$^{+11.1}_{-10.6}$, W$_{20}$=296.9$^{+13.6}_{-12.3}$, and V$_\textsc{{R}}$=23879.90$^{+5.54}_{-5.26}$ km s$^{-1}$. It is possible that they assumed the left peak was part of the HI emission from Mrk 478, leading to the discrepancy between their measurements and our measurements. If we include all three peaks in our measurement, we find a W$_{50}$ value of 476$^{+12}_{-10}$ km s$^{-1}$, a W$_{20}$ value of 477$^{+12}_{-12}$ km s$^{-1}$, and a V$_\textsc{{R}}$ value of 23752$^{+6}_{-6}$ km s$^{-1}$, closer to those of \cite{teng2013}. 
%Therefore, since the right peak consists of no confusion with nearby galaxies at its systemic velocity, and the center peak is near the reported systemic velocity for Mrk 478 of 23700$\pm$79 km s$^{-1}$ in NED, it is reasonable to assume that the right peak is the redshifted horn of Mrk 478.

%w50:	      476.121   +      12.3163   -      10.0952
%w20:	      477.845   +      12.3296   -      12.3156

	NGC 5940: All previous measurements for NGC 5940 originate from observations with the Arecibo telescope, including a reprocessed measurement for inclusion in the EDD. The measurements conducted by \cite{lewis1983} contain consistent W$_{50}$ and V$_\textsc{{R}}$ values with our values, however their W$_{20}$ value of 240 km s$^{-1}$ is slightly higher than ours of 204.5$^{+5.9}_{-6.5}$ km s$^{-1}$. The lower resolution of their spectrum ($\sim$ 8.2 km s$^{-1}$ compared to ours of 1.3 km s$^{-1}$) contributes to some of the discrepancy between the measurements, because we measure W$_{20}$=214 km s$^{-1}$ when we smooth our spectrum to match their resolution.  However, we expect that the lower S/N in their spectrum also contributes to the difference in W$_{20}$. The remaining measurements from \cite{mw1984}, \cite{lewis1987}, \cite{haynes2011}, \cite{edd2005}, and \cite{hyperleda} are consistent with our measurements within the uncertainties.
    
    Mrk 493: The HI profile of Mrk 493 exhibits a strong, narrow Gaussian shape (see Fig.\ \ref{batmen2}). As standard for fitting a Gaussian-shaped profile, the $\textsc{BusyFit}$ parameters controlling the central trough amplitude, offset, and polynomial degree ($c$, $x_{p}$, and $n$, respectively) were set to 0, and we found that we also needed to hold the half-width parameter ($w$) fixed at 10 to allow the profile fit to converge. All previous observations utilized the 305\,m Arecibo telescope. The V$_\textsc{{R}}$ reported by \cite{hg1984} is consistent with our measurement, and their W$_{50}$ value agrees with our measurement at the $\sim$ 2 $\sigma$ level. \cite{lewis1987} also reports a consistent V$_\textsc{{R}}$ value. They define their W$_{50}$ measurement of 35.7 km s$^{-1}$ as an un-smoothed width, which might account for the discrepancy, yet it is consistent with our W$_{50}$ of 54.6$^{+6.5}_{-8.4}$ km s$^{-1}$ at the $\sim$ 2 $\sigma$ level. Values by \cite{mw1984} are consistent with our measurements within the uncertainties.
    
    1H1934-063: \cite{hyperleda} report an HI line width of 248.4 $\pm$ 16.5 km s$^{-1}$ from observations with the Nan\c cay telescope. Their spectrum contains significant baseline fluctuations, causing a discrepancy both between our width measurement and theirs and their V$_{\textsc{R}}$ measurement of 3070 $\pm$ 7 km s$^{-1}$ compared to our value of  V$_{\textsc{R}}$ = 3191.42$^{+0.06}_{-0.09}$ km s$^{-1}$.
    
    NGC 6814: \cite{mw1984} observed this object with the NRAO 91\,m telescope and the 192 channel autocorrelation spectrometer. The low spectral resolution of 22 km s$^{-1}$ and channel spacing of 11 km s$^{-1}$ possibly account for the slightly larger value of W$_{20}$ that they report of 134 km s$^{-1}$ compared to our value of 105.1$^{+11.4}_{-8.4}$ km s$^{-1}$. All other previous measurements from \cite{shostak1978}, \cite{koribalski2004}, \cite{edd2005}, \cite{hr1989} are consistent with our measurements. \cite{liszt1995} estimate the HI diameter of NGC 6814 to be $\sim$ 7$'$ based on their resolved HI map, thus it is likely that the $9\farcm1$ GBT beam encompassed the total HI distribution of NGC 6814.
         
	NGC 7469: Our data for NGC 7469 contained large baseline fluctuations across the whole continuum in a significant amount of the scans which were not included in the final, accumulated spectrum. The absorption profile present in the HI spectrum of NGC 7469 (see Fig.\ \ref{batmen}) persists throughout the literature, causing significant uncertainty in the line flux measurements. \cite{rh1982} utilized the Effelsberg 100\,m telescope with spectral resolution of 13.2 km s$^{-1}$ and channel spacing of 11 km s$^{-1}$, near insufficient to identify the emission profile separate from the noise level. Observations with the Arecibo telescope (e.g., \citealt{mw1984, bf1979, ho2008_data}) all show self-absorption which is commented on in their analyses. \cite{bf1979}, \cite{rh1982}, and \cite{mw1984} also comment on the galaxy companion IC 5283. NED lists the radial velocity of IC 5283 as 4804 km s$^{-1}$, very near the velocity of the blueshifted flank of NGC 7469, with an angular separation of only $1\farcm3$ (well within the GBT L-Band beam). Thus it is likely that most previous studies have the emission from this companion galaxy blended with that of NGC 7469. When comparing our spectrum to those of \cite{mw1984}, \cite{ms1988}, and \cite{ho2008_data}, we note that their higher S/N spectra show an emission bump on the blueshifted wing of the profile, while our spectrum does not. It is likely that this feature is lost in the noise since we had to reject a large number of scans for this object. Consequently, we find a significantly different width for NGC 7469 than these previous studies. Reported W$_{50}$ measurements are as follows: 570 km s$^{-1}$ \citep{bf1979} and 515 km s$^{-1}$ \citep{rh1982}; previous W$_{20}$ measurements consist of: 525.1$\pm$6.8 km s$^{-1}$ \citep{ho2008_data}, 583 km s$^{-1}$ \citep{rh1982}, and 395 km s$^{-1}$ \citep{mw1984} We report measurements of W$_{50}=192.6^{+9.8}_{-13.9}$ and W$_{20}=196.2^{+12.7}_{-16.1}$ km s$^{-1}$. The previous V$_{\textsc{R}}$ measurements are consistent with our values within the uncertainties.
    
    The literature on radial velocities and redshifts for the remaining objects in the sample consist of measurement methods that do not rely on 21\,cm spectroscopy. For example, MCG-06-30-015 has a previous radial velocity measurement of 2323 $\pm$ 15 km s$^{-1}$ from the redshifting of infrared emission lines \citep{fisher1995}. Based on the 21\,cm emission, we report a measurement of V$_{\textsc{R}}=2353.53^{+4.15}_{-3.56}$ km s$^{-1}$. Mrk 279 has measurements of V$_{\textsc{R}}$ ranging from low estimates of 8904 $\pm$ 60 km s$^{-1}$ from the redshift of the strongest optical emission lines (e.g., H$\alpha$, [O III]; \citealt{op1987}), to high estimates of 9600 km s$^{-1}$ from the redshift of the H$\alpha$ emission line \citep{arakelian1971}. Our measurement of V$_{\textsc{R}}=9211.71^{+8.29}_{-6.49}$ km s$^{-1}$ is contained within the wide range of previous values. Similarly, Mrk 817 has a range of V$_{\textsc{R}}$ measurements from 9275 km s$^{-1}$ \citep{fouque1992} to 9430 $\pm$ 35 km s$^{-1}$ (IRAS redshift survey; \citealt{sh1988}). Our measurement of V$_{\textsc{R}}=9420.14^{+4.08}_{-3.91}$ km s$^{-1}$ is in agreement with the higher end of these measurements. Table \ref{z} lists the redshifts we have derived from our HI observations alongside previously published redshifts from a variety of observations and analysis methods.

\section{Distances and Masses}\label{sec:mass}
With the detection of HI 21\,cm emission from 31 AGN host galaxies, we can explore the gas properties of these galaxies compared to their stellar and central black hole properties. We also augmented our sample by including the dwarf Seyfert NGC 4395, as it should provide an interesting comparison as the lowest-mass AGN with a direct black hole mass constraint, hosted by a bulgeless low surface brightness galaxy. We describe here our adopted measurements and derived quantities for the baryonic properties of the galaxies. 
\subsection{Distances}\label{distances}
Only five of the galaxies that we detected have distance measurements independent of their redshifts. The sources of the distances to NGC 3227, NGC 3783, NGC 4051, NGC 4151, and NGC 4593 are summarized in \cite{misty2013}, but in brief, the measurements are generally the average of distances to galaxies within the same group, and were retrieved from the EDD. The exception is NGC 3227, which has an adopted distance that is the same as NGC 3226, with which it is interacting and which has a distance from the surface brightness fluctuation method \citep{tonry2001}. These five galaxy distances have been recalibrated with a Hubble constant of H$_{0}$=72 km s$^{-1}$ Mpc$^{-1}$ for consistency with the Hubble Space Telescope (HST) Key Project \citep{hubble_constant}. Additionally, NGC 4395 has a distance from Cepheid variables \citep{thim2004} of 4.1 $\pm$ 0.4 Mpc.

Many of the galaxies in our sample were included in \cite{misty2013} and \cite{misty2018}. For those objects, we adopt the luminosity distances (D$_\textsc{{L}}$) reported in those works, which are derived from the redshifts of each galaxy. Uncertainties of 500 km s$^{-1}$ were adopted due to the typical range of peculiar velocities reported by \cite{tully2008}. 

For Mrk 1044, MCG+08-11-011, Mrk 374, NGC 2617, Mrk 704, Mrk 478, NGC 5940, Mrk 493, and 1H1934-063, we use our redshift measurement from the HI emission line to estimate D$_\textsc{{L}}$. For consistency with the other objects in our sample, we adopt an uncertainty of 500 km s$^{-1}$ to account for peculiar velocities that may affect the distance derived from the redshift. Adopted distances are listed in Table \ref{masses}.

\subsection{HI and Total Gas Mass}\label{hi_gas_mass}
The integrated flux of the HI emission line allows the atomic gas mass to be estimated because the intensity from the spin-flip radiation of optically thin sources is related to the number of HI atoms in a 1\,cm$^{2}$ cross-section column. The measured flux is thus directly related to the total number of HI atoms in the beam, and the mass is given by 
\begin{equation}\label{eq:2}
\frac{M_{HI}}{M_\odot}=[1.2 \times 10^{5}D^{2}]\sum_{i=1}^{n}T_{L}(i)\Delta v
\end{equation}\label{MHI}
\citep{HImass}, where D is the distance in Mpc (see Table \ref{masses}), the summation is over channels spanning the HI emission-line profile, and $\Delta$v is the channel width in km s$^{-1}$. 

The HI mass of a galaxy is then related to total gas mass as
\begin{equation}\label{eq:3}
M_{GAS}=1.4M_{HI}
\end{equation}
\citep{cox2000}. The scale factor of 1.4 accounts for the amount of atomic helium gas in the galaxy assuming solar abundance. As stated previously, H$_{2}$ is the next most prevalent gas phase in disk galaxies. However, \cite{mcgaugh2012} revisited several molecular gas content estimation techniques (e.g., \citealt{yk1989}), and concluded that there is no compelling evidence for significant sources of baryonic matter in disk galaxies other than what can be directly detected through observations, and that the molecular gas contribution (in gas-rich spirals) is usually smaller than the uncertainty in the M$_{\textsc{HI}}$ calculation. The uncertainties on the gas masses are primarily set by the uncertainties on the galaxy distances, but a small contribution also comes from the uncertainty on the integrated HI line flux. The M$_\textsc{{GAS}}$ data used in \cite{mcgaugh2012} included uncertainties between 0.05 - 0.41 dex, consistent with what we find. Therefore, we assume that HI and helium accounts for approximately all significant gas mass contributions, with other phases and molecular gas providing a negligible contribution.

We derived M$_\textsc{{GAS}}$ for NGC 4395 using the HI flux available in the All Digital HI Catalog \citep{edd2005} with Equations \ref{eq:2} and \ref{eq:3}. The 140-foot Green Bank telescope was used to observe NGC 4395 and produce the subsequent HI flux we have used here. The beam size of the 140-foot is 21$'$, large enough to encompass the angular extent of NGC 4395, therefore allowing use of Equation \ref{eq:2}.

\subsection{Stellar and Baryonic Mass}\label{star_bary_masses}
In order to derive baryonic masses, we adopt the stellar masses determined by \cite{misty2018} for the majority of our targets. In that study, images in the $V$ and $H$ passbands from HST and the WIYN High-Resolution Infrared Camera, respectively, were modeled with $\textsc{Galfit}$ to separate the two-dimensional surface brightness components of the galaxy from the background sky and the AGN point source. $V-H$ colors were derived from the fits to the galaxies and were used with the \cite{bd2001} prescriptions to estimate the stellar mass-to-light (M/L) ratios, and therefore the stellar masses, of the galaxies.  We adopt the stellar masses based on the \cite{bd2001} prescriptions for the 22 galaxies in common between this work and that of \cite{misty2018}, and list them in Table \ref{masses}. 
\begin{figure}
\includegraphics[trim={2.9cm 9cm 0cm 8.9cm},clip,scale=0.55]{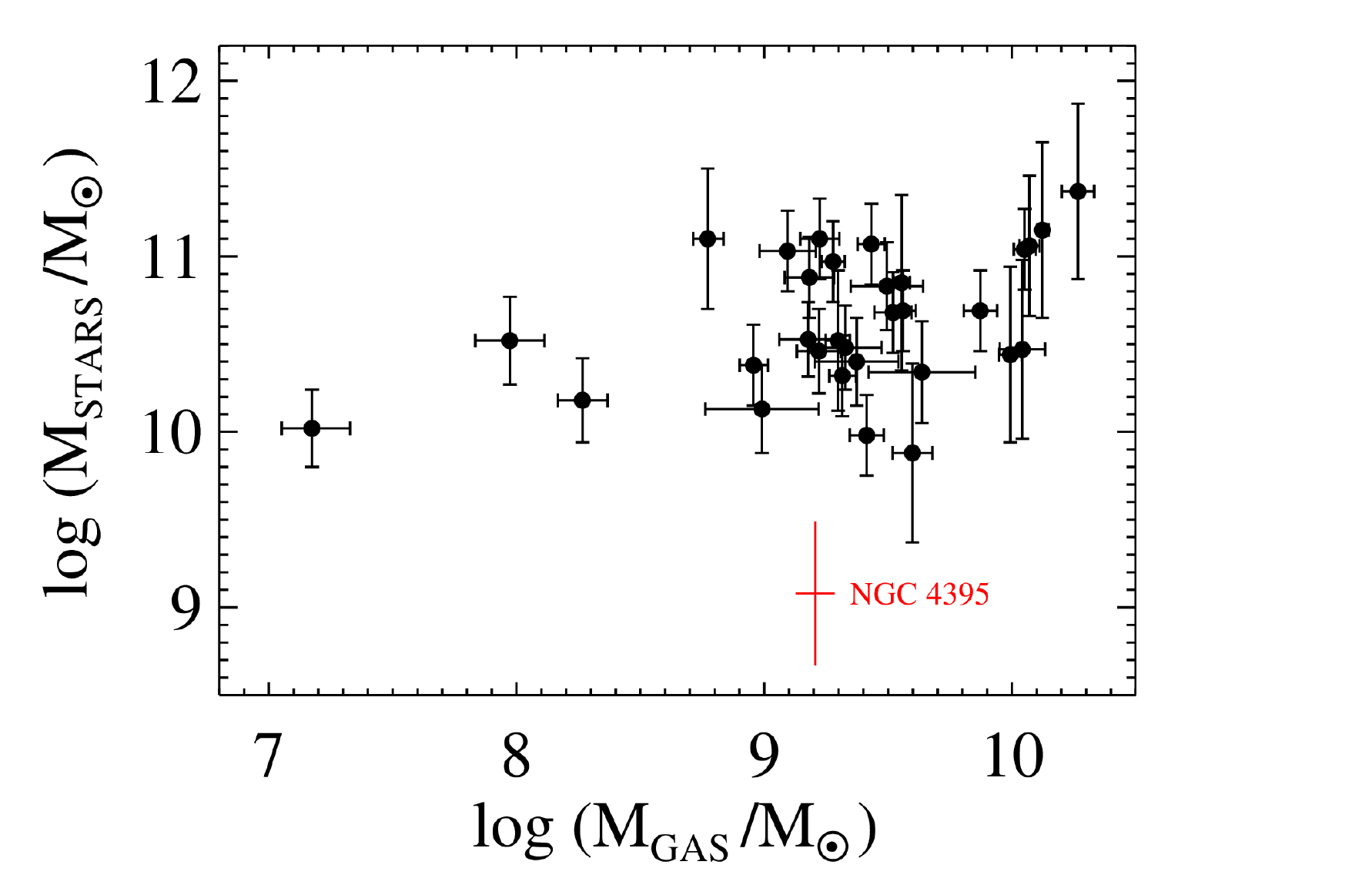}
\caption{Comparison between M$_\textsc{{GAS}}$ and M$_\textsc{{stars}}$. There is a slight preference for gas mass to trace stellar mass, but the range of stellar masses is relatively small and the scatter is quite large.} \label{hi_stars}
\end{figure}

For those objects in our sample that were not included in the \cite{misty2018} study, we estimated the stellar masses based on in-hand data, magnitudes and colors taken from the literature, or a combination of the two. The stellar mass for NGC 5548 was determined from in-hand HST $V$-band and Apache Point Observatory\footnote{Based in part on observations obtained with the Apache Point Observatory
3.5 m telescope, which is owned and operated by the Astrophysical Research
Consortium.} $H$-band images in the same manner as the \cite{misty2018} sample. For MCG-06-30-015 and 1H1934-063, we estimated the stellar masses using the same method and prescription from \cite{bd2001} with our in-hand HST $V$-band images combined with ground-based $K$-band images from the VISTA Hemisphere Survey\footnote{Based on observations obtained as part of the VISTA Hemisphere Survey, ESO Progam, 179.A-2010 (PI: McMahon.)} (VHS; \citealt{mcmahon2013}). 

The stellar masses for Mrk 704, NGC 5940, and Mrk 290 were estimated using the disk $B-V$ colors reported by \cite{granato1993}. The disk colors will not be affected by the central AGN; however, they will also be missing much of the color contributed by the bulge, which influences the total color and therefore the M/L ratio. Due to this limitation, we adopt an uncertainty of 0.28\, mag on the colors. We then combined these colors with in-hand HST $V$-band magnitudes to estimate the stellar M/L ratios and thus stellar masses. We adopted a typical uncertainty of 0.2\,mag for the HST $V$-band magnitudes, consistent with \cite{misty2018}.

The stellar masses for Mrk 1044, MCG-08-11-011, Mrk 374, NGC 2617, Mrk 478, and Mrk 493 were derived by first estimating the mean galaxy color based on their morphological types (\citealt{buta1994}, Table 6). For MCG-08-11-011, Mrk 374, NGC 2617, and Mrk 478, we then combined the estimated color with in-hand HST $V$-band magnitudes to estimate the stellar M/L ratio using the \cite{bd2001} prescriptions, and thus constrain the total stellar mass. In the cases of Mrk 1044 and Mrk 493, where we did not have $V$-band HST imaging in hand, we utilized the surface brightness decompositions of \cite{wang2014} to determine the fraction of the AGN luminosity to the total galaxy luminosity. We then corrected the total $V$-band magnitudes of these galaxies reported by \cite{mackenty1990} for the estimated contributions of the central AGNs. Combined with the galaxy color estimated from the morphological type, we were able to estimate the stellar M/L and thus the total stellar mass. For the HST $V$-band magnitudes, we adopted a typical uncertainty of 0.2\,mag. For the mean galaxy colors and magnitudes from \cite{mackenty1990}, we conservatively adopted uncertainties of 0.35\,mag and 0.25\,mag, respectively. 

We derived the stellar mass for NGC 4395 using the integrated $B$-band magnitude and $B-V$ color from \cite{prugniel1998}, and the stellar M/L ratio from \cite{bd2001}. The contamination from the AGN hosted by NGC 4395 is negligible, yet we conservatively adopt uncertainties of 0.28\,mag for the color and 0.2\,mag for the magnitude.

Finally, baryonic masses were calculated simply as 
\begin{equation}
M_{BARY}=M_{STARS}+M_{GAS}
\end{equation}
and are reported in Table \ref{masses}.
\pagebreak
\subsection{Black Hole Mass}\label{bhm}
The parent sample for the galaxies in this study is the reverberation sample of AGNs with direct black hole mass constraints.  Reverberation mapping \citep{{bm1982},{peterson1993}} is a light-echo technique in which the radius of the spatially-unresolved broad line region (R$_\textsc{{BLR}}$) is measured. The BLR of an AGN is composed of optically thick gas moving at large Doppler velocities deep in the potential well of the black hole. The BLR gas is photoionized by the continuum emission source (likely the accretion disk) and gives rise to the characteristic broad emission lines seen in the spectra of Type 1 AGNs. Variations in the continuum flux drive variations in the broad emission line flux, but delayed in time from the observer's point of view due to the extra path length most of the BLR light must travel to the observer. This time lag value provides a direct measure of the BLR size. Combining R$_\textsc{{BLR}}$ and the Doppler-broadened emission line width through the virial theorem yields a constraint on the total mass enclosed within the BLR, the vast majority of which is the mass of the SMBH. 
\begin{figure}
\includegraphics[trim={2.7cm 9cm 0cm 8.9cm},clip,scale=0.55]{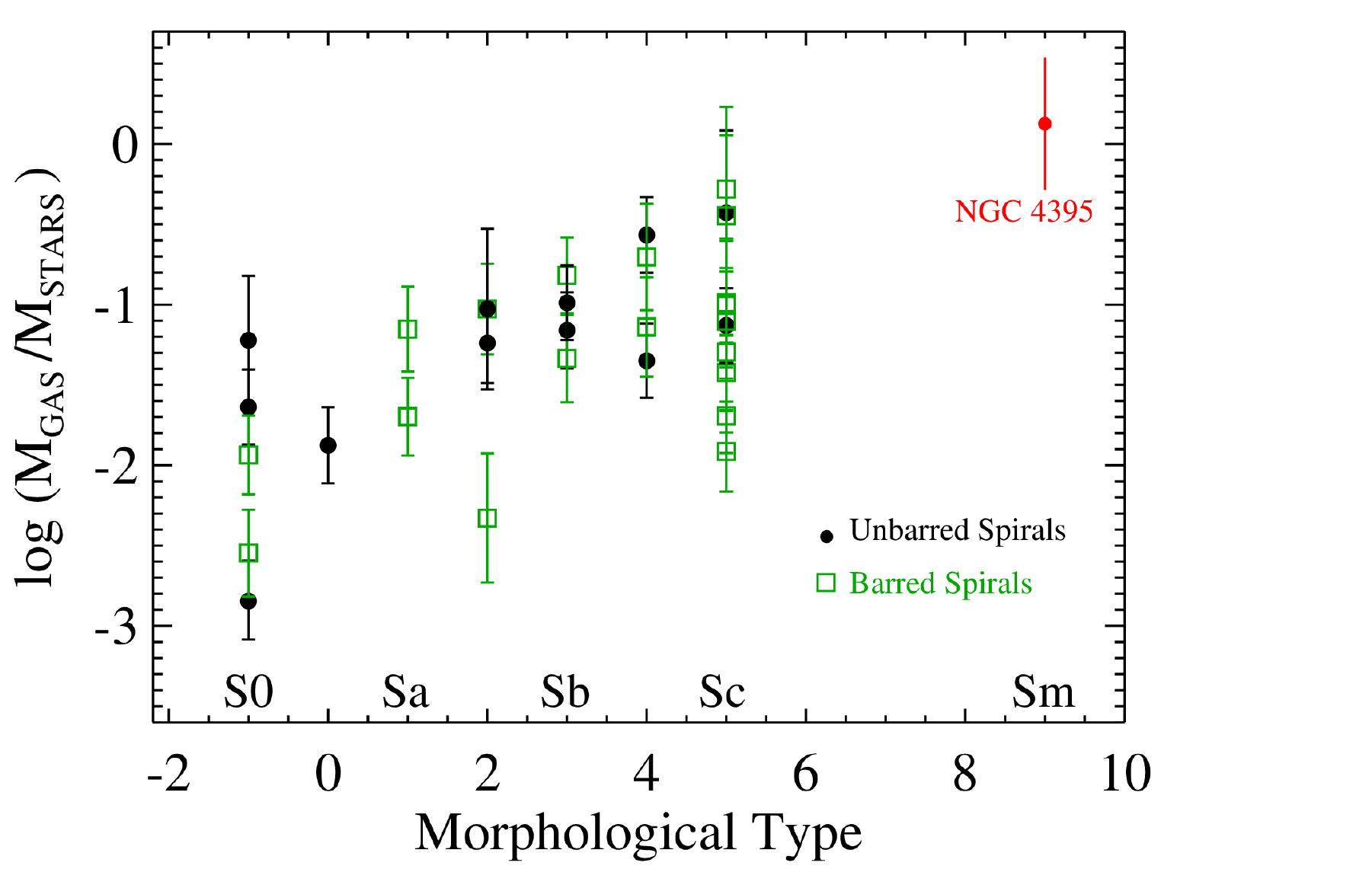}
\caption{Gas fraction as a function of galaxy morphological type. Morphologies are either those listed in NED or from the derived B/T ratios, which were the results of the surface brightness fits carried out by \cite{misty2009a}, \cite{misty2013}, \cite{misty2016}, and \cite{misty2018} (see Sec.\ \ref{gas/stars}, Table \ref{masses}). Morphologies based on B/T values were assigned according to the mean of the distributions in Figure 6 of \cite{kent1985}. The black circles are unbarred spirals, the green squares are barred spirals.} \label{morphologies}
\end{figure}

The black hole masses adopted here are from the AGN Black Hole Mass Database \citep{bhdatabase}, which is a compilation of reverberation-based M$_{\textsc{BH
}}$values. Thus, the masses are calculated as 
\begin{equation}\label{bh_eq}
M_{\textsc{BH}}=f\frac{c\tau V^{2}}{G}
\end{equation}
where $c\tau$ is the time delay for a broad emission line, $V$ is the width of the line, and $G$ is the gravitational constant. The $f$ term is an order-unity scaling factor which accounts for the unknown kinematics and geometry of the BLR gas in AGNs. It is generally derived by assuming that the AGN black hole mass - stellar velocity dispersion relationship (M$_\textsc{{BH}} -\sigma_\textsc{{stars}}$; \citealt{{fm2000},{gebhardt2000}}) is the same as that of nearby galaxies with black hole masses from dynamical modeling. We adopt $\langle f \rangle = 4.3$ which was determined by \cite{grier2013}.
\begin{figure*}
\includegraphics[trim={1.2cm 10.5cm 0cm 10.3cm},clip,scale=0.97]{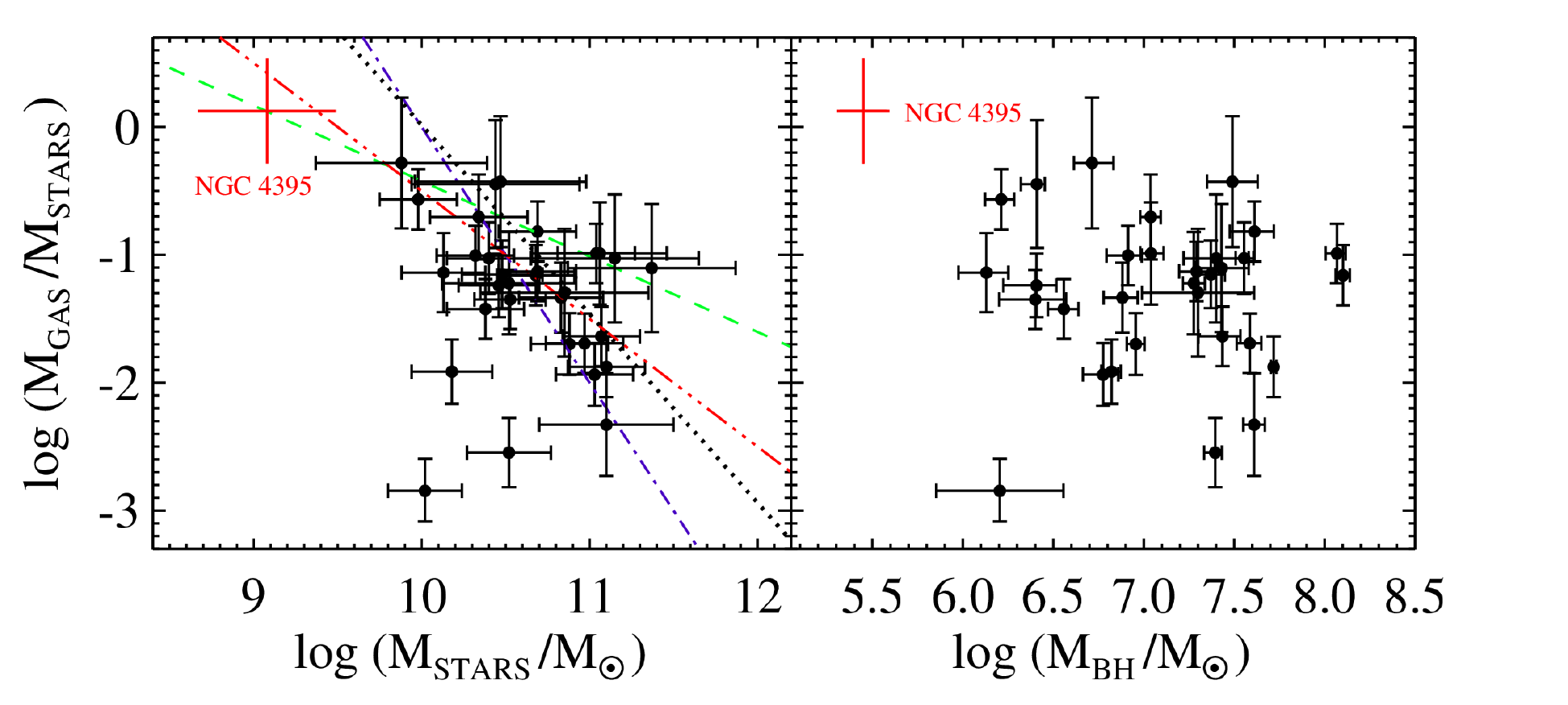}
\caption{M$_\textsc{{GAS}}$/M$_\textsc{{stars}}$ as a function of M$_\textsc{{stars}}$ (left) and M$_\textsc{{BH}}$ (right). The trend in the left plot shows decreasing mass fraction with increasing values of M$_\textsc{{stars}}$. The black dotted line is our best fit with a scatter of (0.41 $\pm$ 0.24) dex. NGC 4395 is not included in the fit. The green dash line is the best-fit line found by \cite{stewart2009} which characterizes the gas mass fraction of data from \cite{mcgaugh2005}. The sample used by \cite{stewart2009} is biased towards gas-rich spirals, while our sample is stellar-rich, which is a possible explanation to the data lying beneath the line. The blue dot-dash line shows a "closed box" relation that demonstrates direct conversion from M$_\textsc{{GAS}}$ to M$_\textsc{{stars}}$. The red dot-dot-dot-dash line shows a constant M$_\textsc{{GAS}}$ relation, where M$_\textsc{{GAS}}$ remains fixed at the approximate average value of our sample at M$_\textsc{{GAS}} \sim$ 10$^{9.5}$ M$_{\odot}$ while M$_\textsc{{stars}}$ varies. On the right plot, there is no evidence of a trend for M$_\textsc{{GAS}}$/M$_\textsc{{stars}}$ as a function of M$_\textsc{{BH}}$.}\label{fractions}
\end{figure*}

A few of the AGNs in our sample are included in the in-progress updates to the AGN Black Hole Mass Database, and thus require a little more explanation. For MCG+08-11-011, Mrk 374, and NGC 2617, we employed the virial M$_\textsc{BH}$ from \cite{Fausnaugh2017}, but scaled to match our adopted $\langle f \rangle$ value. Similarly, we employed the virial M$_\textsc{BH}$ for Mrk 704 from \cite{derosa2018} and rescaled it with our adopted $\langle f \rangle$ value. For Mrk 1044 and Mrk 493, we utilized the H$\beta$ time delay ($\tau_{H\beta}$) measurement from \cite{hu2015} and the width of the H$\beta$ emission line in the variable, root mean square (rms) spectrum ($\sigma_{line}$) from \cite{du2016} together with our adopted $\langle f \rangle$ value to calculate M$_\textsc{BH}$. For NGC 5940 we utilized the $\sigma_{line}$ value from \cite{barth2015} and the $\tau_{H\beta}$ value from \cite{barth2013} to determine a black hole mass.

We have adopted a preliminary black hole mass for Mrk 478 based on early analysis of in-hand reverberation-mapping data (de Rosa, private communication). The black hole mass for 1H1934-063 is based on current work on in-hand reverberation-mapping data from Bentz et al. (2019, in prep). 

\section{Discussion}\label{sec:discussion}
In the past two decades there has been a surge of studies focusing on scaling relationships between host galaxy and SMBH characteristics, and these relationships seem to suggest galaxy $-$ black hole coevolution. Such empirical scaling relations include the relationship between M$_\textsc{{BH}}$ and the luminosity of the bulge (M$_\textsc{{BH}}$ $-$ L$_\textsc{{bulge}}$; \citealt{kr1995}), the M$_\textsc{{BH}}$ $-$ stellar velocity dispersion relation \citep{{fm2000},{gebhardt2000}}, and the most recent calibration of the M$_\textsc{{BH}}$ $-$ M$_\textsc{{stars}}$ relation found by \cite{misty2018}. Many of these scaling relationships are used as inputs or constraints to cosmological galaxy simulations in an attempt to further understand the details of the symbiotic relationship between galaxies and black holes (e.g., \citealt{steinborn2015,volonteri2016,mutlu2018}). Here, we explore potential scaling relationships that include M$_\textsc{GAS}$.

Linear regressions were carried out using the Bayesian method-based algorithm \textlcsc{linmix$\_$err} \citep{linmix_err}, which accounts for measurement uncertainties in both variables and includes an element of random scatter. We report the median and 1-$\sigma$ deviations of large, random samples from the posterior probability distribution as the measurement and uncertainty for the slope, intercept, and scatter of each relationship.

\subsection{Gas Mass - Stellar Mass Relationship}\label{gas/stars}
We first explored the relationship between M$_\textsc{{GAS}}$ and M$_\textsc{{stars}}$ in our sample. We might expect to see smaller amounts of gas in galaxies with higher stellar mass if the gas content of these galaxies is not replenished as quickly as it is used up for star formation.

Fig.\ \ref{hi_stars} displays M$_\textsc{{stars}}$ vs.\ M$_\textsc{{GAS}}$. There is a slight tendency for lower stellar mass to correspond to lower gas mass. The range of stellar and gas masses covered by our sample is fairly small, however, given that most of the points are clumped together at 9\,$<$\,log\,(M$_\textsc{{GAS}}$/M$_{\odot}$)\,$<$ 10 and 10\,$<$\,log\,(M$_\textsc{{stars}}$/M$_{\odot}$)\,$<$\,11, and the scatter is quite large.

We next examined whether the fraction of gas-to-stellar content in the galaxies, M$_\textsc{{GAS}}$/M$_\textsc{{stars}}$, might serve as an indicator of morphological type. One might expect the gas content to change as a function of morphology, with lower M$_\textsc{{GAS}}$ for early-type spirals, and higher values for later-types. 

To explore this, we adopted galaxy morphologies from NED for those galaxies where previous ground-based data provided sufficient angular resolution to determine the morphology (most of the NGC objects, for example). For the more distant and compact galaxies, we determined the morphological type based on the bulge-to-total (B/T) luminosity ratio as follows. For most of the objects, the surface brightness decompositions of \cite{misty2009a}, \cite{misty2013}, \cite{misty2016}, and \cite{misty2018} were used to calculate B/T values, and these were compared to the distributions of B/T relative to galaxy morphology presented by \cite{kent1985} in their Figure 6.  We then adopted the morphology associated with the mean B/T value that most closely matched the B/T measured for each galaxy. For MCG+08-11-011 Mrk 374, NGC 2617, Mrk 704, Mrk 478, NGC 5940, and 1H1934-063, we calculated the B/T ratios from surface brightness decompositions of in-hand HST $V$-band images. Finally, for Mrk 1044 and Mrk 493, we used the surface brightness decompositions from \cite{wang2014} in order to derive the B/T ratios. 

Fig.\ \ref{morphologies} shows the gas-to-stellar fraction as a function of morphology. The black circles in Fig.\ \ref{morphologies} denote unbarred spirals, and the green squares denote barred spirals. The gas-to-stellar fraction appears to be approximately constant as a function of morphology for the barred spirals, albeit with a large scatter and with a lower typical M$_\textsc{{GAS}}$/M$_\textsc{{stars}}$ for SB0 galaxies. However the unbarred spirals show a slight preference for a higher gas fraction at later types, especially when the unbarred Sm galaxy NGC 4395 is included.

In addition, we explored the gas-to-stellar fraction as a function of M$_\textsc{{stars}}$. If the gas reservoir of the galaxy is never refueled by accretion onto the disk, we might expect to see evidence of decreasing gas for galaxies of higher stellar content. However, there would be little evidence of this trend if accretion is ongoing and the reservoir is steadily refueled. In the left panel of Fig.\ \ref{fractions} we plot the fraction M$_\textsc{{GAS}}$/M$_\textsc{{stars}}$ versus M$_\textsc{{stars}}$ and compare to several simple scenarios. The black dotted line shows the best formal fit to the data points and has the form 
\begin{equation}
\begin{split}
\hspace*{-0.7cm}\text{log}\frac{M_{GAS}}{M_{STARS}}=(-1.48 \pm 1.44)\text{log}\bigg(\frac{M_{STARS}}{10^{11}M_{\odot}}\bigg)\\-(1.46 \pm 0.14)
\end{split}
\end{equation}
with a scatter of (0.41 $\pm$ 0.24) dex. NGC 4395 is not included in our formal fit. The blue dot-dash line represents a "closed-box" relation, where there is a one-to-one correlation between the decrease in M$_\textsc{{GAS}}$ and increase in M$_\textsc{{stars}}$, therefore no cold gas accretion. The slope is slightly steeper than the trend of the data, suggesting that some refueling must be occurring on average for the galaxies in our sample. The red dot-dot-dot-dash line represents a constant M$_\textsc{{GAS}}$ relation, where gas is assumed to be replenished at the rate that it is used up for star formation. The slope of this relationship is shallower than our formal fit (which has large uncertainties), suggesting that the average galaxy in this sample is replenishing its gas reservoir, but at a rate that is slower than the gas is being used up. If NGC 4395 is included in the fit, we find a shallower slope of ($-$ 0.77 $\pm$ 0.42), much closer to that of the constant M$_\textsc{{GAS}}$ relation.

\begin{figure*}
\includegraphics[trim={1.2cm 10.5cm 0.5cm 10cm},clip,scale=0.95]{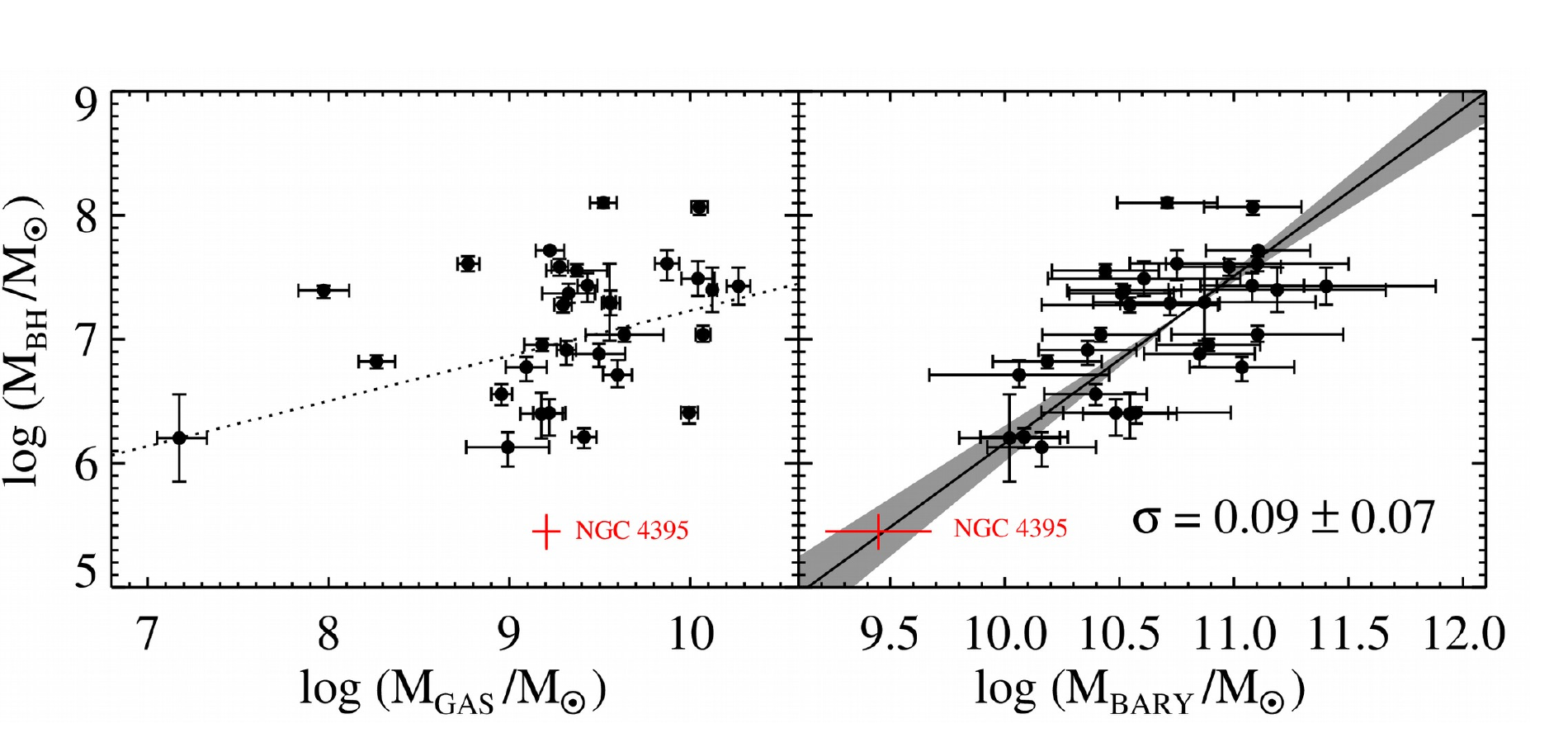}
\caption{M$_\textsc{{BH}}$ as a function of M$_\textsc{{GAS}}$ (left) and M$_\textsc{{BARY}}$ (right). For the data in the left plot, the formal fit for the M$_\textsc{{GAS}}$-M$_\textsc{{BH}}$ relation includes a scatter of (0.32 $\pm$ 0.09) dex. On the right, the M$_\textsc{{BARY}}$-M$_\textsc{{BH}}$ relation seems to exhibit a stronger correlation with less scatter. The red cross in the right panel is a derived baryonic mass for NGC 4395.  NGC 4395 is not inluded in the fit to the black points, nevertheless it seems to follow the same relationship demonstrated by more massive galaxies.}\label{relationships}
\end{figure*}
\cite{mcgaugh2005} compiled a sample of galaxies with extended 21\,cm rotation curves and derived gas and stellar masses for an in-depth baryonic Tully-Fisher relation study. \cite{stewart2009}, in their simulation of the baryonic content of galaxy mergers, assigned gas to the simulated galaxies by quantifying the relation between M$_\textsc{{GAS}}$/M$_\textsc{{stars}}$ as a function of M$_\textsc{{stars}}$ using the results from \cite{mcgaugh2005}. Their M$_\textsc{{GAS}}$/M$_\textsc{{stars}}$ as a function of M$_\textsc{{stars}}$ at $z$ $=$ 0 is represented by the green dashed line in Fig.\ \ref{fractions}. The sample from \cite{mcgaugh2005} used by \cite{stewart2009} is biased towards gas-rich galaxies, which explains why their relationship appears to serve as an upper limit to our sample of galaxies. The general trend of decreasing gas fraction as M$_\textsc{{stars}}$ increases, however, is consistent between their sample and ours. 

A recent study by \cite{calette2018} included a large sample of early and late-type galaxies from the literature, attempting to homogenize their sample as much as possible, checking against potential biases in the process (i.e., selection effects, upper limits, distances which correct for peculiar motions). They found an overall decrease in M$_\textsc{{GAS}}$/M$_\textsc{{stars}}$ as a function of M$_\textsc{{stars}}$ similar to \cite{stewart2009}. However, their average slope from the single- and double-power law best fits is much shallower than ours and the sample of \cite{mcgaugh2005}. \cite{bradford2015} conducted a study of the baryonic content of galaxies selected from the Sloan Digital Sky Survey DR8 using isolated HI galaxy detections from the 40\% ALFALFA survey \citep{haynes2011}, which is a publicly available blind, drift-scan HI survey from the Arecibo Observatory. They also find a decreasing atomic-gas-to-stellar fraction as a function of stellar mass, with a break into a steeper slope at M$_\textsc{{stars}}$ $\approx$ 10$^{9}$ M$_{\odot}$. The slope of the relationship they find after the break is very similar to our best-fit slope.

In addition, we have also explored M$_\textsc{{GAS}}$/M$_\textsc{{stars}}$ relative to M$_\textsc{{BH}}$, as shown in the right-hand panel of Fig.\ \ref{fractions}. We might expect M$_\textsc{{stars}}$ and M$_\textsc{{BH}}$ to increase together, since the gas reservoir is used to create stars and fuel the growth of the SMBH. However, we find no evidence for a trend with this sample of AGNs. Instead, our sample demonstrates a relatively constant value of M$_\textsc{{GAS}}$/M$_\textsc{{stars}}$ as a function of M$_\textsc{{BH}}$, albeit with a large scatter. A formal fit finds a slope of ($-$ 0.02 $\pm$ 0.38), which is consistent with zero.

\subsection{Gas Mass - Black Hole Mass Relationship}
There have been many previous attempts to explore the connection between gas in galaxies and the central supermassive black hole. The vast majority of these studies have focused on AGN characteristics, such as luminosity or accretion rate, instead of the black hole mass itself. Such studies have included, for example, exploring correlations between Seyfert nucleus luminosity and HI emission peculiarities \citep{heckman1978}, the link between the cold gas reservoir and AGN accretion \citep{gorkom1989,pt1998}, and the connection between M$_{\textsc{HI}}$/M$_{\textsc{stars}}$ and black hole accretion rate \citep{fabello2011}. The literature is inconclusive on these themes with other studies finding no evidence of mass transfer from the outer galactic regions to the central AGN when comparing gas content to near-infrared nuclear activity \citep{bieging1983} and finding no discernable connection between global gas content and AGN presence \citep{ho2008_analysis}. However, even with the myriad of studies that do exist, it appears that no one has yet examined the relationship between gas mass and black hole mass.

The stellar content of galaxies seems to correlate with the black hole mass, for example the M$_\textsc{BH}$ $-$ L$_\textsc{bulge}$ \citep{kr1995}, M$_\textsc{BH}$ $-$ $\sigma_\textsc{stars}$ \citep{fm2000}, and M$_\textsc{BH}$ $-$ M$_\textsc{stars}$ \citep{misty2018} relations. Here we explore whether the gas content also demonstrates a relation to M$_\textsc{BH}$.

We plot the reverberation-based black hole masses vs.\ the gas masses in Fig.\ \ref{relationships}.  There is a weak correlation, with a slight preference for more massive black holes to live in galaxies with larger gas reservoirs, but there is also a large scatter. A formal fit between M$_\textsc{{GAS}}$ vs M$_\textsc{{BH}}$ finds:
\begin{equation}
\text{log}\frac{M_{BH}}{M_{\odot}}=(0.36 \pm 0.10)\text{log}\bigg(\frac{M_{GAS}}{10^{9}M_{\odot}}\bigg)+(6.87 \pm 0.03)
\end{equation}
with a scatter of (0.32 $\pm$ 0.09) dex. We also examined whether morphological type played a role in where objects fell in Fig.\ \ref{relationships}, but found no obvious trend. 

\subsection{Baryonic Mass - Black Hole Mass Relationship}
\cite{misty2018} recently calibrated the scaling relationship of M$_\textsc{{BH}}$ to M$_\textsc{{stars}}$ for the AGN hosts in the reverberation sample. Their best fit based on the M/L ratio predictions of \cite{bd2001} has a slope of (1.69$\pm$0.46) and an intercept of (8.05$\pm$0.18), with a scatter of (0.38$\pm$0.13) dex. In the right panel of Fig.\ \ref{relationships}, we show the relationship between M$_\textsc{{BH}}$ and total baryonic masses for galaxies in our sample, many of which were included in the study by \cite{misty2018}. The best fit is given by:
\begin{equation}
\text{log}\frac{M_{BH}}{M_{\odot}}=(1.35 \pm 0.18)\text{log}\bigg(\frac{M_{BARY}}{10^{11}M_{\odot}}\bigg)+(7.51 \pm 0.04)
\end{equation}
with a scatter of (0.09 $\pm$ 0.07) dex. The fits to both M$_\textsc{{BH}}$ $-$ M$_\textsc{{BARY}}$ and M$_\textsc{{BH}}$ $-$ M$_\textsc{{stars}}$ are normalized at 10$^{11}$ M$_{\odot}$, allowing for easier comparison. The stellar content accounts for the majority of the baryonic mass in these galaxies, so the slopes of the two relationships are formally indistinguishable.  The typical fraction of M$_\textsc{{GAS}}$/M$_\textsc{{stars}}$ is about 10\% for these galaxies, but reaches as high as 52\% for Mrk 1044 and 134\% for NGC 4395. The slight increase in baryonic mass over stellar mass accounts for the 0.5 dex shift in the intercept between the two relationships.

Interestingly, even though NGC 4395 has a significantly larger M$_\textsc{{GAS}}$/M$_\textsc{{stars}}$ than any of the other galaxies in our sample, it appears to follow the same relationship between black hole mass and baryonic mass as the other galaxies in our sample. NGC 4395 is not included in our fit to the rest of the data, but if we include it, we find a very similar slope of (1.34 $\pm$ 0.09). This initial study suggests that the SMBHs of AGNs are not only correlated to their stellar content, but the total baryonic mass.

\section{Summary} \label{summary}
We present results from HI spectroscopy of 44 AGNs with reverberation-mapped black hole masses. We detect HI 21\,cm emission in 31 of them, 12 of which are the first reported 21\,cm detections. Measurements of the integrated HI fluxes, W$_{50}$, W$_{20}$, and V$_{\textsc{R}}$ values are determined with two independent methods and are found to be generally consistent. From the HI fluxes, we determine M$_{\textsc{GAS}}$ for each galaxy. Using the stellar masses provided by \cite{misty2018} as well as derived M$_{\textsc{stars}}$ values from in-hand data and the literature, we also produce total M$_{\textsc{BARY}}$ values for the galaxies in our sample. 

We have explored a number of relationships involving M$_{\textsc{GAS}}$. We find no evidence for a correlation between M$_{\textsc{stars}}$ and M$_{\textsc{GAS}}$. We find a weak correlation between M$_{\textsc{BH}}$ and M$_{\textsc{GAS}}$, albeit with a large scatter, and with no obvious trends based on morphological type. We find that the typical M$_{\textsc{GAS}}$/M$_{\textsc{stars}}$ value for our sample is $\sim$ 10\%. For unbarred spirals, there is a slight preference for later morphological types to have larger M$_{\textsc{GAS}}$/M$_{\textsc{stars}}$. For barred spirals, on the other hand, the gas fraction appears to be mostly constant as a function of morphology except for SB0 galaxies, where M$_{\textsc{GAS}}$/M$_{\textsc{stars}}$ is decidedly lower. 

We find evidence of a trend of decreasing M$_{\textsc{GAS}}$/M$_{\textsc{stars}}$ as a function of M$_{\textsc{stars}}$, consistent with findings by other groups, yet we detect no trend with M$_{\textsc{GAS}}$/M$_{\textsc{stars}}$ as a function of M$_{\textsc{BH}}$. Finally, we find a significant correlation between M$_{\textsc{BH}}$ vs M$_{\textsc{BARY}}$, with similar slope to the recalibrated M$_\textsc{{BH}}$ $-$ M$_\textsc{{stars}}$ relation by \cite{misty2018}. The dwarf Seyfert NGC 4395 (which hosts the lowest reverberation-mapped black hole mass) is significantly more gas dominated than the other galaxies in our study, with M$_{\textsc{GAS}}$/M$_{\textsc{stars}}$ $=$ 134\%, but it appears to follow the same trend in M$_{\textsc{BH}}$ vs M$_{\textsc{BARY}}$ defined by the other galaxies in our sample.

\acknowledgements
We thank the referee for helpful suggestions that
improved the clarity of this paper. We thank Joy Skipper for helpful comments on observing scripts and data reduction.

M.C.B gratefully acknowledges support from the NSF
through CAREER grant AST-1253702. 

H.M.C. acknowledges support by the CNES and Institut Universitaire de France.

This research has made use of the NASA/IPAC Extragalactic Database (NED), which is operated by the Jet Propulsion Laboratory, California Institute of Technology, under contract with the National Aeronautics and Space Administration.

\bibliography{references}

%\cleardoublepage
%\cleardoublepage
%\vspace*{19cm}

\begin{center}
\begin{ThreePartTable}
\begin{TableNotes}
\textbf{Note.} $--$ Columns (1-3): Galaxy names in increasing right ascension. Column (4): Listed redshift value from literature. Column (5): Session number of observation during observing block. Column (6): Scan number range of a given observation session. Column (7): Date. Columns (8-10): Universal time, local standard time, and hour angle values at the midpoint of the observation on each date. Column (11): Backend of instrument used for observation, G denotes GBT Spectrometer, V denotes VEGAS.
\end{TableNotes}

\clearpage
\begin{longtable*}{llllllcllrc}
\caption{\\ Target Observations}\\
\toprule
\multicolumn{1}{c}{Target} & \multicolumn{1}{c}{RA}  & \multicolumn{1}{c}{Dec} & \multicolumn{1}{c}{$z$} & \multicolumn{1}{c}{Session} & \multicolumn{1}{c}{Scans} & \multicolumn{1}{c}{Date} & \multicolumn{1}{c}{UT} &  \multicolumn{1}{c}{LST}  & \multicolumn{1}{c}{Hour} & \multicolumn{1}{c}{Backend}\\
& \multicolumn{1}{c}{(hh mm ss.s)} & \multicolumn{1}{c}{(dd mm ss)} & & \multicolumn{1}{c}{Number} & & \multicolumn{1}{c}{(yyyy-mm-dd)} & \multicolumn{1}{c}{(hh mm ss)} &  \multicolumn{1}{c}{(hh:mm:ss)} & \multicolumn{1}{c}{Angle}\\
\multicolumn{1}{c}{(1)} & \multicolumn{1}{c}{(2)} & \multicolumn{1}{c}{(3)} & \multicolumn{1}{c}{(4)} & \multicolumn{1}{c}{(5)} & \multicolumn{1}{c}{(6)} & \multicolumn{1}{c}{(7)} & \multicolumn{1}{c}{(8)} & \multicolumn{1}{c}{(9)} & \multicolumn{1}{c}{(10)} & \multicolumn{1}{c}{(11)}\\
\midrule
\endhead\hline
\endfoot 
\bottomrule 
%\insertTableNotes 
\endlastfoot
Mrk 1501 & 00 10 31.0 & +10 58 30 & 0.08934 & 2 & 6-52 & 2013-02-05  & 22 57 55.0  & 02 43 24.8 & 2.90 & G\\
& &   &&  4 & 6-162 & 2013-02-08 & 21 14 15.0 &  02 11 00.5 & 1.01 & G\\
& &   && 9 & 6-23 & 2013-02-15 & 18 06 40.0 &  22 30 56.7 & -1.64 & G\\
& &   && 54 & 6-43 & 2013-06-30 & 09 07 41.0 & 22 22 31.8 & -1.74 & G\\
& &   && 55 & 6-33  &2013-07-01 & 09 52 59.0 &  23 11 57.8 & -0.92 & G\\
& &   && 62 & 6-19  & 2013-08-10 &05 53 18.0 &  21 49 25.4 & -2.31 & G\\
& &    && 64 & 6-22  & 2013-08-12 &05 16 04.0 &  21 19 57.3 & -2.83 & G\\
Mrk 1044 & 02 30 05.5 & -08 59 53 & 0.01645 & 38 & 6-132 &2018-09-06 & 09 06 28.5 &  02 48 47.0 & 0.31 & V \\
& & & & 44 & 6-139 & 2018-09-16 &06 59 59.5 &  01 21 22.7 & -1.15 & V\\
3C120 &  04 33 11.1 & +05 21 16 & 0.03301 & 20 & 7-106  &2013-03-13 & 01 37 35.0 &  06 45 10.8 & 2.21 & G\\
&   &  && 23 & 6-90  &  2013-03-27 &23 10 33.0 & 06 13 09.1 & 1.67 & G\\
&   &   && 27 & 1-52 &2013-04-03 & 21 28 57.0 &  04 58 58.4 & 0.44 & G\\
&  &  && 59 & 6-25  & 2013-07-27 &18 18 09.0 &  09 21 04.7 & 4.83 & G\\
&   &  && 65 & 6-20  & 2013-08-30 &17 19 56.0 &  10 36 46.1 & -8.01 & G\\
&  & && 69 & 6-96  & 2013-09-03 &11 16 53.0 &  04 48 00.1 & 0.26 & G\\
Ark 120 & 05 16 11.4 & -00 08 66 & 0.03271 & 45 & 146-277 &2018-09-16 & 12 38 53.5 &  07 01 12.4 & 1.75 & V\\
& & & & 52 & 6-135 &2018-10-16 & 09 20 48.5 &  05 40 51.4 & 0.41 & V\\
MCG+08-11-011 & 05 54 53.6 & +46 26 22 & 0.02048 & 46 & 6-135 &2018-09-20 &  08 15 37.5 & 02 52 59.3 & -3.03 & V\\
& & & & 53 & 6-129 &2018-10-17 & 09 19 05.5 &  05 43 04.7 & -0.20 & V\\
Mrk 6 &  06 52 12.2 & +74 25 37 & 0.01881 & 6 & 6-105  & 2013-02-26 &02 26 11.0 &  07 34 35.3 & 1.88  & G\\
&   &  && 10 & 6-91  & 2013-02-21 & 04 36 55.0 & 09 22 25.2 & 2.52 & G\\
&   &  && 12 & 6-48  & 2013-02-26 &23 09 41.0 &  04 18 05.3 & -2.53 & G\\
Mrk 374 & 06 59 38.1 & +54 11 48 & 0.04263 & 32 & 6-137 &2018-09-01 & 09 22 29.5 &  02 45 07.8 & -4.24 & V\\
& & & & 35 & 6-84 &2018-09-04 & 16 52 10.5 &  10 27 52.4 & 3.47 & V\\
& & & & 36 & 6-50 & 2018-09-05 &11 05 51.5 &  04 44 33.0& -2.25 & V\\
Mrk 79 & 07 42 32.8 & +49 48 35 & 0.02219 & 3 & 32-53  &2013-02-07 &  06 33 43.0 & 10 24 32.6 & 2.72 & G\\
NGC 2617 & 08 35 38.8 & -04 05 18 & 0.01421 & 51 & 6-133 &2018-10-15 & 12 23 45.5 &  08 40 21.9 & 0.08 & V\\
& & & & 55 & 6-135 & 2018-10-25 &13 22 04.5 &  10 18 16.0 & 1.71 & V\\
Mrk 704 & 09 18 26.0 & +16 18 19 & 0.02923 & 37 & 56-181 &2018-09-05 & 14 47 00.5 &  08 26 18.3 & -0.87 & V\\
& & & & 49 & 6-135 &2018-10-14 & 12 13 26.5 &  08 26 04.6 & -0.87 & V\\
& & & & 68 & 8-135 &2019-01-02 & 05 18 08.5 &  06 45 02.8 & -2.56 & V\\
& & & & 71 & 6-59 &2019-01-08 & 08 51 54.5 &  10 43 03.3 & 1.27 & V\\
Mrk 110 & 09 25 12.9 & +52 17 11 & 0.03529 & 31 & 142-275 &2018-08-31 & 11 17 01.5 &  04 36 02.1 & -4.82 & V\\
& & & & 33 & 143-270 & 2018-09-01 &14 42 26.5 &  08 05 57.4 & -1.32 & V\\
NGC 3227 & 10 23 30.6 & +19 51 54 & 0.00386 & 8 & 6-13  &2013-02-15 &  05 56 56.0 & 10 19 14.6 & -0.05 & G\\
Mrk 142 & 10 25 31.3 & +51 40 36 & 0.04494 & 5 & 6-147  & 2013-02-09 & 04 00 06.0 & 11 55 48.0 & 0.27 & G\\
&  & && 11 & 6-91  & 2013-02-23 & 06 16 32.0 &  11 10 10.6 & 0.83 & G\\
NGC 3516 & 11 06 47.5 & +72 34 07& 0.00884 & 3 & 22-31  &2013-02-07 & 05 53 49.0 &  09 44 34.4 & -1.35 & G\\
&  &  && 53 & 6-76  &2013-06-24 & 02 08 21.0 &  14 58 25.5 & 3.87 & G\\
& & & & 41 & 6-137 &2018-09-09 & 02 36 23.5 &  20 29 27.6 & 9.38 & V\\
& & & & 43 & 142-274 & 2018-09-11 & 02 48 40.5 & 20 49 39.7 & 9.71 & V\\
& & & & 61 & 6-119 &2018-12-21 & 02 15 07.5 &  02 54 13.1 & -8.65 & V\\
& & & & 62 & 6-75 &2018-12-21 & 07 35 51.5 &  08 15 49.7 & -3.29 & V\\
& & & & 63 & 6-161 & 2018-12-22 &  00 07 18.5 &00 49 59.6 & -10.28 & V\\
& & & & 64 & 6-81 & 2018-12-22 & 21 28 52.5 & 22 15 04.2 & 11.14 & V\\
& & & & 65-66 & 6-138 &2019-01-01 & 01 19 46.5 &  02 42 05.1 & -8.86 & V\\
& & & & 69 & 6-99 & 2019-01-03 & 23 34 38.5 & 01 08 29.5 & -10.42 & V\\
& & & & 70 & 6-73 &2019-01-05 & 22 07 14.5 &  23 48 44.3 & -11.30 & V\\
& & & & 72 & 7-32 &2019-01-08 & 10 41 31.5 &  12 32 58.3 & 0.99 & V\\
SBS 1116+583A & 11 18 57.7 & +58 03 24 & 0.02787& 29 & 6-84  & 2013-04-05 &01 46 15.0 &  08 24 36.5 & -2.91 & G\\
&  & && 36 & 6-93  &2013-05-06 & 22 28 41.0 &  08 08 51.4 & -3.15 & G\\
&  &  && 39 & 6-69  & 2013-05-09 &21 26 51.0 &  07 18 47.8 & -3.74 & G\\
Arp 151 & 11 25 36.2 & +54 22 57 & 0.02109 & 31 & 66-243  & 2013-04-30 &04 02 57.0 &  13 16 01.1 & 1.85 & G\\
NGC 3783 & 11 39 01.7 & -37 44 19 & 0.00973& 50 & 6-76  &2013-06-01 & 01 07	14.0 &  11 30 14.5 & -1.47 & G\\
UGC 06728 & 11 45 16.0 & +79 40 53 & 0.00652 & 29 & 6-284 &2018-08-16 &  09 54 44.5& 02 14 23.3 & -9.52 & V\\
Mrk 1310 & 12 01 14.3  & -03 40 41 & 0.01956& 59 & 26-98  &2013-07-27 & 20 50 37.0   &  11 53 52.0 & -1.48 & G\\
&  &   && 60 & 36-96 & 2013-07-29 &01 40 05.0 &  16 39 50.4 & 3.04 & G\\
NGC 4051 & 12 03 09.6 & +44 31 53 & 0.00234& 3 & 6-13  &2013-02-07 & 05 18 08.0 &  09 18 48.4& -2.98 & G\\
&  &  && 46 & 36-119  &2013-05-23 & 00 15 46.0 &  10 59 19.0 & -2.72 & G\\
&  &  & &47 & 6-95  &2013-05-25 & 23 36 13.0 &  10 31 28.2 & -3.29 & G\\
&  &  && 49 & 2270-2352  & 2013-05-27 & 23 52 16.0& 10 55 28.1 & -2.78 & G\\
NGC 4151 & 12 10 32.6 & +39 24 19&0.00332 & 3 & 14-21  &2013-02-07 & 05 33 14.0 &  09 23 57.6 & -2.78 & G\\
NGC 4253 & 12 18 26.5 & +29 48 46 &0.01293& 45 & 6-241  & 2013-05-22  &01 52 39.0 &  12 32 47.8 & 0.24 & G\\
&  &  && 49 & 2240-2269 &2013-05-27 & 21 36 54.0 &  08 40 00.1 & -3.64 & G\\
& & & & 47 & 6-109 &  2018-09-25 &12 18 27.3 & 09 20 27.3 & -2.97 & V\\
Mrk 50 & 12 23 24.1 & +02 40 45 & 0.02343 & 48 & 6-133 &2018-10-12 & 14 51 38.5 &  10 56 49.5 & -1.44 & V\\
& & & & 50 & 142-271 &  2018-10-14 &17 51 50.5 & 14 05 24.2 & 1.70 & V\\
PG 1229+204 & 12 32 03.6 & +20 09 28&0.06301 &17 &  6-15  &2013-03-05 & 03 59 27.0 &  09 32 25.6 & -2.99 & G\\
&  &  && 42 & 6-81  &2013-05-19 & 00 52 09.0  &  11 20 18.2 & -1.20 & G\\
&  &  && 43 & 6-99  &2013-05-20 & 03 06 54.0 &  13 39 21.9 & 1.12 & G\\
&  &  && 46 & 6-35  & 2013-05-22 &22 00 35.0 &  08 44 02.2 & -3.80  & G\\
NGC 4593 & 12 39 39.4 & -05 20 39&0.00900 & 38 & 6-77  &2013-05-09 & 03 06 56.0 &  12 56 01.8 & 0.27 & G\\
& &  && 41 & 6-116  &2013-05-18 & 01 07 31.0 &  11 31 46.1 & -1.13 & G\\ 
NGC 4748 & 12 52 12.4 & -13 24 53 &0.01463& 7 & 6-87  & 2013-02-14 &07 27 03.0 &  11 45 41.2 & -1.11 & G\\
&  &  && 37 & 10-93  & 2013-05-08 & 03 56 07.0 & 13 41 24.3 & 0.82 & G\\
&  &  && 44 & 6-36  & 2013-05-21 &01 21 09.0 &  11 57 16.0 & -0.92 & G\\
&  &  && 47 & 96-147  & 2013-05-26 & 02 29 44.0 & 13 25 45.1 &0.56& G \\
MCG-06-30-015 & 13 35 53.7 & -34 17 44 &0.00775& 8 & 14-27  &2013-02-15 & 06 27 25.0 &  10 49 49.9 & -2.77 & G\\
&  &  && 59 & 99-151  &2013-07-27 &  23 13 36.0 & 14 17 27.9 & 0.69 & G\\
&  &  && 60 & 6-35 & 2013-07-28 &23 19 44.0 &  14 27 33.5 & 0.86 & G\\
&  &  && 61 & 6-117 &2013-08-03 & 22 35 46.0 &  14 07 07.6 & 0.52 & G\\
MCG-05-33-019 & 13 49 19.2 & -30 18 34 &0.01514& 24 & 6-86  & 2013-04-01 &07 19 54.0 &  14 39 52.3 & 0.84 & G\\
&  &  && 30 & 6-87 &2013-04-29 &05 24 26.0 &  14 34 28.8 & 0.79 & G\\
& &   && 48 & 6-69  &2013-05-27  & 01 59 39.0 &  12 59 31.7 & -0.83 & G\\
Mrk 279 & 13 53 03.4 & +69 18 31 &0.03045& 13 & 31-82  &2013-03-01 & 00 55 27.0 &  06 12 09.1 & -7.68 & G\\
&  &  && 14 & 6-99  & 2013-03-02 &23 49 04.0 &  05 13 28.3 & -8.66 & G\\
&  &  && 22 & 6-75  & 2013-03-24 & 12 08 09.0 & 18 57 22.3 & 5.07 & G\\
&  &  && 25 & 6-64  &2013-04-01 & 10 24 57.0 &  17 45 25.7 & 3.87 & G\\
&  &  && 31 & 6-65  &2013-04-29 &10 34 25.0 &  19 45 18.7 & 5.87 & G\\
NGC 5548 & 14 17 59.5 & +25 08 12 &0.01718& 28 & 6-102  &2013-04-04 & 09 46 21.0 &  17 18 33.1 & 3.01 & G\\
&  &  & & 33 & 6-75 & 2013-05-01 &07 38 06.0 &  16 56 23.9 & 2.64 & G\\
&  & && 37 & 94-161  & 2013-05-08 &06 55 11.0 &  16 40 57.7 & 2.38& G \\
PG 1426+015 & 14 29 06.6 &  +01 17 06 & 0.08657 & 58 & 6-133 &2018-11-26 &  13 52 30.5 & 12 54 56.7 & -1.57 & V\\
& & & & 59 & 6-133 &  2018-11-27 &16 24 05.5 & 15 30 53.2 & 1.03 & V\\
& & & & 67 & 6-123 &2019-01-01 & 14 37 48.5 &  16 02 18.2 & 1.42 & V\\
Mrk 817 & 14 36 22.1 & +58 47 39 &0.03146& 32 & 7-113  &2013-04-30 & 09 56 12.0 &  19 10 56.0 & 4.58 & G\\
&  &  && 35 & 7-66  &2013-05-06 &  10 06 48.0 & 19 45 13.1 & 5.15 & G\\
&  &  && 40 & 6-98  & 2013-05-11 &07 41 23.0 &  17 39 06.9 & 3.05 & G\\
&  & && 56 & 9-54 & 2013-07-07 &18 22 10.0 &  08 06 22.9 & -6.50 & G\\
Mrk 478 & 14 42 07.5 & +35 26 23 & 0.07906 & 42 & 6-135 &2018-09-10 & 21 23 08.5 &  15 23 14.2 & 0.69 & V\\
& & & & 54 & 6-139 &2018-10-17 & 14 48 36.5 &  11 13 29.8 & -3.48 & V\\
NGC 5940 & 15 31 18.1 & +07 27 28 & 0.03393 & 57 & 6-137 &2018-11-07 & 20 06 12.5 &  17 54 45.6 & 2.39 & V\\
& & & & 60 & 6-135 & 2018-11-28 & 15 17 07.5 & 14 27 40.7 & -1.06 & V\\
Mrk 290 & 15 35 52.3 & +57 54 09 &0.02958& 1 & 6-177  & 2013-02-05 & 13 44 31.0 & 17 28 42.2 & 1.88 & G\\
&  & & &17 & 16-51  &2013-03-05 & 04 54 53.0 &  10 28 00.7 & -5.13 & G\\
Mrk 493 & 15 59 09.6 & +35 01 47 & 0.03133 & 34 & 6-131 &2018-09-02 &20 54 26.5 &  14 22 55.0 & -1.60 & V\\
& & & & 56 & 6-139 &2018-10-30 & 14 48 53.5 &  12 05 02.0 & -3.90 & V\\
3C390.3 & 18 42 09.0 &+79 46 17 &0.05610 & 13 & 6-25  & 2013-02-28 &23 15 04.0 &  04 31 29.6 & 9.82 & G\\
&  &  && 26 & 6-64  &  2013-04-01 &11 30 26.0 & 18 58 58.6 & 0.28 & G\\
&  &  && 52 & 6-70 &2013-06-23 & 13 39 28.0 &  02 27 42.7 & 7.76 & G\\
&  &  && 60 & 97-108  &2013-07-29 & 03 29 26.0 &  18 37 56.5 & -0.07 & G\\
& &  && 63 & 6-21 &2013-08-10 & 21 53 39.0 &  13 52 29.5 & -4.83 & G\\
&  &   && 66 & 6-33 & 2013-08-31 & 04 07 46.0 & 21 26 29.1 & 2.74 & G\\
&  &  && 67 & 6-12 &2013-09-01 & 17 54 36.0 &  11 19 31.4 & -7.38 & G\\
& &  && 68 & 6-19 &2013-09-02 & 21 10 18.0 &  14 39 42.1 & -4.04 & G\\
Zw 229-015 & 19 05 25.9 & +42 27 40 & 0.02788 & 30 & 6-135 &2018-08-31 & 05 52 11.5 &  23 10 18.7 & 4.08 & V\\
& & & & 39 & 6-132 &2018-09-08 & 00 39 51.5 &  18 28 39.9 & -0.61 & V\\
PGC 090334 & 19 37 33.0 & -06 13 05 &0.01031& 18 & 6-79  & 2013-03-10 &15 31 13.0 &  21 25 48.0 & 1.80 & G\\
NGC 6814 & 19 42 40.6 & -10 19 25 & 0.00521 & 40 & 6-55 &2018-09-08 & 04 43 35.5 &  22 33 03.9 & 2.84 & V\\
&  & && 21 & 6-11  &  2013-03-18 &14 15 20.0 & 20 41 14.9 & 1.06 & G\\
PG 2130+099 & 21 32 27.8 & +10 08 19 &0.06298& 15 & 6-87  &2013-03-03 & 15 52 47.0 &  21 19 49.6 & -0.21 & G\\
&  &  && 16 & 6-35  &2013-03-04 & 13 55 03.0 &  19 25 42.8 & -2.11 & G\\
&  &  && 19 & 6-113  & 2013-03-11 &13 51 23.0 &  19 49 38.1 & -1.71 & G\\
&  &  && 58 & 6-37  &2013-07-22 & 04 20 43.0 &  19 01 46.0 & -2.51 & G\\
NGC 7469 & 23 03 15.6 & +08 52 26 &0.01632& 28 & 105-204  &2013-04-04 & 13 51 10.0 &  21 24 02.3 & -1.65 & G\\
\label{1}
\end{longtable*}
\end{ThreePartTable}
\vspace{-50cm}
\begin{figure*}
\end{figure*}
\end{center}
\blfootnote{\parbox[3in]{6.8in}{\vspace{-10.7cm}\textbf{Note.} $--$ Columns (1-3): Galaxy names in increasing right ascension. Column (4): Listed redshift value from literature. Column (5): Session number of observation during observing block. Column (6): Scan number range of a given observation session. Column (7): Date. Columns (8-10): Universal time, local standard time, and hour angle values at the midpoint of the observation on each date. Column (11): Backend of instrument used for observation, G denotes GBT Spectrometer, V denotes VEGAS.}}

%\vspace*{8cm}
%\clearpage

\begin{deluxetable*}{lrrlcc}
\tablecaption{Spectral Characteristics}
\tablewidth{0pc}
\tablehead{
\colhead{Target}&\colhead{Exp Time}&\colhead{S/N}&\colhead{RMS}&\colhead{Final Resolution}&\colhead{Backend}\\
\colhead{}&\colhead{(s)}&\colhead{}&\colhead{(K)}&\colhead{(km s$^{-1}$ chan$^{-1}$)}\\
\colhead{(1)}&\colhead{(2)}&\colhead{(3)}&\colhead{(4)}&\colhead{(5)}
&\colhead{(6)}}
\startdata
Mrk 1044 & 14720.6 & 15.7 & 0.0021 & 1.1 & V\\
Ark 120 & 12684.7 & 9.8 & 0.0015 & 2.4 & V\\
MCG+08-11-011 & 14405.1 & 23.4 & 0.0026 & 0.8 & V\\
Mrk 6&12904.6&8.2&0.0014&3.0&G\\
Mrk 374 & 14163.8 & 4.7 & 0.0008 & 8.8 & V\\
Mrk 79 &1261.4&10.4&0.0038&3.0&G\\
NGC 2617 & 14843.2 & 45.4 & 0.0036 & 0.3 & V\\
Mrk 704 & 20883.8 & 6.5 & 0.0009 & 3.2 & V\\
Mrk 110 & 14160.4 & 8.9 & 0.0014 & 3.2 & V\\
NGC 3227&458.6&13.1&0.0062&3.0&G\\
NGC 3516 & 56214.1 & 8.8 & 0.0006 & 4.0 & V\\
SBS1116+583A&12445.8&4.0&0.0009&6.0&G\\
NGC 3783&4013.1&28.5&0.0059&0.6&G\\
Mrk 1310&7282.3&8.1&0.0017&3.0&G\\
NGC 4051&14965.0&112.0&0.0024&0.6&G\\
NGC 4151&344.1&42.0&0.0151&0.6&G\\
Mrk 766&15252.1&5.1&0.0008&6.0&G\\
NGC 4593&10204.7&19.1&0.0031&0.6&G\\
NGC 4748&14160.5&8.9&0.0013&1.8&G\\
MCG-06-30-015&12096.8&3.4&0.0015&3.0&G\\
Mrk 279&14739.9&8.4&0.0013&3.0&G\\
NGC 5548&13415.8&10.0&0.0011&4.8&G\\
Mrk 817&13879.6&5.8&0.0007&7.2&G\\
Mrk 478 & 13437.6 & 3.4 & 0.0012 & 5.6 & V\\
NGC 5940 & 13917.5 & 10.2 & 0.0020 & 1.3 & V\\
Mrk 290&11126.6&5.4&0.0009&6.0&G\\
Mrk 493 & 13051.6 & 38.9 & 0.0024 & 0.8 & V\\
Zw 229-015 & 12750.2 & 6.8 & 0.0009 & 6.4 & V\\
1H1934-063&4586.8&11.5&0.0027&1.8&G\\
NGC 6814 & 2812.5 & 81.0 & 0.0080 & 0.5 & V\\
NGC 7469&4243.3&4.9&0.0022&3.0&G\\
\enddata
\tablecomments{Column (2) lists the total time spent on source after removal of contaminated scans. Column (3) lists approximate S/N, and the values were calculated as either a) the average value of the peak fluxes of the horns and mid-profile peak fluxes divided by the RMS of the noise or b) the peak value of the Gaussian-shaped profile divided by the RMS of the noise. Column (4) lists values for the root mean square of the noise in each spectra. Column (5) denotes the final velocity resolution per channel after spectral smoothing was applied (other than initial Hanning smoothing). Column (6) lists the backend of instrument used for observation, G denotes GBT Spectrometer, V denotes VEGAS. The default velocity resolutions of the GBT Spectrometer and VEGAS are 0.3 and 0.08 km s$^{-1}$ channel$^{-1}$, respectively. }
\label{characteristics}
\end{deluxetable*}
%\vspace*{8cm}
%\clearpage

\begin{deluxetable*}{lllllllllc}\tablecaption{HI Spectroscopic Measurements}
\tablehead{
\colhead{Target}&\colhead{T$_{\textsc{L}}$}&\colhead{T$_{\textsc{L}}$}&\colhead{W$_{50}$}&\colhead{W$_{50}$}&\colhead{W$_{20}$}&\colhead{W$_{20}$}&\colhead{V$_{\textsc{R}}$}&\colhead{V$_{\textsc{R}}$}&\colhead{Backend}\\
\colhead{}&\colhead{\texttt{gmeasure}}&\colhead{$\textsc{BusyFit}$}&\colhead{\texttt{gmeasure}}&\colhead{$\textsc{BusyFit}$}&\colhead{\texttt{gmeasure}}&\colhead{$\textsc{BusyFit}$}&\colhead{\texttt{gmeasure}}&\colhead{$\textsc{BusyFit}$}\\
\colhead{}&\colhead{(K km s$^{-1}$)}&\colhead{(K km s$^{-1}$)}&\colhead{(km s$^{-1}$)}&\colhead{(km s$^{-1}$)}&\colhead{(km s$^{-1}$)}&\colhead{(km s$^{-1}$)}&\colhead{(km s$^{-1}$)}&\colhead{(km s$^{-1}$)}
}
\startdata
Mrk 1044 &4.95$^{+0.06}_{-0.03}$ &$5.01 \pm 0.05$ &183.2$^{+4.6}_{-2.9}$ & $172.2 \pm 5.2$&196.2$^{+4.3}_{-10.4}$ & $202.9 \pm 3.0$&4910.77$^{+0.69}_{-1.35}$ &$4912.00 \pm 0.90$ & V\\
Ark 120 & 3.43$^{+0.15}_{-0.11}$ &$3.63 \pm 0.76$ &337.3$^{+17.0}_{-13.1}$ & $315.9 \pm	20.8$&344.4$^{+18.4}_{-17.9}$& $372.6 \pm 10.3$ &9806.38$^{+9.22}_{-4.51}$ & $9826.00 \pm 26.96$ & V\\
MCG+08-11-011 & 14.71$^{+0.32}_{-0.19}$ &$14.92 \pm	1.93$ &310.8$^{+6.7}_{-5.1}$ &$293.0 \pm 4.6$ &327.3$^{+12.6}_{-8.6}$ & $322.4 \pm 3.9$&6133.26$^{+1.31}_{-1.04}$ &$6141.00 \pm 4.01$ & V\\
Mrk 6	&3.04   $^{+    0.19   }_{-   0.19}$&$3.18 \pm 0.57$&435.8   $^{+      6.4   }_{-     0.5}$&$440.3 \pm 2.8$&447.7   $^{+      9.7   }_{-      12.9}$&$463.7 \pm 3.4$&5631.35$^{+0.48}_{-2.98}$ & $5621.00 \pm 5.12$ & G\\
Mrk 374 &0.59$^{+0.02}_{-0.01}$ & $0.57 \pm 0.27$&263.8$^{+0.9}_{-1.0}$ & $252.3 \pm 59.9$&276.3$^{+6.6}_{-0.9}$ &$272.8 \pm 31.1$ &13250.00$^{+0.07}_{-0.06}$ &$13240.00 \pm 15.18$ & V\\
Mrk 79	&5.02   $^{+  0.34  }_{-0.33}$&$5.51 \pm 0.85$&154.4 $^{+ 9.7  }_{-     4.5}$&$159.6 \pm 4.2$&160.3  $^{+12.9  }_{-     14.3}$&$182.8 \pm 3.5$&6657.41$^{+4.76}_{-3.88}$ & $6656.00 \pm 2.82$ & G\\
NGC 2617 &18.27$^{+0.65}_{-0.31}$ &$18.51 \pm 0.58$ &126.3$^{+5.9}_{-4.2}$ & $111.6 \pm 0.5$&143.8$^{+12.1}_{-9.7}$ &$134.9 \pm 0.5$ &4265.06$^{+0.63}_{-0.28}$ &$4269.00 \pm 0.33$ & V\\
Mrk 704 & 0.19$^{+0.02}_{-0.02}$ & $0.23 \pm 0.05$ & 46.9$^{+6.6}_{-3.2}$ & $52.2 \pm 14.2$ & 57.8$^{+12.8}_{-12.9}$ & $77.1 \pm	21.1$ & 9525.87$^{+1.60}_{-2.55}$ & $9530.00 \pm 9.03$ & V\\
Mrk 110 &0.95$^{+0.08}_{-0.07}$ &$1.14 \pm 0.74$ &127.2$^{+16.6}_{-19.2}$ & $104.9 \pm 8.6$&145.0$^{+18.8}_{-18.5}$ &$171.2 \pm 10.6$ &10558.90$^{+6.56}_{-9.49}$ &$10570.00 \pm 5.83$ & V\\
NGC 3227	&28.51   $^{+ 0.53 }_{-0.34}$&$29.13 \pm 5.35$&430.1  $^{+10.8}_{-     2.8}$&$135.9 \pm 117.8$&441.2  $^{+ 12.9}_{-     15.4}$&$436.3 \pm 3.8$&1144.74$^{+4.33}_{-0.82}$ & $1192.00 \pm 20.21$ & G\\
NGC 3516 & 0.41	$^{+0.02}_{-0.02}$ & $0.46 \pm	0.02$ & 143.4	$^{+5.4}_{-9.6}$ & $134.6 \pm 7.4$ & 156.5 	$^{+15.4}_{-12.5}$ & $183.9 \pm	10.0$ & 2627.31	$^{+3.76}_{-4.67}$ & $2635.00 \pm 4.25$ & V\\
SBS1116+583A	&0.38   $^{+   0.03   }_{- 0.02}$&$0.42 \pm 0.19$&142.2$^{+     9.3   }_{-      10.5}$&$145.1 \pm 37.4$&148.2   $^{+12.9  }_{- 17.3}$&$167.6 \pm 18.8$&8376.82$^{+7.62}_{-2.28}$ & $8378.00 \pm 10.44$ & G\\
NGC 3783	&20.08   $^{+ 1.14 }_{-0.48}$&$20.34 \pm 1.90$&145.9 $^{+ 9.5 }_{-     2.2}$&$138.0 \pm 1.7$&155.4   $^{+12.4}_{-     9.3}$&$152.5 \pm 1.5$&2916.08$^{+3.88}_{-0.64}$ & $2913.00 \pm 0.92$ & G\\
Mrk 1310	&2.26   $^{+   0.08   }_{-   0.03}$&$2.30 \pm 0.42$&243.2   $^{+     8.0   }_{-     6.8}$&$238.3 \pm 2.1$&244.8   $^{+     9.7   }_{-     10.0}$&$250.5 \pm 2.6$&5837.72$^{+5.55}_{-2.50}$ & $5838.00 \pm 1.27$ & G\\
NGC 4051	&60.69   $^{+0.83}_{-0.41}$&$61.37 \pm 0.17$&247.9  $^{+ 2.5 }_{-     1.1}$&$236.3 \pm 0.1$&264.5   $^{+9.5}_{-2.9}$&$264.5 \pm 0.1$&703.56$^{+0.40}_{-0.08}$ & $703.40 \pm 0.05$ & G\\
NGC 4151	&72.73 $^{+1.26}_{- 0.82}$&$72.57 \pm 1.85$&134.1  $^{+ 1.7 }_{-     1.7}$&$120.3 \pm 0.5$&152.5 $ ^{+ 8.5 }_{-     0.7}$&$139.6 \pm 0.4$&998.56$^{+0.14}_{-0.41}$ & $997.90 \pm 0.32$ & G\\
Mrk 766	&0.37   $^{+    0.02  }_{-     0.01}$&$0.37 \pm 0.23$&120.8   $^{+  7.9   }_{- 1.9}$&$104.2 \pm 14.0$&138.6  $^{+   10.1  }_{-     0.6}$&$126.1 \pm 13.9$&3899.68$^{+0.42}_{-2.95}$ & $3904.00 \pm 2.65$ & G\\
NGC 4593	&13.10  $^{+ 0.49}_{- 0.17}$&$13.20 \pm 1.07$&361.5 $ ^{+ 11.7  }_{-     3.3}$&$357.2 \pm 1.8$&367.0  $^{+ 12.7 }_{-     13.6}$&$367.6 \pm 1.6$&2501.80$^{+1.14}_{-5.46}$ & $2501.00 \pm 1.57$ & G\\
NGC 4748	&2.61 $^{+ 0.09 }_{-0.05}$&$2.68 \pm 0.33$&315.4 $^{+9.4}_{-     2.0}$&$306.9 \pm 5.6$&323.3   $^{+ 12.2}_{-     12.3}$&$332.2 \pm 3.6$&4183.19$^{+3.60}_{-2.25}$ & $4196.00 \pm 10.92$ & G\\
MCG-06-30-015	&0.14   $^{+    0.04   }_{-   0.02}$&$0.18 \pm 0.16$&28.3   $^{+      9.0   }_{-     6.5}$&$18.6 \pm 28.3$&29.0   $^{+      6.4   }_{-      8.3}$&$43.1 \pm 16.1$&2353.53$^{+4.15}_{-3.56}$& $2358.00 \pm 8.98
$ & G\\
Mrk 279	&0.96   $^{+    0.07   }_{-    0.08}$&$1.23 \pm 0.41$&146.1   $^{+     18.1   }_{-      22.7}$&$142.7 \pm 12.0$&154.6   $^{+      19.3   }_{-      19.3}$&$219.5 \pm 16.2$&9211.71$^{+8.29}_{-6.49}$ & $9210.00 \pm 4.94$ & G\\
NGC 5548	&1.77 $^{+0.07}_{-0.06}$&$2.03 \pm 0.93$&189.1 $^{+10.3 }_{-     3.0}$&$197.5 \pm 15.7$&197.3  $^{+11.8}_{-14.0}$&$251.2 \pm 9.3$&5145.78$^{+5.15}_{-1.50}$ & $5159.00 \pm 3.79$ & G\\
Mrk 817	&0.63   $^{+0.03   }_{-0.02}$&$0.67 \pm 0.09$&293.5  $^{+7.8 }_{-     7.8}$&$299.1 \pm 54.5$&303.7  $^{+ 10.0 }_{-     7.8}$&$331.0 \pm 17.7$&9420.14$^{+4.08}_{-3.91}$ & $9438.00 \pm 14.24$ & G\\
Mrk 478 &0.64$^{+0.03}_{-0.01}$ &$0.67 \pm 0.27$ &294.5$^{+11.1}_{-10.6}$ & $296.8 \pm 65.6$&296.9$^{+13.6}_{-12.3}$ &$327.0 \pm 53.7$ &23879.90$^{+5.54}_{-5.26}$ &$23870.00 \pm 24.61$ & V\\
NGC 5940 &3.31$^{+0.04}_{-0.02}$ &$3.30 \pm 0.33$ &188.8$^{+0.7}_{-0.5}$ &$183.7 \pm 1.9$ &204.5$^{+5.9}_{-6.5}$ &$201.8 \pm 1.9$ &10209.40$^{+0.06}_{-0.08}$ &$10210.00 \pm 0.65$ & V\\
Mrk 290	&0.70   $^{+   0.02   }_{-      0.03}$&$0.73 \pm 0.04$&192.9   $^{+     12.8   }_{-    3.0}$&$194.3 \pm 8.3$&219.0   $^{+    12.9   }_{-     11.1}$&$224.9 \pm 7.4$&9087.17$^{+1.33}_{-6.42}$ & $9091.00 \pm 15.51$ & G\\
Mrk 493 &3.25$^{+0.16}_{-0.10}$ &$3.20 \pm 0.13$ &54.6$^{+6.5}_{-8.4}$ & $35.0 \pm 10.5$&74.7$^{+12.1}_{-10.9}$ &$55.0 \pm 23.7$ &9442.21$^{+1.92}_{-0.31}$ &$9442.00 \pm 3.85$ & V\\
Zw 229-015 &0.85$^{+0.05}_{-0.04}$ &$0.92 \pm 0.09$ &195.7$^{+13.0}_{-10.6}$ & $202.8 \pm 21.7$&203.5$^{+16.0}_{-15.5}$ &$219.9 \pm 33.6$ &8316.11$^{+9.39}_{-2.12}$ &$8319.00 \pm 13.78$ & V\\
1H1934-063	&4.39  $^{+    0.08   }_{-    0.04}$&$4.42 \pm 0.45$&163.6   $^{+     0.9   }_{-     0.6}$&$160.3 \pm 2.0$&186.4   $^{+      8.6   }_{-      7.2}$&$180.4 \pm 1.9$&3191.42$^{+0.06}_{-0.09}$ & $3192.00 \pm 1.29$ & G\\
NGC 6814 &54.24$^{+2.12}_{-1.15}$ &$54.42 \pm 0.16$ &89.0$^{+4.2}_{-3.1}$ & $79.2 \pm 0.1$&105.1$^{+11.4}_{-8.4}$ &$96.2 \pm 0.1$ &1562.34$^{+0.36}_{-0.16}$ &$1561.00 \pm 0.04$ & V\\
NGC 7469	& 1.91 $^{+0.31}_{-0.31}$ &$1.95 \pm 0.59$& 192.6 $^{+ 9.8}_{-13.9} $&$184.6 \pm 34.0$&196.2   $^{+12.7   }_{-16.1}$ &$208.8 \pm 16.9$& 4927.87 $^{+4.99}_{-6.79}$ & $4939.00 \pm 8.38$ & G\\
\enddata
\tablecomments{Spectroscopic measurements from \texttt{gmeasure} and $\textsc{BusyFit}$. Asymmetric error bars for the \texttt{gmeasure} measurements are the result of our bootstrap method discussed in Sec. \ref{analysis}. The last column lists the backend used for observation, G denotes GBT Spectrometer, V denotes VEGAS.}
\label{spectra}
\end{deluxetable*}

\begin{deluxetable*}{lllllll}
\tabletypesize{\scriptsize}
\tablecaption{Previous Measurements}
\tablehead{
\colhead{Target}&\colhead{Flux}&\colhead{W$_{50}$}&\colhead{W$_{20}$}&\colhead{V$_\textsc{{R}}$} &\colhead{S/N}& \colhead{Ref}\\
&\colhead{(Jy km s$^{-1}$)$^{a}$ or (K km s$^{-1}$)$^{b}$}&\colhead{(km s$^{-1}$)}&\colhead{(km s$^{-1}$)} &\colhead{(km s$^{-1}$)} & 
}
\startdata
Mrk 1044 & \nodata & \nodata & 489 & 4932 &\nodata &  1\\
& $2.58 \pm 0.16^{[a]}$ &193.9 &  \nodata &4914 & \nodata &  2\\
Ark 120 &$1.51 \pm 0.71^{[a]}$ & $194 \pm 33$ &  $233 \pm 50$ &$9807 \pm 17$ & 3.8 &  3\\
& 1.51$^{[a]}$ &$97 \pm 80$ &  \nodata &9740 & 2.1 &  4\\
& 1.965$^{[a]}$ &\nodata &  $370.3 \pm 6.8$ &$9809.2 \pm 3.4$ & \nodata &  5\\
MCG+08-11-011 \hspace{0.5cm} & \nodata \hspace{1.5cm} & \nodata \hspace{2cm} & \nodata \hspace{1.5cm} &6146 \hspace{2cm} & \nodata \hspace{0.5cm} &  1\\
&9.53$^{[a]}$ & $310 \pm 15$ &  \nodata &6133 & 8.5 &  6\\
Mrk 374 & $8.54 \pm 1.94^{[a]}$ &$74 \pm 16$ &  $121 \pm 24$ &\nodata & 6.9 &  7\\
Mrk 79 &3.94$^{[a]}$ & $169 \pm 15$ &  \nodata &6657 & 8.5 &  6\\
& 3.95 $\pm$ 0.46$^{[a]}$ &$155 \pm 7$ & \nodata &6659 $\pm$ 5 & 12.1 &  6\\
NGC 2617 & \nodata &115.1 &  138.0 &$4208 \pm 8$ & \nodata &  8\\
& 7.02$^{[a]}$ &$120 \pm 100$ &  \nodata &4267 & 10.1 &  9\\
& 9.3$^{[a]}$ &112.4 &  143.0 &4265.2 & \nodata &  10\\
& 9.3$^{[a]}$ &112.4 &  \nodata &4264.0 & \nodata &  11\\
Mrk 704 & 0.2$^{[a]}$ &\nodata &  250 &9510 & \nodata &  12\\
NGC 3227 & 15.495$^{[a]}$ & \nodata &  $453.4 \pm 6.8$ &$1135.6 \pm 3.4$ & \nodata &  5\\
& 13.1$^{[a]}$ & 430 & \nodata & 1050 $\pm$ 20 & \nodata & 13\\ 
& \nodata &103 &  437 &$1146 \pm 5$ &  \nodata & 14\\
& 10.5$^{[b]}$ &\nodata &  234$^{[c]}$ & 1284 $\pm$ 9 &\nodata &  15\\
& 14$^{[a]}$ & \nodata & \nodata &  1148 &\nodata & 16\\
& \nodata & \nodata & \nodata &1152 $\pm$ 25& \nodata &  17\\
& \nodata & \nodata & 526 &1146 & \nodata &  1\\
& \nodata & \nodata & $293 \pm 442$ & 1183 &\nodata &  18\\
NGC 3783 &8.45$^{[a]}$ & $145 \pm 5$ &  $151 \pm 30$ &2889 & 15.50 &  6\\
&\nodata & 145 &  159 &2901 $\pm$ 20 & \nodata &  6\\
& 8.83$^{[a]}$ &$147 \pm 13$ &  \nodata &\nodata & 10.2 &  9\\
& \nodata & \nodata & $154 \pm 7$ & 2917 & \nodata & 19\\
NGC 4051 & 39.8$^{[b]}$ &\nodata &  268$^{[c]}$ &706 $\pm$ 9 & \nodata &  15\\
&30.82$^{[a]}$ & $246 \pm 8$ &  \nodata & 704 & 46.5 & 4\\
& \nodata & \nodata & $267 \pm 8$ &$704 \pm 7$  & \nodata &  20\\
NGC 4151 &46.0$^{[b]}$ & \nodata &  156$^{[c]}$ &999 $\pm$ 9 & \nodata &  15\\
& \nodata & \nodata & $142 \pm 6$ &996 & \nodata &  21\\ 
NGC 4593 & 11.1 $\pm$ 1$^{[a]}$ &$358 \pm 10$ &  $378 \pm 14$ &2499 $\pm$ 5 & \nodata &  22\\
& $7.55 \pm 0.36^{[a]}$ &355.5 &  \nodata &2531 &  \nodata & 2\\
NGC 5548 &1.384$^{[a]}$ & \nodata &  $321.1 \pm 6.8$ &$5169.8 \pm 3.4$ & \nodata &  5\\
& \nodata &$110$ &  \nodata & 5200 $\pm$ 20 & \nodata & 13\\
& \nodata & \nodata & 472 &5142 & \nodata &  1\\
& \nodata &$218 \pm 25$ &  \nodata &5152 & 4.1 &  23\\
&1.73 $\pm$ 0.1$^{[a]}$ & 303 $\pm$ 15 &  \nodata &5093 & 9.6 &  24\\
Mrk 478 & 0.48 $\pm$ 0.07$^{[a]}$ &395 $\pm$ 26 &  477 $\pm$ 39 &23406 $\pm$ 13 & 4.81 &  25\\
NGC 5940 &1.828$^{[a]}$ & 187 &  240 &10210 $\pm$ 3 & 9.3 &  26\\
& \nodata & \nodata & 215 &10205 &  \nodata & 1\\
&1.729$^{[a]}$ & 181.4 &  199.1 &10211 & 11.6 &  27\\
& 1.80 $\pm$ 0.08$^{[a]}$ &189 $\pm$ 3 &  \nodata &10207 & 12.2 &  28\\
& 1.36$^{[a]}$ &183 $\pm$ 19 &  \nodata & 10214 &3.4 &  6\\
& \nodata & \nodata & \nodata &10203 $\pm$ 32 & \nodata &  8\\
Mrk 493 & \nodata & \nodata & 60 &9442 &  \nodata & 1\\
& 1.61$^{[a]}$ & 67 & \nodata &9430 & \nodata &  29\\
&1.398$^{[a]}$ & 35.7 &  59.8 & 9443 &10.4 &  27\\
1H1934-063 & \nodata & \nodata & \nodata &$3070 \pm 7$ & \nodata &  8\\
NGC 6814 & 29.5$^{[a]}$ &\nodata &  94 & 1565 $\pm$ 8 &\nodata &  30\\
& \nodata & \nodata & 134 & 1561 & \nodata & 1\\
&37.3 $\pm$ 4.0$^{[a]}$ & 82 &  105 &1563 $\pm$ 2 &  \nodata & 31\\
& 33.68$^{[a]}$ &86 $\pm$ 8 &  \nodata &1563 & 40.2 &  6\\
& \nodata & \nodata & 124 $\pm$ 5 & 1562 $\pm$ 5 & \nodata & 18\\
NGC 7469 & 1.85 $\pm$ 0.2$^{[a]}$ &306 &  \nodata &4877 & 15.8 &  6\\
& 3.741$^{[a]}$ &\nodata &  $525.1 \pm 6.8$ &$4899.5 \pm 3.4$ & \nodata &  5\\
& \nodata & 570 & \nodata &5200 $\pm$ 20 & \nodata &  13\\
& 1.90$^{[a]}$ &$358 \pm 100$ &  \nodata &  4860 &3.8 & 4\\
& \nodata &515 &  583 &$4971 \pm 41$ & \nodata &  32\\
& \nodata & \nodata & 395 &4916 & \nodata &  1\\
\enddata
\tablecomments{References are as follows:
1. \cite{mw1984},
2. \cite{kh2009},
3. \cite{theureau2005},
4. \cite{edd},
5. \cite{ho2008_data},
6. \cite{edd2005}, 
7. \cite{dc2004},
8. \cite{hyperleda}, 
9. \cite{theureau2006},
10. \cite{meyer2004},
11. \cite{doyle2005},
12. \cite{hutchings1989},
13. \cite{bf1979}, 
14. \cite{vandriel2001}, 
15. \cite{dr1978}, 
16. \cite{ds1983}, 
17. \cite{peterson1979}, 
18. \cite{hr1989}, 
19. \cite{theureau1998}, 
20. \cite{tf1981}, 
21. \cite{tc1988},
22. \cite{sd1987}, 
23. \cite{stierwalt2005}, 
24. \cite{haynes2013},
25. \cite{teng2013},
26. \cite{lewis1983},
27. \cite{lewis1987},
28. \cite{haynes2011},
29. \cite{hg1984},
30. \cite{shostak1978},
31. \cite{koribalski2004}
32. \cite{rh1982}. 
\footnotetext{\scriptsize Flux measured in Jy km s$^{-1}$.}
\footnotetext{\scriptsize Flux measured in K km s$^{-1}$} 
\footnotetext{\scriptsize Width corrected for resolution of instrument, and defined as half-width at one-quarter peak intensity. Displayed as double the original value.}
}
\label{previous}
\end{deluxetable*}

%\clearpage

\begin{deluxetable*}{llllllll}
%\tabletypesize{\medium}
\tablecaption{$z$ Comparisons}
\tablehead{
\colhead{Target} & \colhead{$z$ (HI)} & \colhead{$z$ (HI)} & \colhead{Ref} & \colhead{$z$ (Opt)} & \colhead{$z$ (IR)} & \colhead{$z$ (UV)}& \colhead{Ref}\\
& \colhead{This Work} & \colhead{Lit} & & \colhead{Lit} & \colhead{Lit} & \colhead{Lit}\\
\colhead{(1)} & \colhead{(2)} & \colhead{(3)} & \colhead{(4)} & \colhead{(5)} & \colhead{(6)} & \colhead{(7)} & \colhead{(8)}
}
\startdata
Mrk 1044 & 0.01638 & \nodata & \nodata & 0.01621 & \nodata & 0.01600& 10,11\\
Ark 120 & 0.03271 & 0.03271 & 1 & 0.03312 & \nodata & \nodata & 12\\
MCG+08-11-011 & 0.02046 & 0.02046 & 2 & 0.02064 & \nodata & \nodata & 12\\
Mrk 6 & 0.01878 & 0.01881 & 3 & 0.01701 - 0.01975 & \nodata & \nodata & 13\\
Mrk 374 & 0.04420 & 0.04263 & 4 & 0.04385 & \nodata & \nodata & 14\\
Mrk 79 & 0.02221 & 0.02221 & 2 & 0.02192 - 0.02242 & 0.02220 & \nodata & 13,15\\
NGC 2617 & 0.01423 & 0.01421 & 5 & 0.01432 & \nodata & \nodata & 13\\
Mrk 704 & 0.03177 & \nodata & \nodata & 0.02923 - 0.02991 &
0.02900 & 0.02900 & 16,12,15,11\\
Mrk 110 & 0.03522 & \nodata & \nodata & 0.03529 & \nodata & \nodata & 17\\
NGC 3227 & 0.00382 & \nodata & \nodata & 0.00371 - 0.00383 & 0.00400 & \nodata & 18,19,15\\
NGC 3516 & 0.00876 & \nodata & \nodata & 0.00872 - 0.00884 & 0.00900 & \nodata & 18,17,15\\
SBS1116+583A & 0.02794 & \nodata & \nodata & 0.02788 & \nodata & \nodata & 20\\
NGC 3783 & 0.00973 & 0.00973 & 6 & 0.00851 - 0.01022 & 0.00970 & \nodata & 13,15\\
Mrk 1310 & 0.01947 & \nodata & \nodata & 0.01956 - 0.02000 & \nodata & \nodata & 21,22\\
NGC 4051 & 0.00235 & 0.00234 & 7 & 0.00209 - 0.00235 & 0.00200 &\nodata & 18,19,15\\
NGC 4151 & 0.00333 & 0.00333 & 8 & 0.00319 - 0.00320 & 0.00300 & \nodata & 23,18,15\\
Mrk 766 & 0.01301 & \nodata & \nodata & 0.01293 - 0.01300 & 0.01293 - 0.01300 & \nodata & 24,22,25,15\\
NGC 4593 & 0.00835 & \nodata & \nodata & 0.00797 - 0.00900 & 0.00900 & \nodata & 13,26\\
NGC 4748 & 0.01395 & \nodata & \nodata & 0.01463 & 0.01500 & \nodata & 27,15\\
MCG-06-30-015 & 0.00785 & \nodata & \nodata & 0.00775 & 0.00775 - 0.00800 & \nodata & 28,29,15\\
Mrk 279 & 0.03073 & \nodata & \nodata & 0.02970 & 0.03025 - 0.03045 & 0.03050 & 30,31,32,33\\
NGC 5548 & 0.01716 & 0.01699 - 0.01727 & 3 & 0.01645 - 0.01651 & 0.01700 - 0.01717 & 0.01720 & 34,35,20,15,31,33\\
Mrk 817 & 0.03142 & \nodata & \nodata & 0.03120 & 0.03100 - 0.03146 & 0.03130 & 30,15,32,33\\
Mrk 478 & 0.07965 & \nodata & \nodata & 0.07500 - 0.07906 & 0.07700 & 0.07700 & 36,12,37,11\\
NGC 5940 & 0.03405 & 0.03408 & 2 & 0.03369 - 0.03400 & & 0.03400 & 38,39,11\\
Mrk 290 & 0.03031 & \nodata & \nodata & 0.03023 - 0.03040 & 0.03000 - 0.03062 & 0.02960 & 40,20,37,31,33\\
Mrk 493 & 0.03150 & \nodata & \nodata & 0.03131 - 0.03133 & \nodata & 0.03100 & 12,16,11\\
Zw 229-015 & 0.02774 & \nodata & \nodata & 0.02660 - 0.02788 & \nodata & \nodata & 41,42\\
1H1934-063 & 0.01065 & 0.01025 & 5 & 0.01060 & 0.01059 & \nodata & 43,26\\
NGC 6814 & 0.00521 & 0.00521 - 0.00522 & 9,2 & 0.00479 - 0.00503 & 0.00567 & \nodata & 18,12,44\\
NGC 7469 & 0.01644 & 0.01627 & 3 & 0.01580 - 0.02000 & 0.01600 & \nodata & 45,16,15\\
\enddata
\tablecomments{The uncertainties on the redshift measurements ranged from $\sim$0.002\%-0.15\%. Column (4) denotes the references for $z$ from HI analysis and are as follows:
1. \cite{theureau2005},
2. \cite{edd2005},
3. \cite{gallimore1999},
4. \cite{dc2004},
5. \cite{hyperleda},
6. \cite{theureau1998},
7. \cite{verheijen2001}, 
8. \cite{wolfinger2013}, 
9. \cite{koribalski2004}. Column (8) denotes the references for all other $z$ values and are as follows:
10. \cite{huchra1993}, 
11. \cite{monroe2016}, 
12. \cite{dv1991},
13. NED, 
14. \cite{rines2000},
15. \cite{caballero2012}, 
16. \cite{falco1999},
17. \cite{keel1996},
18. \cite{humason1956}, 
19. \cite{hakobyan2012}, 
20. \cite{oh2015}, 
21. \cite{ho2009}, 
22. \cite{stocke1991}, 
23. \cite{wong2008}, 
24. \cite{ramella1995},
25. \cite{smith1987}, 
26. \cite{strauss1992}, 
27. \cite{maza1989}, 
28. \cite{kaldare2003}, 
29. \cite{fisher1995}, 
30. \cite{op1987}, 
31. \cite{mendoza2015}, 
32. \cite{sh1988}, 
33. \cite{tilton2012}, 
34. \cite{haynes2011}, 
35. \cite{humason1956}, 
36. \cite{richards2009}, 
37. \cite{shi2014}, 
38. \cite{rines2016}, 
39. \cite{wu2010},
40. \cite{argudo2015}, 
41. \cite{smith2015},
42. \cite{proust1995},
43. \cite{panessa2011}, 
44. \cite{riffel2013},
45. \cite{joshi2012}.}
\label{z}
\end{deluxetable*}

%\clearpage

\begin{deluxetable*}{lllclclll}\tablecaption{Mass Estimates}
\tablehead{
\colhead{Target} 
& \colhead{Morphology}
& \colhead{D$_\textsc{{L}}$} 
& \colhead{Ref}
&\colhead{Log M$_\textsc{{BH}}$}
&\colhead{Ref}
&\colhead{Log M$_\textsc{{stars}}$}  
&\colhead{Log M$_\textsc{{GAS}}$}
&\colhead{Log M$_\textsc{{BARY}}$}\\
& & \colhead{(Mpc)} 
& \colhead{} 
& \colhead{(M$_{\odot}$)} 
& \colhead{}
& \colhead{(M$_{\odot}$)} 
& \colhead{(M$_{\odot}$)}
& \colhead{(M$_{\odot}$)}\\
\colhead{(1)} & \colhead{(2)} & \colhead{(3)} & \colhead{(4)} & \colhead{(5)} & \colhead{(6)} & \colhead{(7)} & \colhead{(8)} & \colhead{(9)}
}
\startdata
Mrk 1044 & SB(s)c$^{b}$ & 69.1 $\pm$ 7.0 & 1 & 6.71 $^{+0.12}_{-0.10}$ & 7 & 9.88 $\pm$ 0.51 & 9.60 $^{+0.08}_{-0.08}$ & 10.06 $\pm$ 0.39\\
Ark 120 & Sb pec$^{a}$ & 139.6 $\pm$ 7.1 & 2 & 8.07 $^{+0.05}_{-0.06}$ & 8 & 11.04 $\pm$ 0.23 & 10.05 $^{+0.05}_{-0.04}$ & 11.08 $\pm$ 0.21\\
MCG+08-11-011 & SBc$^{b}$ & 86.5 $\pm$ 7.0 & 1 & 7.43 $^{+0.15}_{-0.15}$ & 9 & 11.37 $\pm$ 0.50 & 10.27 $^{+0.07}_{-0.07}$ & 11.40 $\pm$ 0.48\\
Mrk 6 & Sb$^{b}$ & 80.6 $\pm$ 7.1& 2 & 8.10 $^{+0.04}_{-0.04}$ & 8 & 10.68 $\pm$ 0.23& 9.52 $^{+0.07}_ {-0.07}$ & 10.71  $\pm$ 0.22\\
Mrk 374 & SBc$^{b}$ & 190.2 $\pm$ 7.2 & 1 & 7.30 $^{+0.31}_{-0.31}$ & 9 & 10.85 $\pm$ 0.50 & 9.55 $^{+0.03}_{-0.03}$ & 10.87 $\pm$ 0.49\\
Mrk 79 & SBb$^{a}$ & 94.0 $\pm$ 7.2& 2 & 7.61 $^{+0.11}_{-0.14}$ & 8 & 10.69 $\pm$ 0.23 &9.87 $^{+0.07}_ {-0.07}$ & 10.75 $\pm$ 0.21\\
NGC 2617 & Sc$^{b}$ & 59.9 $\pm$ 7.0 & 1 & 7.49 $^{+0.14}_{-0.14}$ & 9 & 10.47 $\pm$ 0.51 & 10.04 $^{+0.09}_{-0.09}$ & 10.61 $\pm$ 0.42\\
Mrk 704 & SBab$^{b}$ & 135.5 $\pm$ 7.1 & 1 & 7.61 $^{+0.06}_{-0.06}$ & 10 & 11.10 $\pm$ 0.40 & 8.77 $^{+0.06}_{-0.06}$ & 11.10 $\pm$ 0.40\\
Mrk 110 & Sc$^{b}$ & 150.9 $\pm$ 7.1 & 2 & 7.29 $^{+0.10}_{-0.10}$ & 8 & 10.69 $\pm$ 0.23 & 9.56 $^{+0.05}_{-0.05}$ & 10.72 $\pm$ 0.22\\
NGC 3227 & SAB(s) pec$^{a}$ & 16.1 $\pm$ 2.4& 3 & 6.78 $^{+0.08}_{-0.11}$ & 8 & 11.03 $\pm$ 0.23 &9.09 $^{+0.11}_ {-0.11}$ & 11.04 $\pm$ 0.23\\
NGC 3516 & (R)SB(s)$^{a}$ & 37.1 $\pm$ 7.0 & 2 & 7.40 $^{+0.04}_{-0.06}$ & 8 & 10.52 $\pm$ 0.25 & 7.97 $^{+0.14}_{-0.14}$ & 10.52 $\pm$ 0.25\\
SBS1116+583A & SBc$^{b}$ & 118.5 $\pm$ 7.1& 2 & 6.56 $^{+0.08}_{-0.09}$ & 8 & 10.38 $\pm$ 0.23 &8.96 $^{+0.06}_ {-0.06}$ & 10.40 $\pm$ 0.22\\
NGC 3783 & (R')SB(r)a$^{a}$ & 25.1 $\pm$ 5.0& 4 & 7.37 $^{+0.08}_{-0.08}$ & 8 & 10.48 $\pm$ 0.24 &9.33 $^{+0.15}_ {-0.15}$ & 10.51 $\pm$ 0.23\\
Mrk 1310 & Sbc$^{b}$ & 82.7 $\pm$ 7.0& 2 & 6.21 $^{+0.07}_{-0.09}$ & 8 & 9.98 $\pm$ 0.23 &9.41 $^{+0.07}_{-0.07}$ & 10.08 $\pm$ 0.19\\
NGC 4051 & SAB(rs)bc$^{a}$ & 9.8 $\pm$ 3.4& 4 & 6.13 $^{+0.12}_{-0.16}$ & 8 & 10.13 $\pm$ 0.25 &8.99 $^{+0.23}_ {-0.23}$ & 10.16 $\pm$ 0.24\\
NGC 4151 & (R')SAB(rs)ab$^{a}$ & 13.9 $\pm$ 3.3& 4 & 7.56 $^{+0.05}_{-0.05}$ & 8 & 10.40 $\pm$ 0.25 &9.37 $^{+0.17}_ {-0.17}$ & 10.44 $\pm$ 0.23\\
Mrk 766 & SBc$^{b}$ & 54.4 $\pm$ 7.0& 2 & 6.82 $^{+0.05}_{-0.06}$ & 8 & 10.18 $\pm$ 0.24 &8.27 $^{+0.10}_ {-0.10}$ & 10.19 $\pm$ 0.24\\
NGC 4395 & SA(s)m$^{a}$ & 4.1 $\pm$ 0.4 & 5 & 5.45 $^{+0.13}_{-0.15}$ & 8 & 9.08 $\pm$ 0.41 & 9.21 $^{+0.08}_{-0.08}$ & 9.45 $\pm$ 0.23\\
NGC 4593 & (R)SB(rs)b$^{a}$ & 37.7 $\pm$ 7.5& 4 & 6.88 $^{+0.08}_{-0.10}$ & 8 & 10.83 $\pm$ 0.25 &9.50 $^{+0.15}_ {-0.15}$ & 10.85 $\pm$ 0.24\\
NGC 4748 & Sab$^{b}$ & 61.6 $\pm$ 7.0& 2 & 6.41 $^{+0.11}_{-0.18}$ & 8 & 10.46 $\pm$ 0.24 &9.22 $^{+0.09}_ {-0.09}$ & 10.48 $\pm$ 0.23\\
MCG-06-30-015 & S0$^{b}$ & 25.5 $\pm$ 3.5& 4 & 6.20 $^{+0.35}_{-0.35}$ & 8 & 10.02 $\pm$ 0.22 &7.17 $^{+0.15}_{-0.12}$ & 10.02 $\pm$ 0.22\\
Mrk 279 & S0$^{a}$ & 129.7 $\pm$ 7.1& 2 & 7.44 $^{+0.10}_{-0.13}$ & 8 & 11.07 $\pm$ 0.23  & 9.43 $^{+0.05}_ {-0.05}$ & 11.08 $\pm$ 0.23\\
NGC 5548 & (R')SA(s)0/a$^{a}$ & 75.0 $\pm$ 7.3& 6 & 7.72 $^{+0.02}_{-0.02}$ & 8 & 11.10 $\pm$ 0.23 &9.22 $^{+0.08}_ {-0.08}$ & 11.11 $\pm$ 0.23\\
Mrk 817 & SBc$^{a}$ & 134.2 $\pm$ 7.1& 2 & 7.59 $^{+0.06}_{-0.07}$ & 8 & 10.97 $\pm$ 0.23 &9.28 $^{+0.05}_ {-0.04}$ & 10.98 $\pm$ 0.23\\
Mrk 478 & Sab$^{b}$ & 351.6 $\pm$ 7.4 & 1 & 7.40 $^{+0.18}_{-0.18}$ & 11 & 11.15 $\pm$ 0.50 & 10.12 $^{+0.03}_{-0.02}$ & 11.19 $\pm$ 0.47\\
NGC 5940 & SBc$^{b}$ & 145.5 $\pm$ 7.1 & 1 & 7.04 $^{+0.07}_{-0.06}$ & 12 & 11.06 $\pm$ 0.40 & 10.07 $^{+0.04}_{-0.04}$ & 11.10 $\pm$ 0.37\\
Mrk 290 & S0$^{b}$ & 130.0 $\pm$ 7.3& 6 & 7.28 $^{+0.06}_{-0.06}$ & 8 & 10.52 $\pm$ 0.40 & 9.30 $^{+0.05}_ {-0.05}$ & 10.54 $\pm$ 0.38\\
Mrk 493 & SB(r)c$^{b}$ & 134.3 $\pm$ 7.1 & 1 & 6.41 $^{+0.04}_{-0.09}$ & 7 & 10.44 $\pm$ 0.50 & 9.99 $^{+0.05}_{-0.05}$ & 10.57 $\pm$ 0.41\\
Zw 229-015 & (R)SBc$^{b}$ & 120.2 $\pm$ 7.2 & 2 & 6.91 $^{+0.08}_{-0.12}$ & 8 & 10.32 $\pm$ 0.23 & 9.31 $^{+0.05}_{-0.05}$ & 10.36 $\pm$ 0.21\\
1H1934-063 & Sbc$^{b}$ & 45.2 $\pm$ 7.0& 1 & 6.40 $^{+0.17}_{-0.20}$ & 11 & 10.53 $\pm$ 0.21 &9.18 $^{+0.12}_{-0.12}$ & 10.55 $\pm$ 0.20\\
NGC 6814 & SAB(rs)bc$^{a}$ & 21.8 $\pm$ 7.0 & 2 & 7.04 $^{+0.06}_{-0.06}$ & 8 & 10.34 $\pm$ 0.29 & 9.64 $^{+0.22}_{-0.22}$ & 10.42 $\pm$ 0.26\\
NGC 7469 & (R')SAB(rs)a$^{a}$ & 68.8 $\pm$ 7.0& 2 & 6.96 $^{+0.05}_{-0.05}$ & 8 & 10.88 $\pm$ 0.23 &9.18 $^{+0.10}_ {-0.10}$ & 10.89 $\pm$ 0.23\\
\enddata
\tablecomments{Mass estimates and morphologies for the AGNs in this study. Morphological classifications in column (2) are from NED or the B/T ratios from the results of the surface brightness decomposition parameters from \cite{misty2009a}, \cite{misty2013}, \cite{misty2016}, and \cite{misty2018}. Classifications were assigned to the B/T values based on Figure 6 of \cite{kent1985} (see Sec.\ \ref{gas/stars}). Column (3) lists the distances employed for each galaxy and are described in Sec.\ \ref{distances}. The references listed in column (4) are for the sources of the distance values and are as follows: 
1. estimated from redshift of HI emission line; this work,
2. \cite{misty2018},
3. \cite{tonry2001},
4. \cite{helene_edd},
5. \cite{thim2004},
6. \cite{misty2013}. 
Column (5) lists the SMBH values of each galaxy and are discussed in Sec.\ \ref{bhm}. The references listed in column (6) are for the sources of the M$_\textsc{{BH}}$ values and are as follows:
7. mass calculated with $\tau_{H\beta}$ by \cite{hu2015}, $\sigma_{line}$ by \cite{du2016}, and scaled with $\langle f \rangle = 4.3$,
8. the AGN Black Hole Mass Database \citep{bhdatabase},
9. the virial mass from \cite{Fausnaugh2017} scaled with $\langle f \rangle = 4.3$,
10. the virial mass from \cite{derosa2018} scaled with $\langle f \rangle = 4.3$,
11. preliminary mass estimate from in-hand reverberation mapping data,
12. mass calculated from the $\sigma_{line}$ by \cite{barth2015}, $\tau_{H\beta}$ by \cite{barth2013}, and scaled with $\langle f \rangle = 4.3$. M$_\textsc{{stars}}$ estimates are listed in column (7) and the calculations are described in Sec.\ \ref{star_bary_masses}.
The calculations for M$_\textsc{{GAS}}$ estimates listed in column (8) are described in Sec.\ \ref{hi_gas_mass}. M$_\textsc{{BARY}}$ values in column (9) were calculated as M$_\textsc{{GAS}}$ + M$_\textsc{{stars}}$. 
\footnotetext{Classification from NED.}
\footnotetext{Derived classification from surface brightness decompositions.}
}
\label{masses}
\end{deluxetable*}

\end{document}